\documentclass[a4paper,11pt]{article}

\usepackage{jcappub} 
\usepackage{caption}
\usepackage{subcaption}
\usepackage{wrapfig}
\usepackage{aas_macros}
\usepackage[skip=0.5ex]{subcaption}
\usepackage{mathbbol}
\usepackage{slashed}
\usepackage{xcolor}
\usepackage{enumitem}
\title{\boldmath Signatures of composite dark matter in the Cosmic Microwave Background spectral distortions}

\author{Anoma Ganguly,}
\author{Rishi Khatri,}
\author{Tuhin S. Roy}

\affiliation{Department of Theoretical Physics, Tata Institute of Fundamental Research,\\Homi Bhabha Road,
	Mumbai 400005, India}
\emailAdd{anoma@theory.tifr.res.in}
\emailAdd{khatri@theory.tifr.res.in}
\emailAdd{tuhin@theory.tifr.res.in}
\abstract{We compute the spectral distortions of the Cosmic Microwave Background (CMB) created by an exotic process that extracts or injects photons of a particular frequency into the CMB. Such signatures are a natural prediction of a class of composite dark matter models characterized by electrically neutral states but with non-zero higher order electromagnetic moments. We consider a simplified model where dark matter exists as a two state system separated by a fixed transition frequency, which can range from radio waves to gamma rays. The electromagnetic transitions between the two states due to CMB photons give rise to thermal distortion, namely, the $\mu$-type distortion in the redshift range $10^5\lesssim z \lesssim 2\times 10^6$ and the $y$-type distortion as well as non-thermal distortions at redshifts $z \lesssim 10^5$. The nature of spectral distortions depends sensitively on the dark matter transition frequency and the strength of couplings of dark matter with visible sector particles as well as its self-interactions, thus opening a new window to probe the nature of dark matter. Non-thermal distortions have unique spectral shapes making them distinguishable from the standard $\mu$ and $y$-type distortions and  potentially detectable in the next-generation experiments such as Primordial Inflation Explorer (PIXIE). We also find that the spectral distortion limits from the COsmic Background Explorer/Far-Infrared Absolute Spectrophotometer (COBE/FIRAS) already give a constraint on the electromagnetic coupling of dark matter which is three orders of magnitude stronger compared to the current direct detection limits for $\sim$ MeV mass dark matter with transition energy in $\sim 1$-$10$ eV range.}
\begin{document}
	{\tiny {\tiny}}	\maketitle
	\flushbottom
	
	\section{Introduction}
	\label{sec:intro}
     Dark matter plays an indispensable role in the success of the standard Lambda Cold Dark Matter ($\Lambda$CDM) model of cosmology. While a large set of observations ranging from galactic \cite{1970ApJ...159..379R,2009GReGr..41..207Z,2006ApJ...648L.109C} to cosmological scales \cite{2020A&A...641A...6P,2022PhRvD.105b3520A} have confirmed the existence of dark matter through gravity,  \emph{all} of our efforts to detect dark matter through non-gravitational interactions have returned empty-handed. Even though time and again interesting anomalies have been reported, which have given rise to excitement and creativity, resulting in numerous models of dark matter interactions, any definitive indication for particle physics interactions of dark matter is yet to be seen.  It is tempting to immediately conclude that all these constraints from various direct-indirect experiments/observables when considered together imply rather vanishingly small couplings between the dark matter and the visible sector. However, one must be cautious in drawing such conclusions since even depicting results from different experiments in a single graph should be done with great prudence.  To be specific, one needs to think of inherent uncertainties associated with each of these experiments, which often gets downplayed since such uncertainties, once taken into consideration, simply downgrade the importance of these results.  Take, for example, the earth-based direct-detection experiments,  where null results get translated into bounds on a product of the dark-visible cross-section and the local dark matter density in the vicinity of the earth.   After taking various simplifying assumptions (such as the standard halo model which assumes an isotropic and isothermal spherical dark matter halo \cite{2017JPhG...44h4001G,2019PhRvD.100b3002M}) one arrives at the local dark matter density and velocity distribution, and thereafter, obtains the direct-detection bounds on dark-visible cross-sections. All such exercises are fine except one should not discount the fact that this lack of observations in the solar neighbourhood can, in principle, simply refer to an underabundance of dark matter in the solar system. On the other hand, conclusions based on the Cosmic Microwave Background (CMB) data, or conclusions based on all sky and/or redshift integrated observations or experiments where we probe models of dark matter by directly producing these in a controlled environment (such as colliders) seem far more robust. It is, therefore, imperative to look for or to constrain models of the dark-sector using experiments/observations that depend mostly on far reliable ``global''  quantities, while taking an agnostic approach to the constraints/bounds from direct detection experiments.
In this work, we take a similar attitude and provide ``robust" constraints on the electromagnetic moments of dark matter by considering data from CMB experiments only.  

A systematic way to understand electromagnetic interactions of dark matter (or for any other particle), is to think through its electromagnetic moments, which is also a systematic expansion in the spirit of an Effective Field Theory (EFT) \cite{1994PhLB..320...99B}. The leading moment is the charge itself, which is heavily constrained and the charge of the dark matter can at best be infinitesimal.  Often referred to as the millicharged dark matter in the literature \cite{1986PhLB..166..196H,1986PhLB..174..151G,2004PhLB..586..366K,2007PhRvD..75k5001F,2007JHEP...03..120C}, these scenarios are well studied. The bounds on the charge of dark matter have been estimated using observations like the Big Bang Nucleosynthesis (BBN) \cite{2000JHEP...05..003D, 2013JCAP...02..010B}, the energy spectrum and anisotropies of the CMB \cite{2004JETPL..79....1D,2013PhRvD..88k7701D, 2018PhRvD..98l3506B},  shapes of galaxies \cite{2011PhRvD..83f3509M}, beam dump experiments\cite{1998PhRvL..81.1175P}, stellar and supernovae cooling \cite{2017JHEP...02..033H} etc. all of which place stringent constraints on the electric charge and the relic abundance of millicharged dark matter. 

For dark matter candidates with vanishing total charge, its interaction with the radiation field may arise because of its non-zero higher-order electromagnetic moments. In the systematic expansion, in the framework of EFT, the charge radius operator followed by the polarizability operator represent the two most significant operators if dark matter is scalar.  For dark matter with non-zero spin, various dipole moments become important. See \cite{2000PhLB..480..181P} for a systematic expansion in moments and \cite{1997PhRvD..55.4129L} for the power counting in the spirit of the Heavy Quark Effective Theory (HQET) \cite{Isgur:1989vq, Georgi:1990um, 1994NuPhB.412..181J, Mehen:2005hc, Wise:1992hn, Yan:1992gz, 1997PhR...281..145C}. 

Note, however, that an electrically neutral dark matter that couples to electromagnetism via non-zero higher-order moments naturally brings into question (at least partial) compositeness of the dark matter. An immediate consequence of such compositeness is the existence of additional transient states in the dark sector in the vicinity of the dark matter itself in the spectrum. For a spectrum where the excited states are nearly degenerate with the ground state, additional operators that involve both the excited state as well as the ground state need to be considered. In fact, these operators may give rise to inelastic collisions and/or transitions among the dark states. Consequently, these can play a vital role in the detection and understanding of the dark matter itself, since these may give rise to inelastic collisions between dark matter and visible particles with a different recoil energy profile compared to the elastic collisions usually assumed in interpreting the results of direct detection experiments \cite{2001PhRvD..64d3502S, 2010PhRvD..82l5011C, 2009PhRvD..80k5005A, 2011PhRvD..83k2002A, 2011PhRvD..84f1101A, PandaX-II:2017zex, XENON:2020fgj}). These new operators may also give rise to radiative transitions yielding  absorption/emission lines in the spectrum of bright compact sources and CMB \cite{2007PhRvD..75b3521P,2010PhRvD..81i5001K, 2024PhRvD.109f3512G}. To be specific, we focus our work precisely in this area, where these inelastic collisions and radiative transitions dominate the phenomenology of dark matter.  

We begin with the construction of a minimal framework of such a dark sector. This framework should be thought of as an effective theory in the sense that a wide class of ultraviolet dark-matter models reduce to this minimal set-up once all additional degrees-of-freedom that cannot be excited (say at CMB frequencies) are integrated out. In our set-up, the two dark states are nearly degenerate and have identical quantum numbers, except for the spins. To be specific, we take the lowest lying state to be a spin-zero pseudo scalar, whereas the heavier state is taken to be a spin-one vector. For an example construction where a more fundamental theory would result in this minimal scenario at CMB frequencies see Ref.~\cite{2024PhRvD.109f3512G}. Even though in \cite{2024PhRvD.109f3512G} both the dark states are composite states, these can also be elementary in general. However, the compositeness or at least partial compositeness is a useful paradigm to keep in mind since in these models one naturally gets closely spaced states in the dark sector. The spin difference between the states allows for radiative transitions between states with emissions/absorption of photons. In this paper, we consider the magnetic transition operator that generates this electromagnetic transition between the two lowest lying dark states. Our scenario, in fact, mimics closely the famed hyper-fine transitions in the Hydrogen atom at 21 cm, which is poised to revolutionize the next generation cosmology. Unlike the hyper-fine transition in the hydrogen atom though, the frequency of transition is not known (even though subjected to various constraints) and can lie anywhere from radio waves to gamma rays, thus motivating the search for dark matter through lines or features in the spectrum of sources across the full electromagnetic spectrum. Our framework is however more general and can be easily extended to include additional higher order moments and/or different transitions.

The CMB, which is the relic radiation from the Big Bang and a ubiquitous source of photons present throughout the history of the Universe, provides a probing ground for any cosmologically relevant sector that couples to radiation.  The precise measurements of the CMB spectrum by the COsmic Background Explorer/Far-InfraRed Absolute Spectrophotometer (COBE/FIRAS) \cite{1996ApJ...473..576F} instrument almost thirty years back has made CMB a precise observable to constrain any non-standard physics that can distort its spectrum from a perfect blackbody at the minuscule level of $\sim 1$ part in $10^5$. Previous works in the literature have extensively investigated the effect of pure energy injection mechanisms on the spectrum of the CMB. Such energy injection processes can occur due to standard processes like dissipation of sound waves \cite{1991ApJ...371...14D,2012A&A...543A.136K}, and other exotic processes like annihilating and decaying dark matter \cite{1993PhRvL..70.2661H,2003PhRvD..68f3504F,2019PhRvD..99l3510A} (see \cite{2013IJMPD..2230014S} for a review). In our framework, the effective dark sector couples to radiation and gives rise to the CMB spectral distortions by injecting/removing photons at fixed frequencies. The change in the number of CMB photons is sourced by the electromagnetic transitions among dark states at a frequency equal to the transition frequency of the dark matter states. In this work, we derive analytic and numerical solutions for the CMB $\mu$-type, $y$-type, and non-thermal spectral distortions given such scenarios. Note that  Ref.~\cite{2015MNRAS.454.4182C}) considered the $\mu$-type distortions from similar photon injections previously and gave an approximate formula for the $\mu$-type spectral distortions. Our analytic result closely matches and is consistent with the formula given in \cite{2015MNRAS.454.4182C}.

Many aspects of the physics of CMB distortions sourced by photons in a given (unknown) frequency are qualitatively similar to the phenomenology of hyperfine transitions in neutral hydrogen which gives rise to global spectral signatures in the CMB spectrum \cite{2004MNRAS.352..142B, 2004ApJ...608..622Z,2006PhR...433..181F,2012RPPh...75h6901P}, even though there exist many important differences.  A net emission or absorption of CMB photons can only happen if the population of dark matter particles in the two states is not in equilibrium with the CMB. This is only possible when, apart from the CMB photons, the transition between the two dark matter states can also be caused by interactions with other particles that are not in equilibrium with the CMB. Therefore, the baryons cannot play this role since they are coupled to the CMB till late redshifts $z \sim 150$. The only other possibilities are neutrinos and dark matter itself. In this work, we focus on dark matter self-interactions only. The inelastic collisions between the dark matter particles can cause collisional excitations or de-excitations which can couple the population of the two dark matter states with the dark matter temperature. 

The subsequent signature of such a scenario on the CMB spectrum is a strong function of the photon frequency as well as the epoch in which the dark matter transitions take place. The redshift $z\sim 2 \times 10^6$ marks the epoch above which the standard processes in the baryon photon plasma, namely, the double Compton scattering, bremsstrahlung, and Compton scattering are successful in establishing a blackbody spectrum of the CMB \cite{1969Ap&SS...4..301Z, 1970Ap&SS...7...20S,1957JETP....4..730K,1982A&A...107...39D} and any deviations from the blackbody spectrum are exponentially suppressed. Any modification from the equilibrium photon number density or energy density at $z \lesssim 2 \times 10^6$ will give rise to deviations in the spectrum of the CMB from a perfect blackbody, referred to as the spectral distortions of the CMB. Such departures, caused by dark matter transitions throughout the evolution of the Universe, are a unique probe of dark matter and its electromagnetic and collisional properties. At $z \lesssim 2 \times 10^6$, the photon number changing processes like bremsstrahlung and double Compton scattering become inefficient and are unable to create a Planck spectrum. In the redshift range $10^5 \lesssim z \lesssim 2\times 10^6$, Compton scattering is still very efficient and equilibrates the CMB to a Bose-Einstein spectrum instead of a Planck spectrum. Thus, the dark matter transitions occurring in the redshift range $10^5 \lesssim z \lesssim 2\times 10^6$ give rise to $\mu$-type distortions.  At $z\lesssim 10^5$, Compton scattering becomes incapable of achieving the Bose-Einstein spectrum in case of any photon injection or absorption. In such cases, the spectral distortions can be thermal or non-thermal\footnote{The spectral distortions in the CMB spectrum arising due to interaction of the CMB with a thermal distribution of electrons is referred to as thermal spectral distortion e.g. $\mu$-type and $y$-type distortions.} depending on the frequency of the photon being absorbed or emitted. If the transition energy of dark matter lies in the Rayleigh-Jeans part of the CMB spectrum, the bremsstrahlung process before recombination can erase modifications in the CMB spectrum by, for example, borrowing energy from the baryonic plasma and cooling it in case of photon absorption by dark matter. The subsequent inefficient Compton scattering between electrons and photons which are at slightly different temperatures w.r.t. each other gives rise to $y$-type distortions in the CMB. If the dark matter transitions occur at higher frequencies, where bremsstrahlung is weaker compared to the Hubble expansion rate, any change in the CMB spectrum due to line emission or absorption stays unaffected and will be preserved till today. Such non-thermal signatures have unique features that distinguish them from the standard thermal $\mu$ and $y$-type distortions. This enhances the prospects of detecting dark matter using such signatures in next-generation experiments like PIXIE \cite{2011JCAP...07..025K} and opens a new window into the properties of dark matter. 
Apart from transitions, the electromagnetic scattering of dark matter particles with electrons and ions can couple dark matter to the baryon-photon fluid. Due to the fact that the ratio of the baryon to photon number $(n_b/n_\gamma \approx 6 \times 10^{10})$ \cite{2020A&A...641A...6P} is small, almost all the entropy of the baryon-photon fluid is in the photons. Thus, any energy gained (lost) by the baryon-photon fluid is gained (lost) by just the photons. In particular, any energy added (lost) to the baryons ends up almost entirely in heating (cooling) the photons, as long as they are thermally coupled by Compton scattering until about $z\approx 200$ \cite{1968ApJ...153....1P,1969JETP...28..146Z}. 
The spectral distortion limits from the thirty-year-old COBE data \cite{1996ApJ...473..576F} can already put stringent constraints on the strength of the magnetic transition operator $\epsilon$ of MeV mass dark matter which is significantly stronger than the bounds from direct detection when dark matter transition energy is in $1$ to $10$ eV range.
	
We use the Planck 2018 \cite{2020A&A...641A...6P} cosmological parameters (Hubble constant: $H_0 = 100\,h = 67.66$ km s$^{-1}$ Mpc$^{-1}$, $\Omega_{m} = 0.3111$, and $\Omega_{b} = 0.049$). 
We also use the publicly available codes Recfast++ \cite{2011MNRAS.412..748C,2010MNRAS.407..599C}, Colossus \cite{2018ApJS..239...35D} in our analysis.

	\section{A minimal framework for a radiatively transitioning dark matter}
	\label{sec:dm_model}
	Any construction that can accommodate a radiatively transitioning dark matter requires a slightly involved dark sector. Such a construction should contain at least two or more states having the same quantum numbers (apart from the spin), where the lowest lying states are identified as dark matter. In subsection~\ref{subsec:2.1} we chalk out a minimal framework of such a scenario along with the essential interactions that enable radiative transitions among the dark-states. Note that the ``model" we present in subsection~\ref{subsec:2.1} should be viewed as an effective description of more fundamental theories. For an elegant ultra-violet complete scenario, where the minimal set-up presented in the following subsection emerges in the far infrared, see Ref.~\cite{2024PhRvD.109f3512G}. Since we do not want to be biased by any specific origin of the effective framework, we consider all the parameters given in subsection~\ref{subsec:2.1} to be independent. Note, however, that the effect of the radiative transitions of dark matter on the CMB depends on a whole new set of macro-parameters (such as the Einstein coefficients, collisional parameters etc.) and their temperature dependence. Even though, in principle, we should be able to determine these macro-parameters from more fundamental interactions present in the effective theory, the procedure is highly non-trivial, tedious, and may depend significantly on the macroscopic physics of the medium. In this work, we will rather take a phenomenological approach and approximate these macro-parameters of the dark sector as scaled-up (or down) version of the macro-parameters of the hydrogen atom system. In subsection~\ref{subsec:2.2}, we define and summarize these parameters.

\subsection{The set-up}
\label{subsec:2.1}
\begin{figure}[t]
		\centering
		\includegraphics[width=0.3\textwidth]{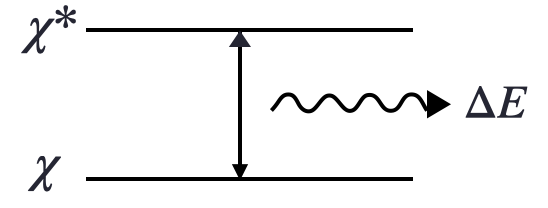}
		\caption{Electromagnetic transition in the simplified dark matter model characterized by two states $\chi$ and $\chi^*$ with an energy splitting $\Delta E$.}
		\label{fig:scales}
    \end{figure}

Our minimal dark-sector set-up is effectively a two-state system along with a conserved global abelian $U(1)$ symmetry transformation. For the rest of this work, we designate the $U(1)$ factor by $U(1)_D$ or dark-$U(1)$ and refer the charges of the various multiplets under $U(1)_D$ to be their dark-charges. Both states in our set-up carry identical and non-zero dark charges. The lowest lying state (call it $\chi$) being the lightest state with a non-zero dark-charge is cosmologically stable. The heavier state (call it $\chi^*$) is approximately degenerate with $\chi$ and carries an identical dark-charge. Additionally, we take $\chi$ to be a spin 0 pseudo scalar and $\chi^*$ to be a spin 1 vector. The fact that we take $\chi$ to be a pseudo scalar is rather a choice and we consider this case for the sake of simplicity. Radiative transitions go through between these states as long as the spin of $\chi*$ and $\chi$ differ by one. A schematic of the two-state system is given in Fig.~\ref{fig:scales}. We describe the set-up in the following bullet points: 
\begin{itemize}
    \item The dark matter $\chi$ is a pseudo scalar with a mass $m_\chi$. It is electrically neutral but has a non-zero dark charge $Q_D$. The other state in our set-up is a spin-1 vector $\chi^*_\mu$ with mass $m_{\chi^*}$, also electrically neutral, and carries the same dark-charge $Q_D$. We consider a case where the states are nearly degenerate, \textit{i.e.}, $m_{\chi^*} - m_\chi \ll m_\chi$, which suggests a compact way to represent both these states in a single multiplet,
    \begin{equation}
        \mathcal{X} \ \equiv \ \chi_\mu^*\gamma^\mu - \chi\gamma^5 \, . 
    \end{equation}
    \item Since the energy of transition is far less than the masses of the states themselves,  the effective theory of this radiative transition is non-relativistic. The primary degrees-of-freedom are the velocity ($v$) eigenstates of the fields  $\chi$ and $\chi^*$ (or rather of $\mathcal{X}$) 
    \begin{equation}
        \mathcal{X}_v \ \equiv \ \sqrt{2 m_\chi} \  e^{i m_\chi \left( v\cdot x\right) } 
            \ P_{+} \mathcal{X}, 
        \quad \text{ where }  \quad
            P_{+} \ = \ \frac{1}{2} \left( 1 + v_\mu\gamma^\mu \right) \;  
        \text{ and } v^2 = 1.
    \end{equation}\label{2.2a}
    The projection operator $P_{+}$ captures only the small fluctuations $\left( \ll m_{\chi}\right)$  \cite{Isgur:1989vq, Georgi:1990um, 1994NuPhB.412..181J, Mehen:2005hc, Wise:1992hn, Yan:1992gz, 1997PhR...281..145C}. The extra $\sqrt{2m_\chi}$ factor in Eq.~\eqref{2.2a} gives the correct non-relativistic normalization for $\mathcal{X}_v$.  
      
   \item The magnetic transition operator \cite{2000PhLB..480..181P} between these two states is given by,
   \begin{equation}
       \frac{1}{m_\chi} \: \epsilon e \ \text{tr}\left(\bar{\mathcal{X}}_v \sigma^{\mu \nu} \mathcal{X}_v F_{\mu \nu} \right) \, ,\label{C23}
   \end{equation}
   where the parameter $\epsilon$ denotes the strength of the magnetic transition operator,  $e$ is the electromagnetic coupling constant (namely $e = \sqrt{4\pi \alpha}$, where $\alpha$ is the fine structure constant). Note that we have taken $m_\chi$ to be the scale of irrelevant operators in the above equation keeping with the spirit of the heavy quark effective theory (HQET). In the UV complete theory of \cite{2024PhRvD.109f3512G},  $\chi$ and $\chi^*$ are electrically neutral composite states of constituents which themselves are electrically charged ($\epsilon$ is proportional to charges of constituents).  
   \item The charge radius operator responsible for elastic scattering between dark matter and visible particles \cite{2000PhLB..480..181P} is given by,
	\begin{equation}
		\frac{r_\chi^2}{6m_\chi}\: e \ \text{tr}\left(\bar{\mathcal{X}}_{v}\,(\partial_{\mu}\mathcal{X}_{v})\,\partial_{\nu}F_{\mu\nu}\right)\,,\label{C32}
	\end{equation}
	where $r_\chi$ is the charge radius of dark matter. In order to reduce the number of parameters, we take the effective strength of both the magnetic transition and the charge radius operator to be of the same order. Therefore, we approximate $r_\chi^2 = \epsilon/m_\chi^2$ in rest of this work. This operator plays an important role in direct detection and in kinetically coupling the baryons and dark matter in the early Universe. There could be host of other operators involving dark matter particles only giving rise to self-interactions of dark matter. The macro-parameters for self-interactions are defined in the next subsection.
\end{itemize}

	\subsection{Macroscopic features}
	\label{subsec:2.2}
	The minimal set of parameters required to describe the macroscopic properties of dark matter in our setup are summarized below:
	\begin{itemize}[leftmargin=18pt]
		\item The total number density of dark matter particles $n_\chi$ is given by the sum total of dark matter particles in the ground state ($n_0$) and the excited state ($n_1$),
		\begin{align}
			n_\chi = n_0 + n_1. \label{2.1}
		\end{align}
		\item The energy/mass splitting between the ground (spin 0) state and the excited (spin 1) state can be expressed in terms of frequency $\nu_0$ or alternatively in temperature units as $T_{*}$ using the following relation,
		\begin{equation}
			\Delta E \equiv \left(m_{\chi^*}-m_{\chi}\right)c^2 = h\nu_0 = k_{B}T_{*}, \label{2.2}
		\end{equation}
		where $h$ and $k_B$ denote the Planck's constant and the Boltzmann constant respectively, and $c$ denotes the speed of light in vacuum. 
		\item We can use the Boltzmann law to express the relative population of dark matter particles in the two states in temperature units using the excitation temperature $T_{\text{ex}}$, defined by the following equation,
		\begin{align}
			\frac{n_0}{n_1} = \frac{g_0}{g_1}\exp\left(\frac{T_*}{T_{\text{ex}}}\right), \label{2.3}
		\end{align}
		where $g_0$ and $g_1$ denote the degeneracy factors of the ground state and the excited state respectively.  Since all the quantum numbers apart from the spin to be identical for the two states, the ratio of degeneracy factors for the two states becomes $g_1/g_0 = 3$. 
		\end{itemize}
		The dark matter particles in the two states can make transitions from the ground state to the excited state or vice versa by absorbing/emitting a photon or through inelastic collisions between themselves. 
		\begin{itemize}[leftmargin=20pt]
			\item \textbf{Radiative transitions:}
		The radiative transitions from one dark matter state to another can occur by absorption or emission of a photon of energy equal to the transition energy $\Delta E$ (defined in Eq.\eqref{2.2}) of dark matter. They are of three kinds, namely, absorption, spontaneous emission, and stimulated emission and are parameterized by the Einstein A and B coefficients as described below \cite{1986rpa..book.....R}:
			\\\\
			The number of transitions per unit time per unit volume from the ground state to the excited state is proportional to the Einstein coefficient $B_{01}$ and the mean intensity of incident light $\bar{J}$,
			\begin{equation}
				\frac{dn_{0\rightarrow1}}{dt} = n_0B_{01}\bar{J}\,,\label{2.4}
			\end{equation}
			The number of radiative transitions per unit time per unit volume from the excited state to the ground state is a sum total of the spontaneous emission rate $\propto A_{10}$ and the stimulated emission rate $\propto B_{10}$,
			\begin{equation}
				\frac{dn_{1\rightarrow0}}{dt} = n_1(A_{10} + B_{10}\bar{J})\,.\label{2.5}
			\end{equation}
			The three Einstein coefficients are related to each other via the Einstein relations,
			\begin{equation}
				A_{10} = \frac{2h\nu_0^3}{c^2}B_{10}\quad\text{and}\quad
				g_0B_{01} = g_1B_{10}.
			\end{equation}
			The electromagnetic transitions between the spin 0 and spin 1 dark matter states are analogous to the hyper-fine transition in a hydrogen atom. In particular, we are interested in weak transitions similar in strength to the 21 cm transitions in neutral hydrogen. Therefore, we parameterize the Einstein A coefficient in dark matter in terms of the Einstein coefficient for hyper-fine transitions in the hydrogen atom,
			
			\begin{equation}
				A_{10} = \alpha_\text{A}\, A_{10}^{\text{HI}}, \hspace{5pt} \text{where} \hspace{5pt} A_{10}^{\text{HI}} = 2.85 \times 10^{-15} \text{s}^{-1}. \label{2.7?}
			\end{equation}
			The microscopic properties of the dark matter model are embedded inside the Einstein coefficients. In this work, we can express the dimensionless spontaneous emission rate parameter $\alpha_\text{A}$ in terms of  $\epsilon$ and the hyper-fine splitting of dark matter ($\Delta E$). We can thus simply scale the hyper-fine splitting parameters of the hydrogen atom,
			\begin{equation}
				\alpha_\text{A} \approx \epsilon^2 \left(\frac{\Delta E}{\Delta E_{\text{HI}}}\right)^3\left(\frac{m_\text{e}}{m_\chi}\right)^2\,,\label{2.8}
			\end{equation}
			where $\Delta E_{\text{HI}}$ is the hyper-fine splitting in the hydrogen atom and $m_\text{e}$ is the mass of electron. We will be using this scaling law and $\epsilon$, $\Delta E$, $m_\chi$ as free parameters while studying constraints from COBE-FIRAS and PIXIE and comparing them with the direct detection experiments. Most of our theoretical discussion however does not rely on this parameterization and $\alpha_\mathrm{A}$ can be considered as a free parameter in general. 
			\item \textbf{Collisional transitions}\\
			The inelastic collisions between two dark matter particles can cause excitations or de-excitations changing the population of dark matter particles in the two states. We note that the dominant interactions involved in inelastic collisions can be in general non-electromagnetic or new dark interactions. We parameterize the collisional excitation rate by the parameter $C_{01}$ and the collisional de-excitation rate by the parameter $C_{10}$ in the following way:
			\\\\
			The number of collisional transitions per unit volume from the ground state to the excited state and vice-versa are given by, 
			\begin{align}
				\frac{dn_{0\rightarrow 1}}{dt} = n_0C_{01} \hspace{5pt}\text{and}\hspace{5pt}\frac{dn_{1\rightarrow 0}}{dt} = n_1C_{10}\,.\label{2.7}
			\end{align}
			For a thermal velocity (Maxwell Boltzmann) distribution of dark matter particles at temperature $T_\chi$, the two collisional coefficients are related to each other by the following relation,
			\begin{align}
				C_{01}(T_\chi) &=\frac{g_{1}}{g_{0}}\exp(-T_*/T_{\chi})\,C_{10}(T_\chi).
			\end{align}	
			Similar to the Einstein coefficients, the collisional coefficients also encode the microscopic properties that determine dark matter self-interactions. We assume the dark matter collision cross-section to be velocity-independent. We use the bullet cluster observations \cite{2004ApJ...606..819M} and parameterize the dark matter collision cross-section in terms of the bullet cluster limit using the dimensionless parameter $\alpha_\text{C}$ as,
			\begin{align}
				C_{10} = n_{\chi}\langle \sigma v\rangle, \hspace{5pt}
				\text{where}\hspace{5pt}
					\langle \sigma v \rangle = 	\alpha_\text{C}  \sigma_{\mathrm{BC}}  \sqrt{\frac{8k_\text{B}T_\chi}{\pi m_\chi}}\,,\label{2.11}
			\end{align}
			where $v$ denotes the relative velocity of dark matter particles. Since the dark matter collision cross-section denoted by $\sigma$ is velocity independent, the temperature dependence of $v \propto \sqrt{T_\chi}$ can be trivially separated out. The bullet cluster bound on dark matter collision cross-section increases linearly with the dark matter mass as \cite{2004ApJ...606..819M},  
			\begin{equation}
				\sigma_{\text{BC}} = 3.6\times 10^{-27}\left(\frac{m_\chi}{\text{MeV}}\right) \,\text{cm}^2. \label{2.12}
			\end{equation}

		\end{itemize}

	\section{Thermal history of dark matter}
		The thermal evolution of the two-state dark matter system, as introduced in the previous section, is characterized by $T_\chi$ which describes the thermal energy of the dark matter particles, and $T_\text{ex}$ which determines the relative population of dark matter particles in the two states. We divide this section into two parts and discuss the role played by different elastic and inelastic scattering processes in the dark sector and the Standard Model sector in determining $T_\text{ex}$ and $T_\chi$.  
		
		\subsection{Evolution of the excitation temperature}
		The evolution of $T_\text{ex}$ is determined by the competition between two processes, namely, the radiative transitions sourced by the CMB photons which drive $T_{\text{ex}}\rightarrow T_{\text{CMB}}$, and the collisional transitions sourced by the inelastic scattering between dark matter particles which drive $T_{\text{ex}}\rightarrow T_{\chi}$. The change in the population of dark matter particles in the ground state ($n_{0}$) due to collisional and radiative transitions can be derived using Eq.\eqref{2.4}, \eqref{2.5}, and \eqref{2.7}, and is given by,
		\begin{align}
			\frac{dn_0}{dz}-\frac{3n_0}{1+z} &=-\frac{1}{H(1+z)}\left(n_1C_{10}-n_0C_{01}+n_1A_{10}+(n_1B_{10}-n_0B_{01})\bar{J}_\text{CMB}\right)\,,\label{3.1}
		\end{align}
		where $H$ denotes the Hubble expansion rate and $\bar{J}_\text{CMB}$ denotes the black body CMB intensity at $\nu_{0}$. The evolution equation for $T_{\text{ex}}$ can be derived by differentiating Eq.\eqref{2.3} with respect to redshift and substituting Eq.\eqref{2.1} and \eqref{3.1} into it. After simplification, we get,
		\begin{eqnarray}
			\dfrac{dT_\text{ex}}{dz} = \frac{T_{\text{ex}}^2}{T_*H(1+z)}\left(1+\frac{g_1}{g_0}e^{-T_*/T_\text{ex}}\right)
			\Bigg[C_{10}\left(1-e^{-T_*\left(\frac{1}{T_\chi}-\frac{1}{T_\text{ex}}\right)}\right)
			\nonumber\\ +\frac{A_{10}}{1-e^{-T_*/T_\text{CMB}}}\left(1-e^{-T_*\left(\frac{1}{T_\text{CMB}}-\frac{1}{T_\text{ex}}\right)}\right)\Bigg].  \label{3.2}
		\end{eqnarray}
		
		\subsection{Evolution of the dark matter temperature}
		\label{subsec:Tdm}
		At very high redshifts, the electromagnetic interactions between the dark matter particles and the Standard Model particles keep the dark matter in kinetic equilibrium with the Standard Model sector (electrons, ions, and photons). When these electromagnetic interactions freeze out, dark matter kinetically decouples from the baryon photon plasma. Since dark matter is non-relativistic at the time of decoupling, its temperature $T_\chi$ cools adiabatically as $\propto (1+z)^2$ due to Hubble expansion. If there exist momentum exchange interactions between the dark sector and the Standard Model sector at later redshifts, they cause a deviation from the adiabatic fall in the dark matter temperature. The most general equation describing the evolution of the dark matter  temperature in the presence of momentum transfer interactions is given by,
		\begin{equation}
			\frac{dT_{\chi}}{dz} = \frac{2T_{\chi}}{1+z} - \frac{\sum_i\Gamma_{\chi_i}}{(3/2)n_{\chi}k_{B}H(1+z)}\,,\label{3.3}
		\end{equation}
		where the first term on the right represents the adiabatic $\propto (1+z)^2$ fall in the temperature and the second term shows the effect of momentum exchange interaction on $T_\chi$ in terms of $\Gamma_{\chi_i}$ which denotes the volumetric rate of energy transfer between dark matter and $i$th Standard Model degree of freedom, and the sum is over all Standard Model degrees of freedom present at a given epoch. We will be interested in the epochs after the electron-positron annihilation when the only electromagnetically interacting Standard Model particles are baryons\footnote{We will be using baryons to mean ions and electrons in this paper.} (ions and electrons) and photons.
		We can simplify Eq.\eqref{3.3} further by defining the heating efficiency of the momentum exchange process as,
		\begin{eqnarray}
			 \eta_{\chi_i} = \frac{\Gamma_{\chi_i}}{(3/2)n_{\chi}Hk_{\text{B}}T_{\chi}}\,,\label{3.4}
		\end{eqnarray}
		By substituting Eq.\eqref{3.4} to Eq.\eqref{3.3}, the evolution equation for $T_\chi$ simplifies to,
		\begin{equation}
			(1+z)\frac{dT_{\chi}}{dz} = 2T_{\chi} - \sum_i\eta_{\chi_i}T_\chi. \label{3.5}
		\end{equation}
		The momentum transfer between the Standard Model sector and the dark sector can happen directly via elastic scattering of dark matter particles with ions and electrons. In addition, a net collisional excitation or de-excitation can change the kinetic energy of the dark matter particles, thereby affecting $T_\chi$. We describe these interactions in more detail below:
		\begin{itemize}[leftmargin=20pt]
			\item 	\textbf{Scattering between dark matter and ions/electrons} \\
			In our model, the electromagnetic scattering between the dark matter particles and the Standard Model particles can be both elastic as well as inelastic. 
			The relevant cross-section for scattering processes that enter Eq.\eqref{3.3} is the momentum-transfer cross-section which is defined as \cite{2014PhRvD..89b3519D},
			\begin{equation}
				\bar{\sigma} \equiv \int d\Omega \frac{d\sigma}{d\Omega}(1-\cos\theta),
			\end{equation} 
			where $\theta$ is the scattering angle between the incoming and outgoing dark matter particle and $d\sigma/d\Omega$ is the differential cross-section.
			
			The dominant contribution to the elastic scattering comes from the charge radius operator (see Eq.\eqref{C32}) which has the following momentum-transfer cross-section \cite{2000PhLB..480..181P, 2024PhRvD.109f3512G},
			\begin{align}
				\bar{\sigma}^{\text{ES}}_b &= \frac{16\pi\epsilon^2\alpha^2\mu_{\chi_b}^2}{3 m_\chi^4}, \label{3.7} 
			\end{align}
			where the symbol $b$ stands for either ion or electron and $\mu_{\chi_b}$ denotes the reduced mass of the dark matter and ion/electron system. 
			
			The dominant contribution to inelastic scattering comes from the magnetic dipole transition operator (see Eq.\eqref{C23}) which causes dark matter transitions when the dark matter particles interact with the magnetic field generated by baryons/electrons. The origin of the magnetic field can either be the intrinsic spin or the motion of the baryons/electrons. We consider these contributions separately and state both the momentum transfer cross-sections for inelastic scattering below \cite{2000PhLB..480..181P,2024PhRvD.109f3512G},
			\begin{align}
				\bar{\sigma}^{\mathrm{IS}}_{b1} = \frac{8\pi\epsilon^2\alpha^2\mu_{\chi_b}^2}{m_\chi^2m_b^2}, \hspace{10pt}
				\bar{\sigma}^{\mathrm{IS}}_{b2} = \frac{32\pi\epsilon^2\mu_{\chi_b}^2}{m_\chi^2m_b^2}\left(\frac{v_{\chi_b}}{c}\right)^2,
				\label{3.8}
			\end{align}
			where $v_{\chi_b}$ is the relative velocity of dark matter with respect to ions/electrons.
		As discussed in section \ref{sec:dm_model}, for simplicity we assume the strength of all the electromagnetic moments of dark matter to be parameterized by the same parameter $\epsilon$. 
			
		In general, we can parameterize the velocity dependence of the momentum transfer cross-section as a power law, $\bar{\sigma} = \sigma_0\left(v_{\chi_b}/c\right)^n$. The volumetric energy transfer rate for a general interaction cross-section $\bar{\sigma}$ is given by \cite{2014PhRvD..89b3519D},
			\begin{equation}
				\Gamma_{\chi_b} =- \frac{2^{\frac{n+5}{2}}\Gamma(3+n/2)}{\sqrt{\pi}}\frac{m_{\chi}m_b\, n_b\,n_\chi\sigma_0}{(m_\chi+m_b)^2}\left(\frac{T_b}{m_b} + \frac{T_\chi}{m_\chi}\right)^{\frac{n+1}{2}}\left(T_\chi - T_b\right)\,,\label{3.9}
			\end{equation}
			where $n_b$ and $m_b$ denote the number density and mass of the ion/electron respectively and $T_b$ denotes the temperature of baryons.
			
			Substituting Eq.\eqref{3.9} into Eq.\eqref{3.4}, we can obtain a general expression for the heating efficiency due to dark matter baryon scattering,
			\begin{equation}
				\eta_{\chi_b} = - \frac{2^{\frac{n+7}{2}}\Gamma(3+n/2)}{3\sqrt{\pi}}\frac{m_{\chi}m_b\, n_b\,\sigma_0}{H(m_\chi+m_b)^2}\left(\frac{T_b}{m_b} + \frac{T_\chi}{m_\chi}\right)^{\frac{n+1}{2}}\left(T_\chi - T_b\right)\;. \label{3.10}
			\end{equation}
			We can use Eq.\eqref{3.10} to calculate the heating efficiencies for both elastic and inelastic scattering momentum transfer cross-sections (defined in Eq.\eqref{3.7} and Eq.\eqref{3.8} respectively) and substitute it into Eq.\eqref{3.5} to get the evolution of the dark matter temperature.
%
			\\
			\item \textbf{Energy exchange with the CMB through collisional transitions in dark matter} \\
			In addition to scattering with baryons, dark matter can also gain (lose) energy from (to) the CMB photons indirectly. In an inelastic collision between two dark matter particles, collisional transitions transfer the kinetic energy to the internal energy and vice versa. 
			The net number of transitions resulting in energy transfer to the dark sector is the difference between the collisional de-excitation rate and the collisional excitation rate. The volumetric heating rate is equal to the energy change in a single transition i.e. $k_\text{B}T_{*}$ times the net rate of energy transfer per unit volume, and is given by,
			\begin{align}
				\Gamma_{\chi\gamma} 
				&= k_{\text{B}}T_{*}\left(n_1C_{10} - n_0C_{01}\right),\nonumber\\
				&= n_\chi C_{10} k_{\text{B}}T_{*}\frac{1}{\left(1+\left(g_0/g_1\right)e^{T_{*}/T_{\text{ex}}}\right)} \left(1-e^{-T_*\left(\frac{1}{T_\chi}-\frac{1}{T_\text{ex}}\right)}\right)\,.\label{3.11}
			\end{align}
			It is evident from Eq.\eqref{3.11} that the energy transfer due to collisional transitions occurs when the excitation temperature is not equal to the dark matter temperature. The deviation of $T_\text{ex}$ from $T_\chi$ happens because of radiative transitions which couple $T_\text{ex}$ to the CMB temperature $T_\text{CMB}$. Thus, when $T_\mathrm{CMB}\gtrsim T_\mathrm{ex}>T_\chi$, the radiative excitations of dark matter and subsequent collisional de-excitations indirectly transfer energy from the CMB and heat dark matter. The heating efficiency due to collisional transitions is given by,
			\begin{equation}
				\eta_{\chi\gamma} = \frac{2C_{10}\left(T_*/T_\chi\right)}{3H\left(1+\left(g_0/g_1\right)e^{T_{*}/T_{\text{ex}}}\right)}\left(1-e^{-T_*\left(\frac{1}{T_\chi}-\frac{1}{T_\text{ex}}\right)}\right)\,.\label{3.12}
			\end{equation}\\
			The effect of collisional heating of dark matter is similar to the collisional heating of the neutral hydrogen gas during the Dark Ages, which was first discussed in \cite{2018PhRvD..98j3513V}.
		\end{itemize}
		The final evolution equation for dark matter temperature taking into account the scattering with baryons (from Eq.\eqref{3.10}) and collisional transitions (from Eq.\eqref{3.12}) as described above, can be expressed as,
		\begin{equation}
			\frac{dT_{\chi}}{dz} = \frac{2T_{\chi}}{1+z} - \left(\sum_b\eta_{\chi_b} + \eta_{\chi \gamma} \right)\frac{T_{\chi}}{1+z},
		\end{equation}
            where as mentioned before, $b$ runs over ions and electrons.
	\section{Spectral signatures of composite dark matter in the CMB}
	\label{sec:cmb_distortions}
	We now present a framework to compute the signatures of composite dark matter in the CMB spectrum. The electromagnetic transitions in dark matter can be caused by the emission/absorption of CMB photons. Such changes in the number of CMB photons can give rise to deviations in the spectrum of CMB from a perfect blackbody. These deviations are known as the spectral distortions of the CMB. 
	
	In the standard $\Lambda$CDM cosmology, the shape of the CMB spectrum that we observe today is a consequence of different electromagnetic processes occurring between the CMB photons, electrons, and baryons which comprise the well-established Standard Model sector.  Any possible electromagnetic interactions between dark matter and the Standard Model sector have to be weak so as not to conflict with the already existing high-precision cosmological data. In order to compute the spectral signatures of dark matter, we also need to take into account all the leading-order electromagnetic processes in the Standard Model sector.
	
	We begin by introducing the framework used to study the CMB spectrum in the standard model of cosmology in subsection \ref{subsec:cmb_sm}. We then modify this formalism to incorporate the effect of an additional photon number-changing process in subsection \ref{subsec:cmb_dm} and provide solutions for the different spectral distortions created by dark matter in the CMB spectrum in subsection \ref{ssec:soln}.
	
	\subsection{Formation of the CMB spectrum in the standard model of cosmology}
	\label{subsec:cmb_sm}
	We briefly review the different electromagnetic processes in the Standard Model sector that play a key role in establishing the black body spectrum of the CMB. 
	After Big Bang Nucleosynthesis, the baryon-photon plasma is composed of the CMB photons, electrons, and ions (H$^+$, He$^+$, and He$^{++}$). The electromagnetic interactions in the baryon-photon plasma can be categorized into two kinds:  
	\begin{itemize}[leftmargin=12pt]
		\item \textbf{Photon number changing:} The inelastic scattering between electrons and ions is called bremsstrahlung, and the one between electrons and photons is called double Compton scattering. These processes not only transfer energy between electrons/ions and photons but also change the photon number by either emitting a photon in the final state or absorbing a photon in the initial state. 
		\item  \textbf{Photon number preserving:} The elastic scattering between electrons and photons is called Compton scattering. Compton scattering only causes a transfer of energy between electrons and photons. The photon number in the initial and final states remains unchanged. 
	\end{itemize}
	The shape of the CMB spectrum is a consequence of the combined action of different photon energy and/or photon number changing processes listed above \cite{1969Ap&SS...4..301Z, 1970Ap&SS...7...20S,1957JETP....4..730K,1982A&A...107...39D}. These processes try to bring the CMB temperature and the temperature of baryons in equilibrium with each other. To understand this quantitatively, we need to solve the evolution equation for the photon distribution function in the presence of standard electromagnetic processes. For simplicity, we can separate the time-independent part of the photon distribution function from the time-dependent part by defining a new quantity $x_\mathrm{e}$, which is a ratio of the photon frequency $\nu$ w.r.t. the electron temperature $T_\text{e}$, given by,
	\begin{equation}
		x_\text{e} \equiv \frac{h\nu}{k_\text{B}T_\text{e}}.	\label{4.1}
	\end{equation}
	In the absence of any energy injection after electron-positron annihilation, the electron temperature remains in equilibrium with the CMB, with $T_\mathrm{e}\approx T_\mathrm{CMB}$ until $z\approx 200$ \cite{1968ApJ...153....1P,1969JETP...28..146Z}. We note that because of the large photon-to-baryon entropy ratio, the electrons have a thermal Maxwell-Boltzmann distribution established by Compton scattering even when the CMB spectrum deviates from a Bose-Einstein spectrum \cite{1970JETPL..11...35Z}. Since the spectral distortions expected in the CMB are small \cite{1996ApJ...473..576F}, $T_\text{e}\approx T_\text{CMB}$ remains a good approximation as long as electrons and photons are kinetically coupled by Compton scattering. Since both frequency $\nu$ as well as $T_\text{e}$ redshift as $\propto (1+z)$ till $z\approx 200$, the quantity $x_\text{e}$ or the dimensionless frequency becomes time invariant in the absence of any energy injection.
	We can then express the evolution of the photon distribution function or photon occupation number $n$ as a function of the dimensionless frequency $x_\mathrm{e}$ and time $t$. This equation is called the generalized Kompaneets equation \cite{1969Ap&SS...4..301Z,1991A&A...246...49B,1993PhRvD..48..485H,2012JCAP...06..038K}, and is given by,
	\begin{align}
		\frac{\partial n (x_{\text{e}}, t)}{\partial t}&= K_\text{C}\dfrac{1}{x_{e}^2}\dfrac{\partial}{\partial x_{\text{e}}}x_{\text{e}}^4\left(n+n^2+\dfrac{\partial n}{\partial x_{\text{e}}}\right) + \left(K_{\text{br}} + K_{\text{dC}}\right)\dfrac{e^{-x_{\text{e}}}}{x_{\text{e}}^3}\left[1-n(e^{x_{\text{e}}}-1)\right] \nonumber\\& + x_{\text{e}}\dfrac{\partial n}{\partial x_{\text{e}}}\dfrac{\partial}{\partial t}\left[\text{ln}\dfrac{T_{\text{e}}}{T_{\text{CMB}}}\right],
		\label{4.2}
	\end{align}
	where the last term is due to the deviation of the electron temperature from the CMB temperature, and $T_\text{CMB} = 2.725\left(1+z\right)\,$K. The rate coefficients for Compton scattering, bremsstrahlung, and double Compton scattering in terms of the electron number density $n_\text{e}$, baryon number density $n_\text{B}$, the Thomson scattering cross-section $\sigma_\text{T}$, and electron mass $m_\text{e}$ are defined as,
	\begin{align}
		K_\text{C} &= n_{\text{e}}\sigma_\text{T} c\frac{k_\text{B}T_{\text{e}}}{m_{\text{e}} c^2}\,,\label{4.3}\\
		K_{\text{br}} &= n_{\text{e}}\sigma_\text{T} c\frac{\alpha n_\text{B}}{\left(24\pi^3\right)^{1/2}}\left(\frac{k_BT_{\text{e}}}{m_{\text{e}} c^2}\right)^{-7/2}\left(\frac{h}{m_{\text{e}}c}\right)^3g_{\text{br}}(x_{\text{e}}, T_{\text{e}})\,,\label{4.4}\\
		K_{\text{dC}} &= n_{\text{e}}\sigma_\text{T} c\frac{4\alpha}{3\pi}\left(\frac{k_BT_{\text{e}}}{m_{\text{e}} c^2}\right)^{2}g_{\text{dC}}(x_{\text{e}}, T_{\text{e}})\int dx_{\text{e}}\,  x_\text{e}^4\,n\left(n+1\right)\,,\label{4.5}	
	\end{align}
	where $g_{\text{br}}$ and $g_{\text{dC}}$ are the Gaunt factors for double Compton and bremsstrahlung processes respectively. We use the fitting formulas for $g_\mathrm{br}$ and $g_\mathrm{dC}$ from \cite{2014MNRAS.444..420V} and \cite{2012MNRAS.419.1294C} respectively.
	
	We can make a few qualitative deductions about the photon distribution function from the generalized Kompaneets equation given in Eq.\eqref{4.2}:
	\begin{itemize}
		\item[a)] In the presence of Compton scattering only, the equilibrium solution of the photon occupation number is the Bose-Einstein spectrum with a temperature $T_\text{CMB}=T_\text{e}$ and a non-zero chemical potential. 
		
		\item[b)] In the presence of bremsstrahlung and double Compton scattering, the equilibrium solution for the photon occupation number is the Planck spectrum with a temperature $T_\text{CMB}=T_\text{e}$. Note that the Planck spectrum is a special case of the Bose Einstein spectrum with a zero chemical potential.
	\end{itemize}
	We can use the two main features outlined above to understand how the different photon number and energy changing processes keep the CMB spectrum a blackbody at high redshifts ($z\gtrsim 2\times 10^6$). If an exotic source injects/extracts energy from the CMB, the Compton scattering efficiently distributes this energy over the full CMB spectrum giving rise to a Bose-Einstein spectrum with a non-zero chemical potential. The bremsstrahlung and double Compton scattering processes with rates
    \begin{equation}
       \left.\frac{1}{n}\frac{\partial n}{\partial t}\right|_{\text{br+dC}} \propto \frac{(1-e^{-x_\text{e}})e^{-x_\text{e}}}{x_\text{e}^3} \sim \frac{1}{x_\text{e}^2}
    \end{equation}
     in Eq.\eqref{4.2} are efficient in the low frequency tail of the CMB and drive the chemical potential to zero creating a Planck spectrum at low frequencies. The  frequency at which the two inelastic processes (double Compton scattering and bremsstrahlung) are as efficient as the Compton scattering will be denoted here by the critical frequency ($x_\text{c}$) and is defined via the relationship,
	\begin{align}
		\left .\frac{K_\mathrm{br}+K_\mathrm{dC}}{x_\mathrm{e}^2}\right|_{x_\text{e}=x_\text{c}} = K_\mathrm{C}\implies x_\text{c} = \sqrt{\frac{(K_{\text{br}}+K_{\text{dC}})}{K_\text{C}}}\,.\label{4.6}
	\end{align}
	Therefore, at a given redshift, Compton scattering dominates at higher frequencies ($x_\text{e}\gg x_\text{c}$) and is efficient in driving the photon spectrum to a Bose-Einstein spectrum. On the other hand, at low frequencies  ($x_\text{e} \ll x_\text{c}$), the inelastic processes dominate and drive the chemical potential to zero. Since the Compton scattering is efficient at all frequencies ($K_\text{C}>H$), it redistributes the photons in the Rayleigh-Jeans tail ($x_\text{e}< x_\text{c}$) to higher frequencies ($x_\text{e}> x_\text{c}$) and vice-versa, driving the chemical potential at higher frequencies also to zero. In this way, the photon creation/destruction by inelastic processes in combination with the photon redistribution by Compton scattering are successful in establishing the CMB Planck spectrum till redshift of $\sim 2\times10^6$.
	\subsection{Evolution of the CMB spectrum in the presence of dark matter}
	\label{subsec:cmb_dm}
 To study the effect of photon absorption/emission by dark matter, we need to add a new term to the generalized Kompaneets equation (see Eq.\eqref{4.2}) introduced in the last section. We can express the rate of change of the photon occupation number due to dark matter interactions in terms of the rate of change of the total number density of photons, which we call $\dot{N}$, where the dot represents the derivative w.r.t. the proper time. This can be done by recognizing that the total number density can be obtained by integrating the photon occupation number over the momentum space. 
	We can express the photon number density ($N$) for a general photon distribution function ($n$) defined as a function of $x_\text{e}$ in terms of the photon number density ($N_\text{PL}$) of the Planck spectrum at a temperature $T_\text{CMB}$ as, 
	\begin{equation}
		\mathcal{N}\equiv\dfrac{N}{N_{\text{PL}}} = \left(\frac{T_\text{e}}{T_\text{CMB}}\right)^3\dfrac{\int dx_\text{e} \,x_\text{e}^2\,n}{\int dx \,x^2\,n_{\text{PL}}} \equiv \dfrac{1}{\mathcal{I}_2}\left(\frac{T_\text{e}}{T_\text{CMB}}\right)^3\int dx_\text{e} \,x_\text{e}^2\,n\,,\label{4.7} 
	\end{equation}
	where $x = h\nu/(k_BT_\text{CMB})$ and $n_\mathrm{PL}$ denotes the photon occupation number for a Planck distribution. The quantity $\mathcal{I}_2$ is a constant and is given in Appendix \ref{app:mu}. 
	
	In the case of our dark matter model, the absorption/emission of CMB photons happens at the transition frequency of dark matter $\nu_{0}$. At a given redshift, the resultant change in the number density of CMB photons of frequency $\nu_0$ per unit time is given by the number of transitions from spin 0 to spin 1 state minus the transitions from spin 1 to spin 0 states per unit time per unit volume (see Eq.\eqref{3.1}),
	\begin{align}
		\dot{N}_{\chi\gamma}
		&= -\left(n_0B_{01} - n_1B_{10}\right)\bar{J}_\text{CMB} + n_1A_{10},\nonumber\\
		&= -n_\chi A_{10}\frac{1}{\left(1+\left(g_0/g_1\right)e^{h\nu_0/\left(k_BT_{\text{ex}}\right)}\right)}\left(\frac{1+e^{h\nu_0/\left(k_\text{B}T_{\text{ex}}\right)}}{1+e^{h\nu_0/\left(k_\text{B}T_{\text{CMB}}\right)}}-1\right)\,.\label{4.8}
	\end{align}
	We can define a dimensionless form for $\nu_{0}$ using Eq.\eqref{4.1} which is given by,
	\begin{equation}
		x_0(t) \equiv \dfrac{h\nu_0}{k_BT_\text{e}(t)}\,.\label{4.9}
	\end{equation}
	Note that $x_0$ is not time invariant and in the approximation $T_\text{e}=T_\text{CMB}$, $x_0$ varies as $1/(1+z)$ with redshift till $z \approx 200$.
	We can express the change in the photon occupation number $\partial n_{\chi\gamma}/\partial t$ in terms of $\dot{N}_{\chi\gamma}$ as, 
	\begin{equation}
		\dot{N}_{\chi\gamma}(x_0) = \int dx_\text{e} \, \dot{N}_{\chi\gamma}(x_{\text{e}})\,\delta(x_{\text{e}}-x_0(t))= \dfrac{N_\text{PL}}{\mathcal{I}_2}\left(\frac{T_\text{e}}{T_\text{CMB}}\right)^3\int dx_\text{e} \,x_\text{e}^2 \frac{\partial n_{\chi\gamma}}{\partial t}, \label{4.10}
	\end{equation}
	where in the first equality the Dirac delta distribution\footnote{We are assuming that the line profile of dark matter line is well approximated by a Dirac delta distribution} picks out the contribution of the source at $x_\text{e}=x_0$. In the second equality, we have taken the time derivative of Eq.\eqref{4.7} to relate the change in the total number density of CMB photons to the change in the occupation number $n_{\chi\gamma}$ approximating at a leading order that $T_\mathrm{e}/T_\mathrm{CMB}$ and $x_\mathrm{e}$ are independent of time. Using Eq.\eqref{4.10}, we can identify the source term from dark matter to be,
	\begin{equation}
		\frac{\partial n_{\chi\gamma}}{\partial t} = 	\dfrac{1}{x_{\text{e}}^2}\dfrac{\mathcal{I}_2}{N_\text{PL}}\left(\frac{T_\text{CMB}}{T_\text{e}}\right)^3\dot{N_{\chi\gamma}}\,\delta(x_{\text{e}}-x_0(t))\,.\label{4.11}
	\end{equation}
	In the presence of emission/absorption by dark matter, we add the term in Eq.\eqref{4.11} to Eq.\eqref{4.2}. We thus have our final evolution equation, which is the generalized Kompaneets equation with a monochromatic source term,
	\begin{eqnarray}
		\frac{\partial n (x_{\text{e}}, t)}{\partial t}&= K_\text{C}\dfrac{1}{x_{e}^2}\dfrac{\partial}{\partial x_{\text{e}}}x_{\text{e}}^4\left(n+n^2+\dfrac{\partial n}{\partial x_{\text{e}}}\right) + \left(K_{\text{br}} + K_{\text{dC}}\right)\dfrac{e^{-x_{\text{e}}}}{x_{\text{e}}^3}\left[1-n(e^{x_{\text{e}}}-1)\right] \nonumber\\& + x_{\text{e}}\dfrac{\partial n}{\partial x_{\text{e}}}\dfrac{\partial}{\partial t}\left[\text{ln}\dfrac{T_{\text{e}}}{T_\text{CMB}}\right] +\dfrac{1}{x_{\text{e}}^2}\dfrac{\mathcal{I}_2}{b_\text{R}T_\text{e}^3}\dot{N_{\chi\gamma}}\,\delta(x_{\text{e}}-x_0(t)), \label{4.12}
	\end{eqnarray}
	where in the last term we have substituted $N_\text{PL} = b_\text{R}T_\text{CMB}^3$, where $b_\text{R} = 16\pi k_\text{B}^3\zeta(3)/(c^3h^3)$ is a constant. 

	We can identify three frequency scales which will be important to get a qualitative understanding of the CMB spectrum from Eq.\eqref{4.12}, namely, 
	\begin{enumerate}
		\item[(a)] $x_0$: the dark matter transition frequency defined in Eq.\eqref{4.9}.
		\item[(b)] $x_\text{c}$: the critical frequency at which the bremsstrahlung~+~double Compton scattering rates are equal to the Compton scattering rate defined in Eq.\eqref{4.6}. 
		\item[(c)] $x_\text{H}$: the threshold frequency at which the bremsstrahlung~+~double Compton scattering rate is equal to the Hubble rate defined as,
		\begin{equation}
			x_\text{H} = \sqrt{(K_{\text{br}}+K_{\text{dC}})/H}.
		\end{equation}
	\end{enumerate}

    \subsection{Solutions to the generalized Kompaneets equation in the presence of a monochromatic source term}
    \label{ssec:soln}
	The two important epochs in the history of the early Universe with reference to the CMB spectral distortions are:
	\begin{itemize}
		\item The epoch $z_\text{BB}\approx2\times 10^6$ marks the redshift of the CMB blackbody photosphere. As long as the fractional change in the photon number density or energy density due to any exotic process is much less than unity, the inelastic and elastic electromagnetic processes ensure that the CMB has the Planck spectrum at $z>z_\text{BB}$. Therefore, in our case, the spectral distortions get created at redshifts $z<z_\text{BB}$.
		\item The epoch $z_\text{C}\approx 10^5$ marks the redshift above which the Compton scattering rate is stronger than the Hubble expansion rate i.e. $K_\text{C}/H\gtrsim 1$. This ensures that the CMB spectrum in the redshift range $z_\text{C}\lesssim z \lesssim z_\text{BB}$ is the equilibrium Bose-Einstein spectrum.
	\end{itemize}
 We now derive analytic solutions for the $\mu$-type ($z_\text{C}\lesssim z \lesssim z_\text{BB}$) and $y$-type ($z\lesssim z_\text{C}$) spectral features and the evolution equation for the non-thermal spectral features ($z\lesssim z_\text{C}$) imprinted by dark matter into the CMB.
%
%
%
%
%
%
	\subsubsection{$z\gtrsim z_\mathrm{C}: \mu$-type distortion}
  The Bose-Einstein spectrum with $\mu \neq 0$, also called a $\mu$-type distortion, is given by
		\begin{eqnarray}
			n_{\text{BE}}(x_\text{e}) = \frac{1}{e^{x_\text{e}+\mu}-1} \approx \frac{1}{e^{x_\text{e}}-1}-\frac{\mu e^{x_\text{e}}}{\left(e^{x_\text{e}}-1\right)^2}, \label{a11}
		\end{eqnarray}
		where $\mu$ denotes the dimensionless chemical potential. Since we consider small distortions with $\mu\ll 1$, we have expanded $n_\text{BE}$ in a Taylor series to first order in $\mu$.
		The $\mu$ parameter is determined by a combination of two quantities, namely the fractional changes in the total number density and the total energy density of the photons in the CMB. The evolution equation for the $\mu$ parameter (see Appendix \ref{app:mu} for the full derivation) is given by \cite{1970Ap&SS...7...20S},
		\begin{align}
			\dfrac{\text{d}\mu}{\text{d}t} \approx 1.4\left(\dot{\mathcal{E}} - \dfrac{4}{3}\dot{\mathcal{N}}\right)\,,\label{4.17b}
		\end{align}
		where $\mathcal{E}$ and $\mathcal{N}$ represent the ratio of the energy density and number density of the CMB spectrum post photon injection/extraction w.r.t.  the energy density and number density of the Planck spectrum respectively (see Eq.\eqref{a1} and \eqref{a2} in Appendix \ref{app:mu}).
		 
		 \color{black}
		Another way of defining the $\mu$-type distortion in the literature is to express the Bose-Einstein spectrum as a deviation from the Planck spectrum at a temperature $T_\text{ref}$ different from $T_\text{e}$. The temperature $T_\text{ref}$ is defined such that the photon number density of the Bose-Einstein spectrum (at temperature $T_\text{e}$ and the dimensionless chemical potential $\mu$) matches with the photon number density of the Planck spectrum (at temperature $T_\text{ref}$) i.e. $N_\text{PL}(T_\text{ref}) = N_\text{BE}(\mu, T_\text{e})$. The resultant Bose-Einstein spectrum is given by,
		\begin{eqnarray}
			n_\text{BE}(x_\text{ref}) \approx \frac{1}{e^{x_\text{ref}}-1} + \frac{\mu e^{x_\text{ref}}}{\left(e^{x_\text{ref}}-1\right)^2}\left(\frac{x_\text{ref}}{2.19}-1\right)\equiv n_\text{PL}(x_\text{ref}) + \mu n_\mu(x_\text{ref})\,,\label{4.16}
		\end{eqnarray}
		where $x_\text{ref} \equiv h\nu/(k_\text{B}T_\text{ref})$ and $T_\text{e} /T_\text{ref} = 1 + 0.456\mu$. We note that the total spectrum and the observable parameter $\mu$ in Eq.\eqref{a11} and \eqref{4.16} is the same. We have just used a different choice of the reference Planck spectrum in splitting the total spectrum into a Planck spectrum and a distortion component.
		
		In the context of our dark matter model, the electromagnetic transitions in the dark sector would change the CMB photon occupation number at a specific frequency $x_0$ (see Eq.\eqref{4.9} for definition). If $x_0$ lies in the Rayleigh-Jeans tail of the CMB spectrum such that $x_0<x_\text{c}<1$ (see Eq.\eqref{4.6}), the inelastic processes (double Compton~+~bremsstrahlung) will compensate for the change in the photon occupation number due to dark matter in an attempt to erase any departure from the Planck spectrum. Thus, 
		to compute the $\mu$-type distortion, we have to obtain $\dot{\mathcal{E}}$ and $\dot{\mathcal{N}}$ by solving the full generalized Kompaneetz equation Eq.\eqref{4.12} taking into account all number and non-number changing processes. At a given redshift, we expect the CMB to have a Planck spectrum at low frequencies i.e. $\mu\rightarrow 0$ as $x_\text{e}\rightarrow0$ and a Bose Einstein spectrum at high frequencies i.e. a constant $\mu$ at $x_e\gtrsim1$. This motivates the ansatz for the photon occupation number in Eq.\eqref{4.12} to have a Bose-Einstein form with a frequency-dependent chemical potential $\mu \equiv \mu(x_{\text{e}}, t)$ with the boundary condition: $\mu = 0$ at $x_\text{e} = 0$ and $\mu = \mu(t)$ at $x_\text{e} = 1$. This approach is similar to that adopted in \cite{1969Ap&SS...4..301Z,1970Ap&SS...7...20S,2012JCAP...06..038K}. Thus, we have,
		\begin{equation}
			n(x_\text{e}, t) = \dfrac{1}{e^{x_\text{e}+\mu(x_\text{e}, t)}-1}. \label{4.18}
		\end{equation}
		Assuming the $\mu$-type distortion to be small ($\mu\ll1$), we expand Eq.\eqref{4.18} in a Taylor series and take the low frequency limit ($x_\text{e}\ll 1$). We approximate $\mu(x_\text{e}, t) \approx \mu_x(x_\text{e})\,\mu_t(t)$ to decouple the time evolution of $\mu$ from the frequency dependence. Thus, Eq.\eqref{4.18} becomes,
		\begin{align}
			n(x_\text{e}, t) = \dfrac{1}{e^{x_\text{e}}-1} - \mu_x(x_\text{e})\,\mu_t(t)\dfrac{1}{x_\text{e}^2} + \mathcal{O}(\mu^2). \label{4.19}
		\end{align}
		By substituting Eq.\eqref{4.19} into Eq.\eqref{4.12}, we get,
		\begin{eqnarray}
			-\dfrac{\mu_x}{x_\text{e}^2}\frac{\text{d}\mu_t}{\text{d}t}
			&= -K_\text{C}\,\mu_t\,\dfrac{1}{x_\text{e}^2}\dfrac{\text{d}}{\text{d}x_\text{e}}\,x_\text{e}^2\,\dfrac{\text{d}\mu_x}{\text{d}x_\text{e}} + \left(K_{\text{br}} + K_{\text{dC}}\right)\,\dfrac{\mu_x\mu_t}{x_\text{e}^4}\nonumber\\
			&-\dfrac{1}{x_\text{e}}\dfrac{\text{d}}{\text{d}t}\left[\ln\,\dfrac{T_\text{e}}{T_\text{CMB}}\right] + \dfrac{1}{x_\text{e}^2}\dfrac{\mathcal{I}_2}{b_RT_\text{e}^3}\dot{N}_{\chi\gamma}\,\delta(x_\text{e}-x_{0}(t)). \label{4.20}
		\end{eqnarray}
		We can rearrange the terms in Eq.\eqref{4.20} to partially separate $x_\text{e}$ and $t$ dependent terms on the two sides of the equation, which gives,
		\begin{eqnarray}
			-\dfrac{1}{\mu_t}\dfrac{\text{d}\mu_t}{\text{d}t} + \dfrac{x_\text{e}}{\mu_x\mu_t}\dfrac{\text{d}}{\text{d}t}\left[\ln\,\dfrac{T_\text{e}}{T_\text{CMB}}\right] &= -K_\text{C}\,\dfrac{1}{\mu_x}\dfrac{\text{d}}{\text{d}x_\text{e}}\,x_\text{e}^2\,\dfrac{\text{d}\mu_x}{\text{d}x_\text{e}} + \left(K_{\text{br}} + K_{\text{dC}}\right)\,\dfrac{1}{x_\text{e}^2} \nonumber\\ & +  \dfrac{1}{\mu_x\mu_t}\dfrac{\mathcal{I}_2}{b_RT_\text{e}^3}\dot{N}_{\chi\gamma}\,\delta(x_\text{e}-x_{0}(t)) = \lambda(x_\text{e}, t)\,.\label{4.21}
		\end{eqnarray}
		The second term on the LHS being linear in $x_\text{e}$ is suppressed since $x_\text{e}\leq 1$. Thus, to the lowest order, we can approximate the LHS of Eq.\eqref{4.21} to only have time dependence. The terms on the RHS that correspond to different rates (see expressions for $K_\text{C}$, $K_\text{br}$, and $K_{\text{dC}}$ in Eq.\eqref{4.3}, \eqref{4.4}, and \eqref{4.5}) have a slow variation with time in the redshift range of interest and can be considered to solely depend on $x_\text{e}$ at the lowest order. This implies that the LHS and RHS of Eq.\eqref{4.21} are almost entirely functions of $t$ and $x_\text{e}$ respectively (justifying our initial assumption that we can approximate $\mu(x_\text{e}, t) \approx \mu_x(x_\text{e})\,\mu_t(t)$). Thus, $\lambda$ should be independent of both $t$ and $x_\text{e}$ i.e. a constant. We can rewrite the $x_\text{e}$ dependent part of Eq.\eqref{4.21} as,
		\begin{eqnarray}
			\dfrac{\text{d}}{\text{d}x_\text{e}}\,x_\text{e}^2\,\dfrac{\text{d}\mu_x}{\text{d}x_\text{e}} - \dfrac{x_\text{c}^2}{x_\text{e}^2}\,\mu_x + \dfrac{\lambda}{K_\text{C}}\mu_x= \dfrac{\mathcal{I}_2}{K_\text{C}\mu_t}\dfrac{\dot{N}_{\chi\gamma}}{b_RT_\text{e}^3}\delta(x_\text{e}-x_{0}(t)). \label{4.22}
		\end{eqnarray}
		We can simplify Eq.\eqref{4.22} since the Compton scattering rate dominates over all other processes in the redshift range of interest i.e. $\lambda/K_\text{C}\ll 1$. 
		The leading part of the evolution equation of $\mu_x$ is therefore given by,
		\begin{eqnarray}
			\dfrac{\text{d}}{\text{d}x_\text{e}}\,x_\text{e}^2\,\dfrac{\text{d}\mu_x}{\text{d}x_\text{e}} - \dfrac{x_\text{c}^2}{x_\text{e}^2}\,\mu_x = \dfrac{\mathcal{I}_2}{K_\text{C}\mu_t}\dfrac{\dot{N}_{\chi\gamma}}{b_RT_\text{e}^3}\delta(x_\text{e}-x_{0}(t))\,.\label{4.23}
		\end{eqnarray}
		To solve Eq.\eqref{4.23}, we use the Green's function technique to obtain $\mu_x$ for $x_\text{e} \in [0, 1]$ as outlined in Appendix \ref{app:A}. 
		
		The net rate of change of photon number density $\mathcal{\dot{N}}$ can be obtained from Eq.\eqref{4.7}, and is given by,
		\begin{align}
			\mathcal{\dot{N}}
			= \frac{1}{\mathcal{I}_2}\left(\dfrac{T_{\text{e}}}{T_\text{CMB}}\right)^3\int dx_{\text{e}} \,x_{\text{e}}^2 \,\dfrac{\partial n}{\partial t} + \frac{3}{\mathcal{I}_2}\left(\dfrac{T_{\text{e}}}{T_\text{CMB}}\right)^3\dfrac{\partial}{\partial t}\left[\ln \dfrac{T_{\text{e}}}{T_\text{CMB}}\right]\int dx_{\text{e}}\, x_{\text{e}}^2 \,n, \label{4.25}	
		\end{align}
		We plug Eq.\eqref{4.19} into Eq.\eqref{4.25} and evaluate the integrals using the solution for $\mu_x$ to get the final expression for $\mathcal{\dot{N}}$ (for details see Appendix \ref{app:A}),
		\begin{align}
			\mathcal{\dot{N}} \approx \dfrac{\mu_t}{\mathcal{I}_2}\dfrac{\left(K_{\text{br}} + 	K_{\text{dC}}\right)}{x_\text{c}}+ \dfrac{\dot{N}_{\chi\gamma}}{N_{\text{PL}}}e^{x_\text{c}}e^{-x_\text{c}/x_0}\,.\label{4.26}
		\end{align}
		The exponential factor ($e^{-x_\text{c}/x_0}$) in the second term captures the effect of inelastic processes in counteracting the effect from dark matter transitions. In particular, if dark matter absorbs the photons from the Rayleigh Jeans tail ($x_0<x_\text{c}$) of the CMB, where inelastic processes are efficient in creating photons, they can completely erase any change in the photon number created by dark matter. The effect diminishes as transition frequency becomes larger since the inelastic processes (with rates $\propto 1/x_\text{e}^2$) become less efficient at higher frequencies. The approximate solution to the photon injection scenario provided earlier in \cite{2015MNRAS.454.4182C} is close to the analytic solution derived in Eq.\eqref{4.26} and they become identical in the limit $x_c\ll 1$.
		
		The rate at which energy is extracted from CMB is given by the transition energy $\Delta E = k_B T_*$ (see Eq.\eqref{2.2} for definition) times the rate of dark matter transitions per unit volume,  
		\begin{align}
			\mathcal{\dot{\mathcal{E}}} = \dfrac{k_B T_*\dot{N}_{\chi\gamma}}{E_{\text{PL}}} = \left(\frac{b_\text{R}k_B }{a_\text{R}}\right)\frac{T_*}{T_\text{CMB}} \frac{\dot{N}_{\chi\gamma}}{N_{\text{PL}}}\,,\label{4.27}
		\end{align}
		where in the last equality we have used $N_\text{PL}/E_\text{PL} = b_\text{R}/(a_\text{R}T_\text{CMB})$.
		Note that unlike Eq.\eqref{4.26} where the inelastic processes compensate for the photon number change at low frequencies, they cannot compensate for the energy loss. This is because when the baryons and electrons lose energy to create extra photons, the lost energy is retrieved back via Compton scattering with CMB photons from the high energy part of the CMB spectrum creating a negative $y$-type distortion initially. This $y$-type distortion can further Comptonize and evolve towards a $\mu$-type distortion \cite{2012JCAP...09..016K}.
		
		Substituting Eq.\eqref{4.26} and Eq.\eqref{4.27} into \eqref{4.17b}, we get the final evolution equation for $\mu_t$,
		\begin{align}
			\dfrac{\text{d}\mu_t}{\text{d}t} = 1.4\left[- \dfrac{4}{3}\frac{\mu_t}{\mathcal{I}_2}\sqrt{\left(K_{\text{br}} + 		K_{\text{dC}}\right)K_\text{C}} + \dfrac{\dot{N}_{\chi\gamma}}{N_{\text{PL}}}\left(\left(\frac{b_\text{R}k_B }{a_\text{R}}\right)\frac{T_*}{T_\text{CMB}}- \dfrac{4}{3}e^{x_\text{c}}e^{-x_\text{c}/x_0}\right)  \right]\,.\label{4.28}
		\end{align}
		The analytical solution of the differential equation \eqref{4.28} is given by \cite{1982A&A...107...39D, 2012JCAP...06..038K},
		\begin{align}
			\mu_t(z=0) &= \mu(z_{\text{max}})e^{-\mathcal{T}(z_{\text{max}})} \nonumber\\ &+ 1.4\int_{z_\text{c}}^{z_\text{max}}\frac{dz}{\left(1+z\right)H}\frac{\dot{N}_{\chi\gamma}}{N_{\text{PL}}}\left\{\left(\frac{b_\text{R}k_B }{a_\text{R}}\right)\frac{T_*}{T_\text{CMB}}- \dfrac{4}{3}e^{x_\text{c}}e^{-x_\text{c}/x_0}\right\}e^{-\mathcal{T}(z)}\,,\label{4.29}
		\end{align}
		where $e^{-\mathcal{T}}$ denotes the blackbody visibility function (see Eq.\eqref{bb_od} in Appendix \ref{app:A} for the definition of the blackbody optical depth $\mathcal{T}$ derived in \cite{1982A&A...107...39D, 2012JCAP...06..038K}). The blackbody visibility function goes to zero at redshifts $z \gtrsim z_\text{BB}$ implying that any distortion w.r.t. the Planck spectrum is exponentially suppressed. 
		We can set $z_\text{max}=$ few$\times z_\text{BB}$ when the CMB spectrum is a perfect blackbody and $\mu(z_\text{max})=0$. Since the contribution from $z\gg z_\text{BB}$ is exponentially suppressed, setting $z_\text{max}=z_\text{BB}$ is an adequate approximation for most energy release scenarios. We note that the $\mu$-type distortions thus created can be positive or negative depending on which term (energy change or number change) in Eq.\eqref{4.29} dominates, even though we always have net absorption of energy.
\\\\
		Apart from dark matter transitions which can absorb or emit photons, CMB also loses energy if dark matter is in kinetic equilibrium with the baryon-photon fluid and hence with the CMB at very early redshifts. If dark matter is kinetically coupled to the CMB in the redshift range $z_\text{BB}\lesssim z \lesssim z_\text{C}$, then the energy lost by CMB will lead to $\mu$-type distortions. This is because dark matter, being non-relativistic, cools as $T_\chi \propto (1+z)^2$ if uncoupled to the CMB. When kinetically coupled to the CMB, it follows the CMB temperature, $T_\chi \approx T_\text{CMB} \propto (1+z)$. Thus, CMB has to constantly supply energy to dark matter to maintain it at the same temperature. The physics of distortions in the case of adiabatic cooling of dark matter is identical to the distortions created in the CMB due to the adiabatic cooling of baryons. The reference \cite{2012A&A...540A.124K} developed an analytic method to compute the CMB spectral distortions caused by adiabatic cooling of baryons. We use a similar formalism to calculate the spectral distortions for adiabatic cooling of dark matter. In general the scattering of dark matter particles with baryons (see section \ref{subsec:Tdm} for more details) causes a transfer of energy from baryons to dark matter. The energy lost by baryons to dark matter is gained from the CMB through Compton scattering. The rate of change of the fractional energy density of the CMB can be obtained from Eq.\eqref{3.9}, and is given by,
		\begin{equation}
			\dot{\mathcal{E}} = \frac{\Gamma_{\chi_b}}{E_\text{PL}}\,.\label{4.31a}
		\end{equation}
		Note that since the CMB loses energy through Compton scattering, the photon number remains unchanged. In the regime $T_\chi\approx T_\mathrm{CMB}$, the dark matter, baryon, and photon system remains in thermodynamic equilibrium. This allows us to use the conservation of entropy of the dark matter, baryon, and photon system to compute the resultant thermal distortions. This principle was first applied to the case of adiabatic cooling of baryons in \cite{2012A&A...540A.124K}. Extending the same argument to dark matter, the  fractional change in the CMB energy density in the redshift range $z_\text{min}\lesssim z \lesssim z_\text{max}$ is given by,
        \begin{equation}
            \mathcal{E}_\text{adiabatic} = -\frac{3n_\chi k_\mathrm{B}}{2a_\mathrm{R}T_\text{CMB}^3}\ln\left(\frac{1+z_\text{max}}{1+z_\text{min}}\right)\,.\label{4.30}
        \end{equation}
        The resultant $\mu$-type distortion (see Eq.\eqref{4.17b}) is given by $1.4\,\mathcal{E}_\text{adiabatic}$, where we set $z_\text{max} = z_\text{BB}$ and $z_\text{min} = z_\text{C}$ or $z_\text{dec}$ (redshift at which dark matter kinetically decouples from the baryon-photon plasma), whichever is earlier.
		
		\subsubsection{$z\lesssim z_\mathrm{C}$ and $x_0\lesssim x_\mathrm{H}: y$-type distortion}
		At redshifts $z\lesssim z_\text{C} \sim 10^5$, Comptonization becomes inefficient and can no longer establish a Bose-Einstein spectrum. In such cases, the nature of spectral distortion created by dark matter becomes a strong function of the photon frequency dark matter absorbs or emits.
  
		When the dark matter transitions change the photon number in the Rayleigh-Jeans tail such that $x_0\lesssim x_\text{H}$, the Bremsstrahlung and double Compton scattering erase this change in an attempt to keep the photon number unchanged and restore the Planck spectrum in the low energy part of the spectrum. This results in the cooling of the plasma. The Compton scattering brings the plasma temperature back in equilibrium with the CMB creating a $y$-type distortion in the CMB. 
  We note that in the entire process the number density of CMB photons remains unchanged. The net effect is a loss in the energy from CMB to dark matter through photon absorption and subsequent collisional de-excitation of dark matter.

The $y$-type distorted photon spectrum can be obtained by solving Eq.\eqref{4.12} in the presence of Compton scattering term as described in Appendix \ref{app:y}. We use the dimensionless frequency $x= h\nu/(k_\text{B}T_\text{CMB})$ from now on for convenience as is conventional for $y$-type distortions. We note that since $T_\text{e}\approx T_\text{CMB}$ in the early Universe, $x_\text{e}\approx x$. The resultant photon occupation number is given by,
\begin{align}
	n(x) = n_{\text{PL}}(x) + y n_y(x),
\end{align}
where the $y$-type spectrum is given by,
\begin{equation}
	n_y(x) = \frac{xe^{-x}}{(e^x-1)^2}\bigg[x\left(\frac{e^x+1}{e^x-1}\right)-4\bigg].
\end{equation}
Since there is no net change in the number of CMB photons, the $y$ parameter only depends on the fractional change in the energy density of CMB photons. The evolution equation for the $y$-parameter is given by,
		\begin{align}
			\dfrac{\text{d}y}{\text{d}t} = \frac{1}{4}\dot{\mathcal{E}} \, \implies
                y = \int_{z_\text{br}}^{z_{\text{c}}} \frac{dz}{\left(1+z\right)H}\left( \frac{1}{4}\dot{\mathcal{E}}\right)\,.\label{4.36}
		\end{align}
	For a given $x_0$, the lower limit on the integral is at $z_{\text{br}}(x_0)$ which denotes the redshift at which the bremsstrahlung rate is equal to the Hubble expansion rate. Note that at redshifts $z\lesssim z_\text{c}$, bremsstrahlung makes the dominant contribution to inelastic scattering (see Eq.\eqref{4.4} and Eq.\eqref{4.5}). We can estimate $z_{\text{br}}$ as a function of dark matter transition frequency $\nu_{0}$ by solving the equation below for $z$,
		\begin{equation}
			\frac{K_\text{br}(x_0, z)}{x_0^2}\approx H(z).
		\end{equation} 
		If dark matter remains kinetically coupled to the CMB at $z\lesssim z_\text{C}$ the main effect comes from the adiabatic cooling of dark matter. The resulting $y$-type distortions can be computed from Eq.\eqref{4.30} where $z_\text{min}$ is the decoupling redshift and $z_\text{max}=z_\text{c}$. Note that we are ignoring the fact that the transition from $\mu$-type era to $y$-type era is not sharp but there is an intermediate (also called the $i$-type) era \cite{2012JCAP...09..016K,2013MNRAS.434..352C} in between. However, we will see that since the shapes of these different types of distortions are very similar, this is a good approximation, especially for getting constraints from COBE-FIRAS and future COBE-like experiments such as PIXIE. We note that the $y$-type distortions thus created are negative, i.e. $y<0$.
		
		\subsubsection{$z \lesssim z_\mathrm{C}$ and $x_0\gtrsim x_\mathrm{H}$: Non-thermal distortion}
		When the dark matter transitions happen at frequencies $x_0\gtrsim x_\text{H}$ at redshifts $z\lesssim z_\mathrm{C}$, both elastic and inelastic scattering processes are inefficient. In this regime, we have new kinds of spectral distortions which are non-thermal in nature. One such non-thermal signature in the standard model of cosmology expected to occur in the low-frequency tail of the CMB ($x\ll 1$) is the well-known Global 21 cm signal  \cite{2004MNRAS.352..142B, 2004ApJ...608..622Z,2006PhR...433..181F,2012RPPh...75h6901P}. 
		As discussed in the reference \cite{2024PhRvD.109f3512G}, the non-thermal distortions due to dark matter and the global 21 cm absorption feature from neutral hydrogen during the Dark Ages have very similar phenomenology. We note that in the 21 cm cosmology literature, the spectral distortion in the CMB monopole spectrum in the 21 cm band (10 MHz $ \lesssim \nu<1.4$ GHz) is referred to as the global signal.
		
		The peculiar shapes of non-thermal distortions are a unique probe of dark matter. Unlike the thermal $\mu$ and $y$-type distortions which can be degenerate with the signatures expected to occur from interactions between baryons and photons in Standard Cosmology, the non-thermal distortions have unique features making them, in principle, distinguishable from other standard and new physics sources of CMB spectral distortion. The global feature/non-thermal distortion can be parameterized in terms of the brightness temperature which is related to the specific intensity $I_\nu$ of the radiation by the following relation,
		\begin{equation}
			T_\text{b}(\nu) \equiv \frac{c^2}{2\nu^2k_\text{B}}I_\nu, \label{4.34}
		\end{equation} 
		where $T_\text{b}(\nu)$ denotes the brightness temperature which tracks the intensity as a function of frequency in temperature units. Note that Eq.\eqref{4.34} does not necessarily assume an underlying blackbody spectrum. Even if one assumes a Planck spectrum, $T_\text{b}$ is not necessarily the same as the blackbody temperature. For example, in case of the CMB, $T_\text{b}=T_\text{CMB}$ only in the Rayleigh-Jeans tail ($x \ll 1$), however, it is different from $T_\text{CMB}$ at higher frequencies (including $x\sim 1$) as can be seen in the equation below:
        \begin{equation}
            T^\text{Planck}_\mathrm{b}(\nu) = \frac{h\nu}{k_\mathrm{B}}\frac{1}{e^{h\nu/(k_\mathrm{B}T_\mathrm{CMB})}-1}\,.\label{tb}
        \end{equation}
		The brightness temperature is affected when the CMB photons of frequency equal to the transition frequency encounter dark matter and get absorbed or emitted via stimulated or spontaneous emission. Subsequently, the bremsstrahlung scattering between electrons and ions, if efficient, can erase the distortion created by the dark matter. The radiative transfer equation for the CMB brightness temperature including dark matter interactions and the bremsstrahlung process is given by \cite{2024PhRvD.109f3512G},
		\begin{equation}
                \begin{split}
			\dfrac{dT_\text{b}(\nu)}{dz} 
			= \frac{T_\text{b}(\nu)}{1+z} + \dfrac{d\tau_\chi}{dz}\left(-T_\text{b} + \frac{h\nu}{k_B}\frac{1}{\left(e^{h\nu/\left(k_BT_\text{ex}\right)}-1\right)}\right)\\ \quad
			+ \dfrac{d\tau_\text{br}}{dz}\left(-T_\text{b} + \frac{h\nu}{k_B}\frac{1}{\left(e^{h\nu/\left(k_BT_\text{e}\right)}-1\right)}\right),
            \end{split}
		\end{equation}
		where $\nu$ denotes the frequency of the photon at redshift $z$.  The first term on the RHS represents the adiabatic fall $\propto 1/(1+z)$ in the brightness temperature due to Hubble expansion. The other two terms on the RHS capture the absorption and emission effects due to dark matter and bremsstrahlung respectively. If dark matter absorbs/emits photons at some redshift $z_0$, then at later redshifts the distortion created at $z_0$ is transmitted to frequency $\nu(z) = \nu_{0}(1+z)/(1+z_0)$. Note that the observed intensity at a given frequency will also have contributions from the random motion of dark matter particles resulting in the broadening of the line around $\nu(z)$. However, the Doppler line width is negligible compared to the total width of the spectral signature arising due to the continuous absorption/emission of photons over a significantly large redshift range. The quantity $d\tau_\chi/dz$ denotes the optical depth due to dark matter interactions per unit redshift, which is given by,
		\begin{align}
			\dfrac{d\tau_\chi}{dz} = -\frac{g_1/g_0}{1+\left(g_1/g_0\right)e^{-T_*/T_\text{ex}}}\left(1-e^{-T_*/T_\text{ex}}\right)\frac{n_\chi A_{10}c^4}{8\pi\nu_0^3H}\left(\frac{1+z}{1+z_0}\right)\delta(z-z_0).
		\end{align}
		The optical depth per unit redshift due to bremsstrahlung (see Eq.\eqref{4.4}) is given by \cite{2024PhRvD.109f3512G},
		\begin{equation}
			\frac{d\tau_{\text{\text{br}}}}{dz}(x) 
			= - \frac{K_\text{br}}{H(1+z)}\left(\frac{1-e^{-x}}{x^3}\right).
			\label{4.41}
		\end{equation}
		The non-thermal distortion in the observer's frame is parameterized as a change in the intensity of CMB w.r.t. to the black body spectrum,
		\begin{align}
			\delta I\left(\nu_{\mathrm{obs}}, z=0\right)_{\text{observer's frame}} 
			\equiv \frac{2\nu_{\mathrm{obs}}^2k_\mathrm{B}}{c^2}\,\left(T_{b}\left(\nu_{\mathrm{obs}}, z=0\right) -\frac{h\nu_{\mathrm{obs}}}{k_B}\frac{1}{e^{h\nu_{\mathrm{obs}}/\left(k_BT_\text{CMB}(z=0)\right)}-1}\right),
		\end{align}
		where $\nu_{\mathrm{obs}} = \nu_0/(1+z_0)$ denotes the observed frequency of CMB at $z=0$.
	\section{Physics of the non-thermal distortions}
	As shown in previous sections, for small distortions, the thermal $\mu$ and $y$-type distortions have fixed shapes and we just need to calculate their amplitudes. Non-thermal distortions, on the other hand, produce a rich variety of spectral shapes which are sensitive to the dark matter parameters. We devote this section to describing these features as well as the underlying physics in detail.
 
	The non-thermal distortion due to dark matter can occur both as an absorption feature or as an emission feature. The quantity that determines the nature of the non-thermal spectral distortion is the excitation temperature of the two dark matter states. In particular, if $T_{\text{ex}}<T_{\text{CMB}}$, dark matter absorbs the CMB photons. If $T_{\text{ex}}>T_{\text{CMB}}$, dark matter emits photons. Surprisingly, we find that $T_\mathrm{ex}>T_\mathrm{CMB}$ is a possibility even though the kinetic temperature of dark matter is always below the CMB temperature.
  
        To get a qualitative understanding of the interactions that govern the excitation temperature, we rewrite the evolution equation of $T_\text{ex}$ (Eq.\eqref{3.2}) as,
        \begin{equation}
        \begin{split}
            \frac{d\ln{T_\mathrm{ex}}}{d\ln{z}} = \mathcal{R}_{\text{DM}}\left(1-e^{-T_*\left(\frac{1}{T_\chi}-\frac{1}{T_\text{ex}}\right)}\right) &+ \mathcal{R}_{\text{CMB}}\left(1-e^{-T_*\left(\frac{1}{T_\text{CMB}}-\frac{1}{T_\text{ex}}\right)}\right),\\
            \text{where} \qquad 
            \mathcal{R}_{\text{DM}} &\equiv \frac{T_\text{ex}}{T_{*}}\frac{C_{10}}{H}\left(1+ \frac{g_1}{g_0}e^{-T_*/T_\text{ex}}\right)\\
            \text{and} \qquad \mathcal{R}_{\text{CMB}} &\equiv \frac{T_\text{ex}}{T_{*}}\frac{A_{10}}{H}\left(\frac{1+\frac{g_1}{g_0}e^{-T_*/T_\text{ex}}}{1-e^{-T_*/T_\text{CMB}}}\right).\label{5.1}
        \end{split} 
        \end{equation}
	The first term on the right side of Eq.\eqref{5.1} implies that the inelastic collisional transitions between dark matter particles (due to their self-interactions) tend to bring the excitation temperature in equilibrium with the dark matter temperature and $\mathcal{R}_{\text{DM}}$ denotes the ratio of collision rate to the expansion rate. The second term represents the radiative transitions which tend to bring the excitation temperature in equilibrium with the CMB temperature. We use $\mathcal{R}_{\text{CMB}}$ to denote the ratio of the radiative transition rate to the expansion rate.
    
	The fundamental dark matter model properties that govern the rates $\mathcal{R}_{\text{DM}}$ and $\mathcal{R}_{\text{CMB}}$ cover a large region of parameter space allowing interesting dynamics for $T_\text{ex}$. This opens the possibility for a rich set of spectral features that could be detected in the next-generation experiment PIXIE \cite{2011JCAP...07..025K}. To understand the underlying physics, we categorize the spectral signatures into three kinds: absorption, emission, and a combination of both, which are shown in the three panels of Fig.\ref{fig:signal_types}. For this figure, we choose a fixed set of parameters: $m_\chi=1$ MeV, $\epsilon=10^{-7}$, $\Delta E=10$ eV, and vary the collision parameter which takes values $\alpha_\text{C} = \left\{1, 10^{-2}, 10^{-4}\right\}$.  Each row in Fig.\ref{fig:signal_types} consists of three panels, where the first panel shows the evolution of the rates $\mathcal{R}_{\text{DM}}$ and $\mathcal{R}_{\text{CMB}}$, the second panel shows the corresponding evolution of $T_\text{CMB}$, $T_\chi$, and $T_\text{ex}$, and the third panel shows the resultant spectral signature in the CMB band. The three rows of Fig.\ref{fig:signal_types} represent the three different choices of the $\alpha_\text{C}$, namely, $\alpha_\text{C}=1$ for the top row, $\alpha_\text{C}=10^{-2}$ for the middle row, and $\alpha_\text{C}=10^{-4}$ for the bottom row. The top (bottom) row in Fig.\ref{fig:signal_types} demonstrates the case where the non-thermal spectral distortion is characterized by an almost pure absorption (emission) feature. In the middle row, we have a combination of emission and absorption. In order to extract the underlying physics, we divide the full period of evolution into three regions: high redshifts ($10^8<z<10^6$), intermediate redshifts ($10^6<z<10^3$), and low redshifts ($z<10^3$):
    \begin{figure}[t]
	\includegraphics[width=1.0\textwidth]{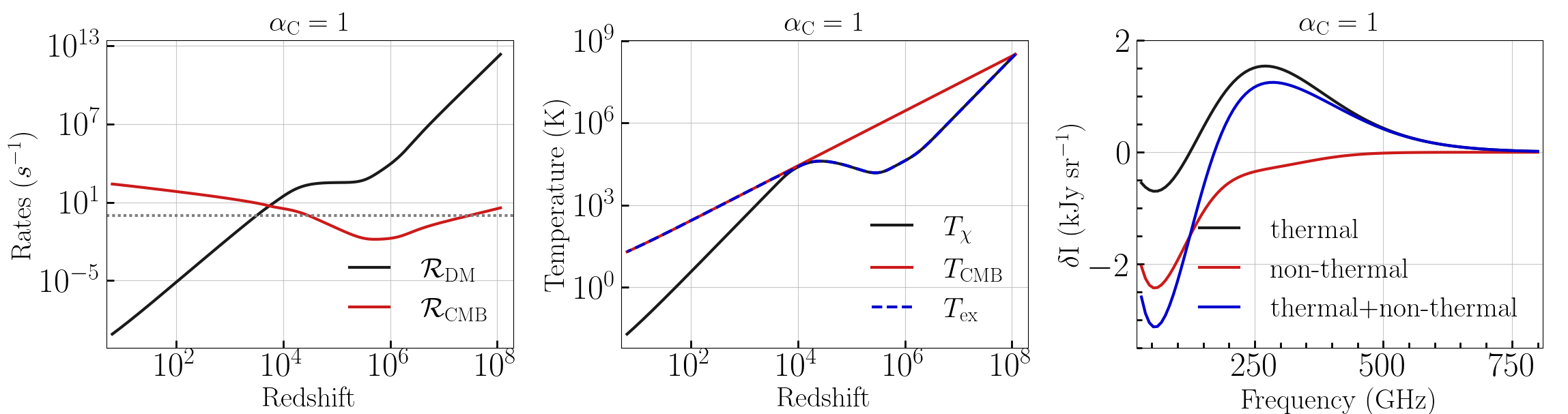}\\
        \includegraphics[width=1.0\textwidth]{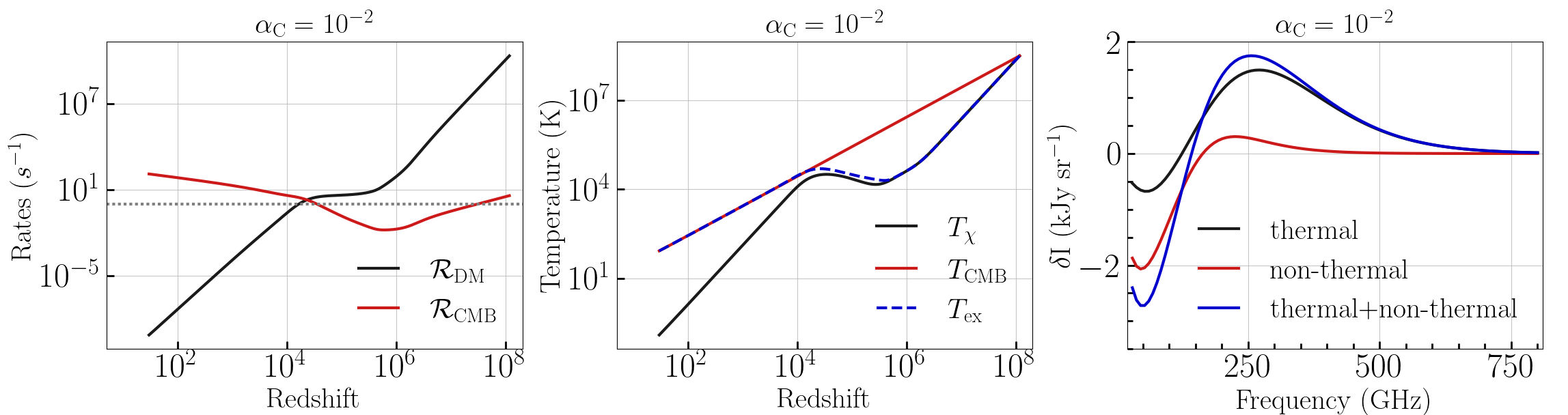}\\
        \includegraphics[width=1.0\textwidth]{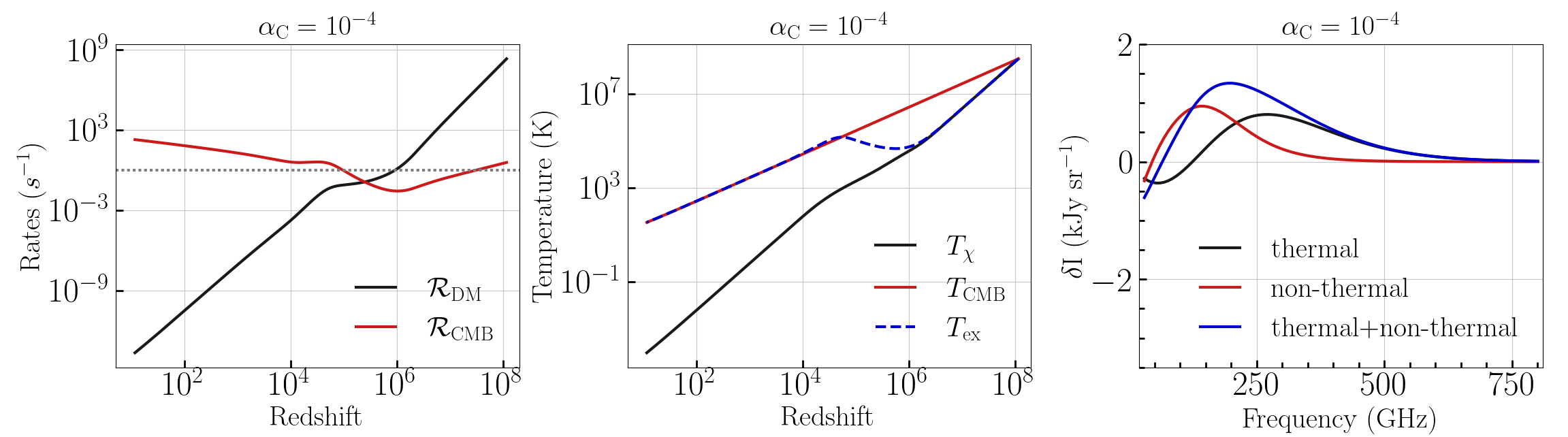}
	\caption{Thermal and non-thermal distortions created by dark matter in the CMB spectrum with model parameters fixed at $m_\chi=1$ MeV, $\epsilon = 10^{-7}$, and $\Delta E = 10$ eV for three different values of $\alpha_\text{C}$.}
		\label{fig:signal_types}
	\end{figure}
    \begin{figure}[t]
		\centering
         \includegraphics[width=0.8\textwidth]{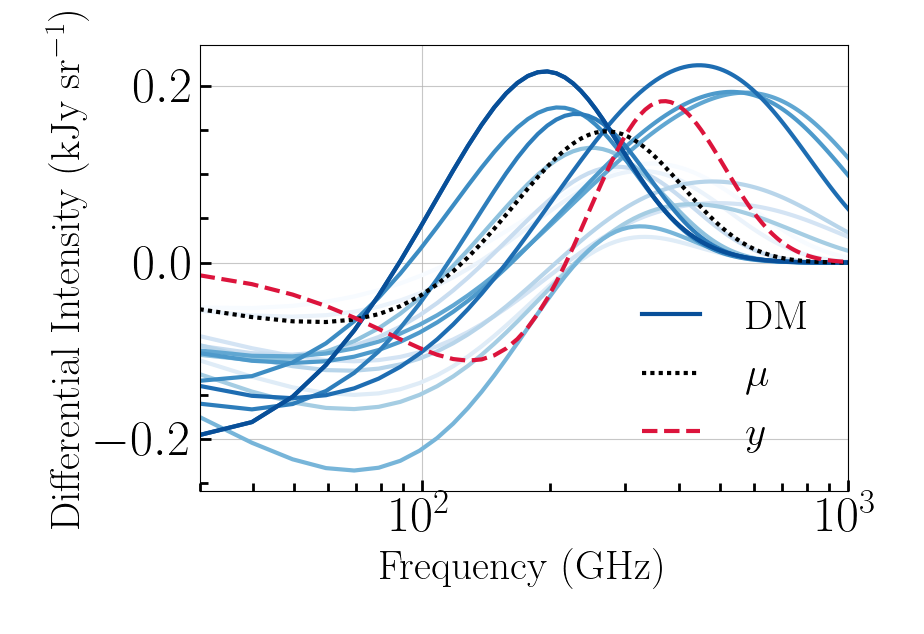}
		\caption{CMB spectral distortions from dark matter scanning the parameter space allowed by COBE and potentially detectable by PIXIE (see Fig.\ref{fig:u1}) are shown in blue. The darkness of the blue color scale increases with the peak-to-peak amplitude of the shape. For comparison, the thermal distortions at the 50-$\sigma$ detection limit of PIXIE ($\mu=5\times10^{-7}$ and $y=1\times 10^{-7}$) are also shown as black dotted and red dashed lines respectively.}
		\label{fig:signal_zoo}
	\end{figure}
	\begin{itemize}[leftmargin=20pt]
		\item \textbf{Absorption feature: $\alpha_\text{C} = 1$}
		
		At high redshifts, the high number density and temperature of dark matter result in a higher collision rate between dark matter particles ($\mathcal{R}_\text{DM}\gg  \mathcal{R}_\text{CMB}$) as shown in Fig.\hyperref[fig:signal_types]{2a}. Since collisions are much stronger than the radiative transitions, they bring $T_\text{ex}$ in equilibrium with $T_\chi$ as shown in Fig.\hyperref[fig:signal_types]{2b}. The dark matter particles start absorbing the CMB photons. As the dark matter number density falls as $\propto (1+z)^3$ and the temperature falls as $\propto (1+z)^2$, the inelastic collision rate parameter falls as $\mathcal{R}_\text{DM} \propto n_\chi T_\text{ex} T_\chi^{1/2}/H \propto T_\text{ex}(1+z)^2 \sim (1+z)^4$, where $H\propto (1+z)^2$ in the radiation dominated era assuming $T_\text{ex}\approx T_\chi \propto (1+z)^2$. Since $T_*/T_\mathrm{CMB}\ll 1$ at high redshifts, the radiative transition rate parameter falls as $\mathcal{R}_\text{CMB} \propto  T_\text{ex}T_\text{CMB}/H \propto T_\text{ex}(1+z)^{-1} \sim(1+z)$ which is less steep compared to $\mathcal{R}_\text{DM}$. 
		
		At intermediate redshifts, the energy transferred from CMB to dark matter in a single collisional de-excitation is of the order of its thermal energy ($k_\text{B}T_* \sim k_\text{B}T_\chi$) which is sufficient to heat the dark matter causing a rise in $T_\chi$ and therefore, in $T_\text{ex}$. Since both $\mathcal{R}_\text{DM}$ and $\mathcal{R}_\text{CMB}$ are $\propto T_\text{ex}$, they fall less steeply and can even start increasing with time. Around $z\sim 10^4$, the radiative transitions due to CMB photons take over the collisional transitions and the Hubble expansion rate, bringing $T_\text{ex}$ in equilibrium with $T_\text{CMB}$. 
		
		At low redshifts, collisional transitions become weaker than the Hubble expansion rate ($\mathcal{R}_\text{DM} < 1$) and can no longer keep dark matter coupled to the CMB. Thereafter, $T_\chi$ cools adiabatically as $(1+z)^2$. Since the Universe is matter dominated $H\propto (1+z)^{3/2}$, the inelastic collision rate falls as $\mathcal{R}_\text{DM} \propto (1+z)^{7/2}$. On the other hand, $T_*/T_\mathrm{CMB}\gg 1$ at low redshifts, so the radiative transition rate rises as $\mathcal{R}_\text{CMB} \propto (1+z)^{-1/2}$ and keeps $T_\text{ex}$ in equilibrium with $T_\text{CMB}$. 
		
		The absorption happening at $10^4\lesssim z \lesssim 10^5$ for $x_\mathrm{0}\gtrsim1$ results in a non-thermal distortion feature, while at higher redshifts we have a $\mu$-type distortion. We note that since the dark matter transitions happen at $x_\mathrm{0}\gtrsim1$ at low redshifts, the $y$-type distortion from bremsstrahlung cooling is absent in this case. The resultant non-thermal distortion occurs as an absorption signal which is shown along with the total thermal $\mu + y$ distortions in Fig.\hyperref[fig:signal_types]{2c}. 
		
		\item \textbf{Emission feature: $\alpha_\text{C} = 10^{-4}$}\\
		To demonstrate the generation of an emission signature, we decrease the collision parameter $\alpha_\text{C}$ to $10^{-4}$ keeping all the other parameters the same as in the previous case. The corresponding figures are shown in the third row of Fig.\ref{fig:signal_types}. 
		
		The main change occurring after decreasing $\alpha_\text{C}$ is the downward shift in the collision rate curve $\mathcal{R}_\text{DM}$ as shown in Fig.\hyperref[fig:signal_types]{2d} compared to the previous case shown in Fig.\hyperref[fig:signal_types]{2a}. The radiative transition rate $\mathcal{R}_\text{CMB}$ stays identical. At redshifts $\gtrsim 10^6$, the collisional transitions maintain $T_\text{ex}$ in equilibrium with $T_{\text{DM}}$ and the dark matter particles absorb CMB photons. 
		
		Unlike the previous case, the collision rate falls below the Hubble expansion rate much earlier, at around $z\sim 10^6$.  The radiative transitions are still weaker than the Hubble expansion rate ($\mathcal{R}_\text{CMB}<\mathcal{R}_\text{DM}<1$), causing $T_\text{ex}$ to freeze out at intermediate redshifts $10^6 \lesssim z \lesssim 10^4$. A constant $T_\text{ex}$ brings an inflection in the rates $\mathcal{R}_\text{CMB}$ and $\mathcal{R}_\text{DM}$ because both rates vary linearly with $T_\text{ex}$. Since $T_\text{ex}$ is frozen, eventually  $T_\text{CMB}$ falls below $T_\text{ex}$ and we have a net emission from dark matter resulting in an emission signature as shown in Fig.\hyperref[fig:signal_types]{2e} and \hyperref[fig:signal_types]{2f}. 
		
		As we go to lower redshifts, the radiative transition rate rises as $\propto (1+z)^{-1/2}$ and takes over the Hubble expansion as well as the collision rate ($\mathcal{R}_\text{CMB}>1>\mathcal{R}_\text{DM}$). This causes $T_\text{ex}$ to come in equilibrium with $T_\text{CMB}$, causing the emission signal to vanish. The collision rate becomes weaker than the Hubble expansion rate. It is therefore unimportant at lower redshifts and the dark matter heating caused by absorption of CMB photons and subsequent collisional de-excitation becomes negligible.
		
		\item \textbf{Absorption followed by an emission feature: $\alpha_\text{C} = 10^{-2}$}\\
		We increase the collision cross-section parameter to an intermediate value $\alpha_\text{C}=10^{-2}$ keeping the other parameters fixed to the previous two cases. The corresponding figures are shown in the second row of Fig.\ref{fig:signal_types}. We see in the first panel that the collision rate stays at an intermediate position between the values of $\mathcal{R}_\text{DM}$ in Fig.\hyperref[fig:signal_types]{2a} and Fig.\hyperref[fig:signal_types]{2c}.
		
		The redshift range in which the collision and radiative transition rates become comparable to the Hubble expansion rate $\mathcal{R}_\text{DM} \sim \mathcal{R}_\text{CMB}\sim 1$ happens at $10^5 \lesssim z \lesssim 10^4$. This shrinks the redshift range for $T_\text{ex}$ freeze-out compared to the previous case. This causes the formation of an absorption feature accompanied by an emission feature as shown in Fig.\hyperref[fig:signal_types]{2i}.
	\end{itemize}
	A random scan over the different dark matter model parameters allowed by COBE \cite{1996ApJ...473..576F} measurements (see Fig.\ref{fig:u1}) reveals a rich set of spectral features comprising emission, absorption, or a mixture of both in the CMB spectrum as shown in Fig.\ref{fig:signal_zoo}. These spectral features have distinct shapes that differ from the standard thermal $\mu$ and $y$ distortions (denoted by the black dotted line and the red dashed line respectively in Fig.\ref{fig:signal_zoo}). These unique spectral shapes are potentially detectable by PIXIE \cite{2011JCAP...07..025K} which would have three orders of magnitude better sensitivity compared to COBE-FIRAS. If detected, such signatures can open a new window into the thermal evolution and the particle properties of dark matter.
	\label{sec:expsign}
	\section{Constraints from COBE-FIRAS and forecasts for PIXIE}
	\label{sec:constraints}
	The precise measurements of the CMB spectrum by COBE/FIRAS have put strong constraints on the amplitude of spectral distortions in the CMB. In this section, we use the COBE measurements to constrain the parameter space of our dark matter model. We first discuss the constraints from the thermal $\mu$ and $y$-type distortions. In general, the thermal spectral distortions from any physical mechanism that is active over a wide redshift range will result in total spectral distortion that is a superposition of $y$-type, $i$-type, and $\mu$-type distortions \cite{2012JCAP...09..016K,2013MNRAS.434..352C}. We define a universal distortion parameter, $u$, which can be used in such cases for COBE-FIRAS-like experiments. This is possible because the shape of the $\mu$, $y$, and $i$-type distortions are very close and the maximum distortion amplitude is approximately the same for a given total energy injection. Irrespective of whether it is a $y$-type, $i$-type or $\mu$-type, the distortion parameter $u$ allows us to directly map the upper limits on $y$-type only or $\mu$-type only distortions into the upper limits for the case where we have a mixture of $y$-type and $\mu$-type (and $i$-type) distortions without explicitly fitting every model to the data. We verify these approximations with a Bayesian analysis with actual spectral distortions calculated for each set of parameter values using COBE data. We also calculate the resultant spectral distortion constraints for total spectral distortions including the non-thermal component. 
	
	\subsection{Thermal distortions}
	We ignore the non-thermal component of the spectral distortions for now and derive constraints on the dark matter model parameters using only the thermal $\mu$- type and $y$-type distortions. Traditionally, the CMB spectral distortion limits are derived by fitting the $\mu$ and $y$ spectrum independently to the observed sky spectrum. However, in realistic cases with continuous energy injection, we will have both $\mu$-type and $y$-type distortions and in addition everything in between ($i$-type). Thus, in order to constrain the model parameters in such cases, we should fit the total distortion to the data. We therefore want to combine $\mu$- type, and $y$-type distortions into a single distortion measure $u$:
    \begin{equation}
		u = \frac{\mu}{1.4} + 4y\,,\label{6.3} 
	\end{equation}
    where $u$ is the total fractional energy added or subtracted from the CMB which goes into the thermal distortions. 
        We note that this fractional change in energy is defined at a constant number density of photons. In particular, for a $\mu$-type distortion, this is the pure energy addition without changing the number density of photons that will create a given amplitude of $\mu$-type distortion.
	Thus, the 2-$\sigma$ limits on $\mu$ and $y$ parameters from COBE\footnote{If we simply fit the $y$-type distortion along with the Galaxy spectrum given in Table 4 of \cite{1996ApJ...473..576F}  and the CMB temperature to the COBE-FIRAS data, we get a 2-$\sigma$ upper bound $|y|<8\times 10^{-6}$. The bounds given in \cite{1996ApJ...473..576F} include additional uncertainty in modelling of Galactic spectrum. A different modelling of our Galaxy can also make the bound on $\mu$ a factor of 2 stronger \cite{2022PhRvD.106f3527B} compared to the one quoted in \cite{1996ApJ...473..576F}.} thus place an upper limit on the fractional change in the energy density of CMB via $\mu$ and $y$ distortions to be,
	\begin{align}
		|\mu| < 9\times10^{-5},\, |y|=0 &\implies |u| <6.4\times 10^{-5},\nonumber\\
		|\mu|=0,\, |y|<1.5\times10^{-5} &\implies |u| <6\times 10^{-5}.
	\end{align}
	Since both the limits come out to be approximately similar, with a difference of $\sim 7\%$. This implies that COBE effectively constrains the total fractional change in the energy density of the CMB. 
	In deriving constraints on our dark matter model parameters, we use the $2$-$\sigma$ limit from COBE,
	\begin{equation}
		|u| \lesssim 6\times 10^{-5}\,.\label{6.4}
	\end{equation}
	The next generation experiment PIXIE \cite{2011JCAP...07..025K} will have three orders of magnitude higher sensitivity than COBE. It will be able to detect $|\mu| \approx 5 \times 10^{-8}$ and $|y| \approx 10^{-8}$ at the $5$-$\sigma$ level. Thus, the expected $2$-$\sigma$ detection sensitivity on $u$ would be,
	\begin{equation}
		|u| \approx  10^{-8}\,.\label{6.5}
	\end{equation}
	To use the distortion parameter $u$, we need to calculate the total thermal energy density added to the CMB, after correction for any change in number density during the $\mu$-type era (see Eq.\eqref{4.17b}) and use Eq.\eqref{6.4} and \eqref{6.5} to get constraints from COBE-FIRAS and forecasts for PIXIE respectively. 
 
\begin{figure}[t]
		\centering
		\includegraphics[width=0.75\textwidth]{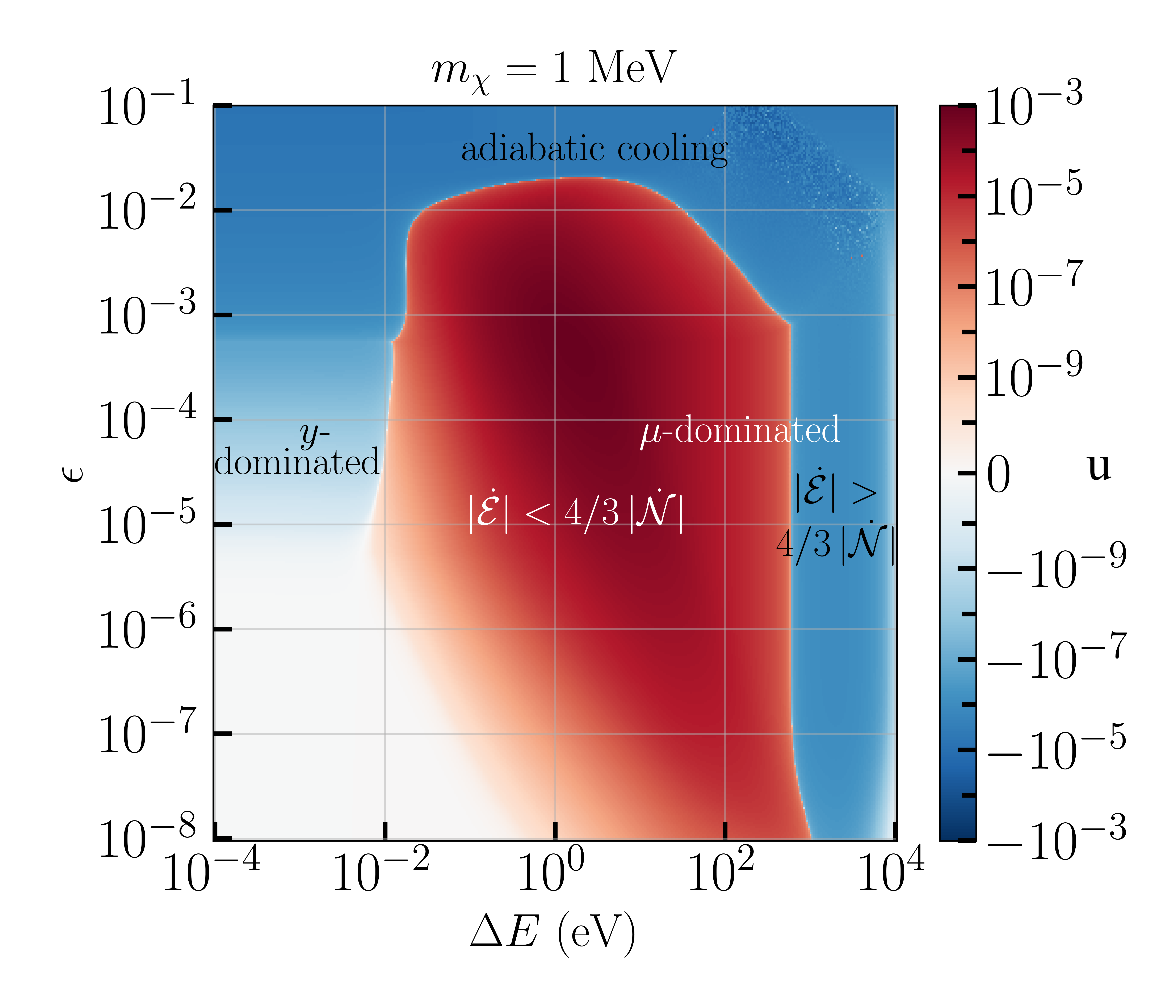}
		\caption{The value of $u$ in the $\epsilon$ vs $\Delta E$ plane for $1$ MeV dark matter. The color scale represents $u$ which can take both negative (blue) and positive values (red).}
		\label{fig:u1_phys}
	\end{figure}

 For a given set of dark matter model parameters ($m_\chi, \epsilon, \Delta E,$ and $\alpha_\text{C}$), one can compute the $\mu$ and $y$ parameters separately from Eq.\eqref{4.29} and Eq.\eqref{4.36} respectively and substitute them to Eq.\eqref{6.3} to get the universal distortion parameter $u$. For a demonstration, we show $u$ as a function of $\epsilon$ and $\Delta E$ in Fig.\ref{fig:u1_phys} after fixing $m_\chi = 1$ MeV and $\alpha_\text{C}=1$. The color scale represents the value of the universal distortion parameter $u$ which can be both negative (blue) and positive (red). 
Note that for this specific choice of $\alpha_\text{C}=1$, we saturate the bullet cluster bound and the dark matter always absorbs the CMB photons. In order to explain the underlying physics of Fig.\ref{fig:u1_phys}, we focus on two distinct regions, namely, $\epsilon \lesssim 10^{-2}$, characterized by radiative transitions; and $\epsilon \gtrsim 10^{-2}$, where the scattering of baryons and dark matter plays the major role.
	\begin{itemize}[leftmargin=20pt]
		 \item $\epsilon\lesssim 10^{-2}$: For small $\epsilon$, line absorption by dark matter is the dominant source of energy transfer from the CMB to dark matter. As shown in Eq.\eqref{6.3}, if $|\mu|/1.4<4|y|$, the thermal distortion is dominated by the $y$-type distortion, whereas if $|\mu|/1.4>4|y|$, the thermal distortion is dominated by the $\mu$-type distortion.
		\begin{itemize}[leftmargin=20pt]
			\item[1.] \textbf{$y$-dominated:} 
			The maximum contribution from $y$-type distortion to $u$ occurs in the middle left blue region of Fig.\ref{fig:u1_phys}. As discussed in section \ref{sec:cmb_distortions}, the $y$-type distortion is created when the bremsstrahlung process cools the baryons and electrons by creating the photons that were absorbed by dark matter in the Rayleigh-Jeans tail of the CMB. The Compton scattering of the cooled electrons with the CMB photons results in a negative $y$-type distortion. The ratio of the energy density absorbed by dark matter to the energy density of CMB is small at high redshifts and increases as we go to $z<z_\text{c}$, resulting in $y$-type distortions dominating over the $\mu$-type distortions for small $\Delta E$ where bremsstrahlung cooling is important. 
            We note that the lower limit ($z_\text{br}$) on Eq.\eqref{4.36} is determined by when the dark matter transition frequency $x_0$ exceeds $x_\text{H}$. For larger $\Delta E$, this transition happens at higher redshifts and thus reducing the redshift range when the $y$-distortion is generated. For $\Delta E\gtrsim 0.01$ eV, the $y$-distortions become negligible compared to $\mu$-type distortions.
            For a fixed $\Delta E$, as we decrease $\epsilon$ the radiative transition rate ($\mathcal{R}_\text{CMB}\propto A_{10}\propto\epsilon^2$) falls resulting in a decrease in the amplitude of distortions as shown in the lower left region in Fig.\ref{fig:u1_phys}. The thermal history of dark matter in the $y$-dominated regime for $\epsilon=10^{-4},$ and $\Delta E = 10^{-3}$ eV is shown in Fig.\ref{fig:y_thermal_history} of Appendix \ref{app:thermal_history}.
			\item[2.]  \textbf{$\mu$-dominated:}
			As we increase $\Delta E$, in the $y$-type era, $\Delta E$ falls in the main CMB band where bremsstrahlung becomes increasingly inefficient in cooling the baryons. Thus, the thermal distortion becomes dominated by the $\mu$-type distortion.

   In the $\mu$-type era, as we increase $\Delta E$, initially $|\dot{\mathcal{E}}| < 4/3\,|\dot{\mathcal{N}}|$, consequently $\mu>0$. 
            The cross-over from the $y$-dominated region to the $\mu$-dominated region happens at $\Delta E= 10^{-2}$ eV when a negative $y$ is cancelled by a positive $\mu$ yielding a negative to positive transition in the $u$ parameter. 
            At $\Delta E\gtrsim 50$ eV, the CMB photons absorbed by dark matter are sufficient to bring dark matter in thermal equilibrium with the CMB. This can be seen in Fig.\ref{fig:mu_thermal_history} of Appendix \ref{app:A} by comparing the thermal history of dark matter as $\Delta E$ is increased from 1 eV (first panel) to 50 eV (second panel).  In such cases, the fractional energy absorbed by dark matter depends only on the entropy density of dark matter particles, which in turn depends on $m_\chi$ and is independent of $\epsilon$ and $\Delta E$. 
            As we increase $\Delta E$ further, eventually we have $|\dot{\mathcal{E}}| > 4/3\,|\dot{\mathcal{N}}|$, and $\mu < 0$. Therefore, we encounter another cross-over where $u$ transitions from positive to negative.
            We can use Eq.\eqref{4.29} to find a critical value of $\Delta E$ at which $\mu\approx 0$ in the thermally coupled regime ($T_\text{ex}=T_\chi=T_\text{CMB}$),
				\begin{align}
					\Delta E \big|_{\mu \approx 0} = \frac{4a_\text{R}}{3b_\text{R}k_\text{B}}e^{x_\text{c}}e^{-x_c/x_0}T_\text{CMB}\approx \frac{4a_\text{R}}{3b_\text{R}k_\text{B}}T_\text{CMB}(z\sim 10^6)\approx 1 \,\text{keV},
				\end{align}
				where we have approximated the exponential terms to be unity because $x_\text{c}\sim 0.01$ and $x_0\gg x_\text{c}$ for the $\Delta E$ range under consideration.

            However, for very large $\Delta E$ in the extreme right of Fig.\ref{fig:u1_phys}, $|u|$ starts decreasing again as there are not enough photons left to be absorbed in the high-frequency Wein tail of the CMB spectrum. This causes dark matter to decouple from the CMB earlier. This can be seen in Fig.\ref{fig:sc_thermal_history} of Appendix \ref{app:thermal_history}, where the excitation temperature decouples from the CMB temperature earlier for $\Delta E = 1$ keV (third panel) compared to $\Delta E = 50$ eV (second panel).
	\end{itemize}
		\begin{figure}[t]
		\centering
		\includegraphics[width=0.495\textwidth]{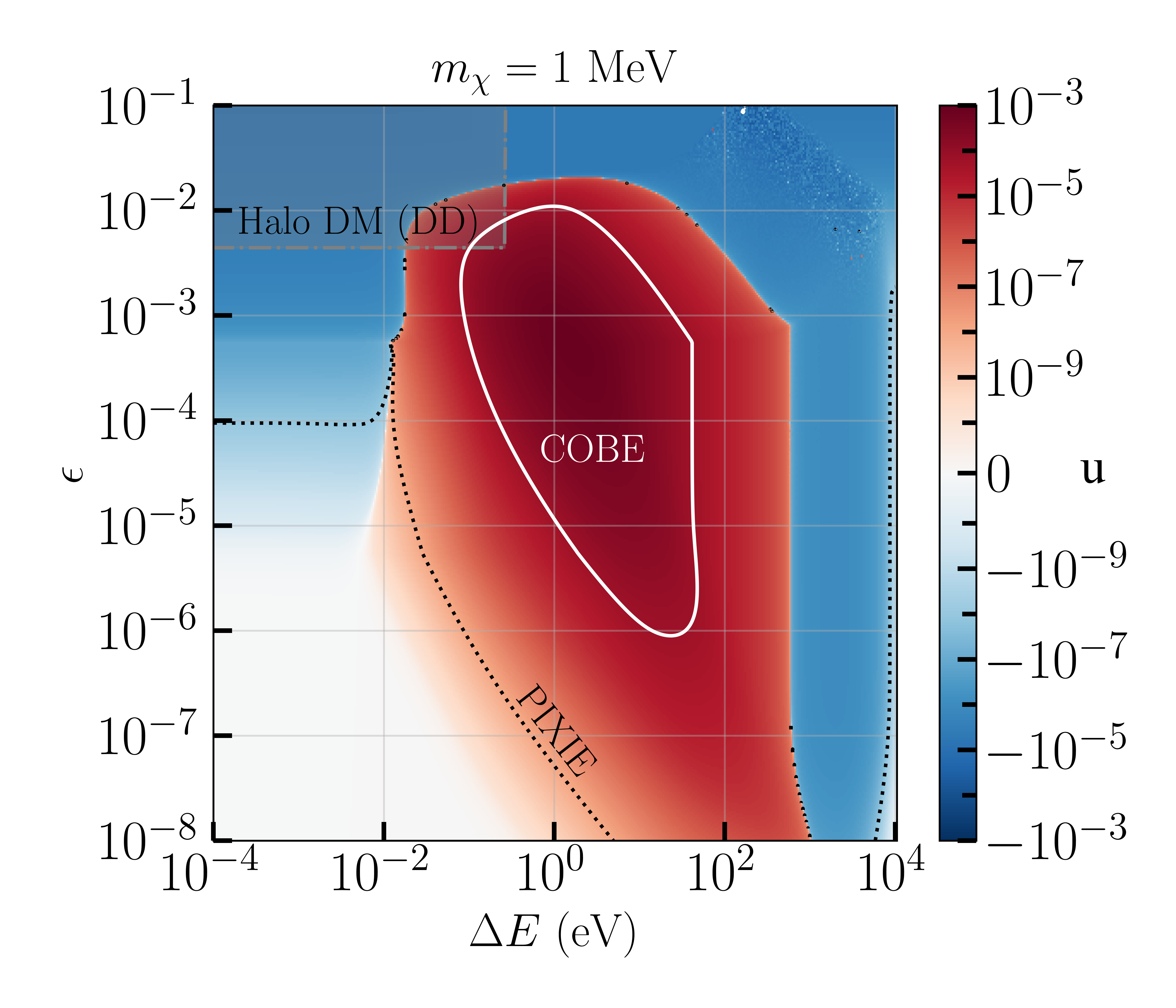}
		\includegraphics[width=0.495\textwidth]{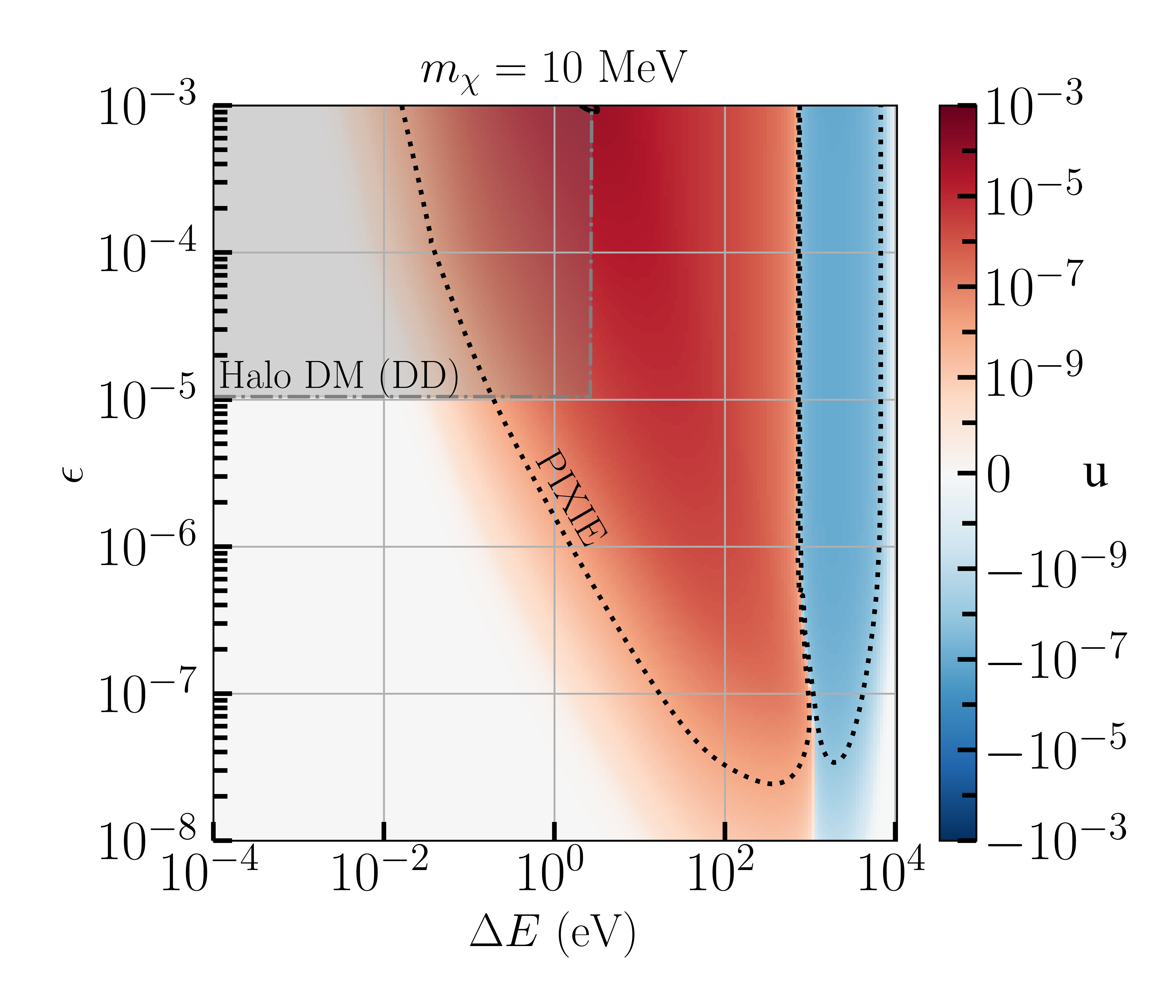}
		\includegraphics[width=0.495\textwidth]{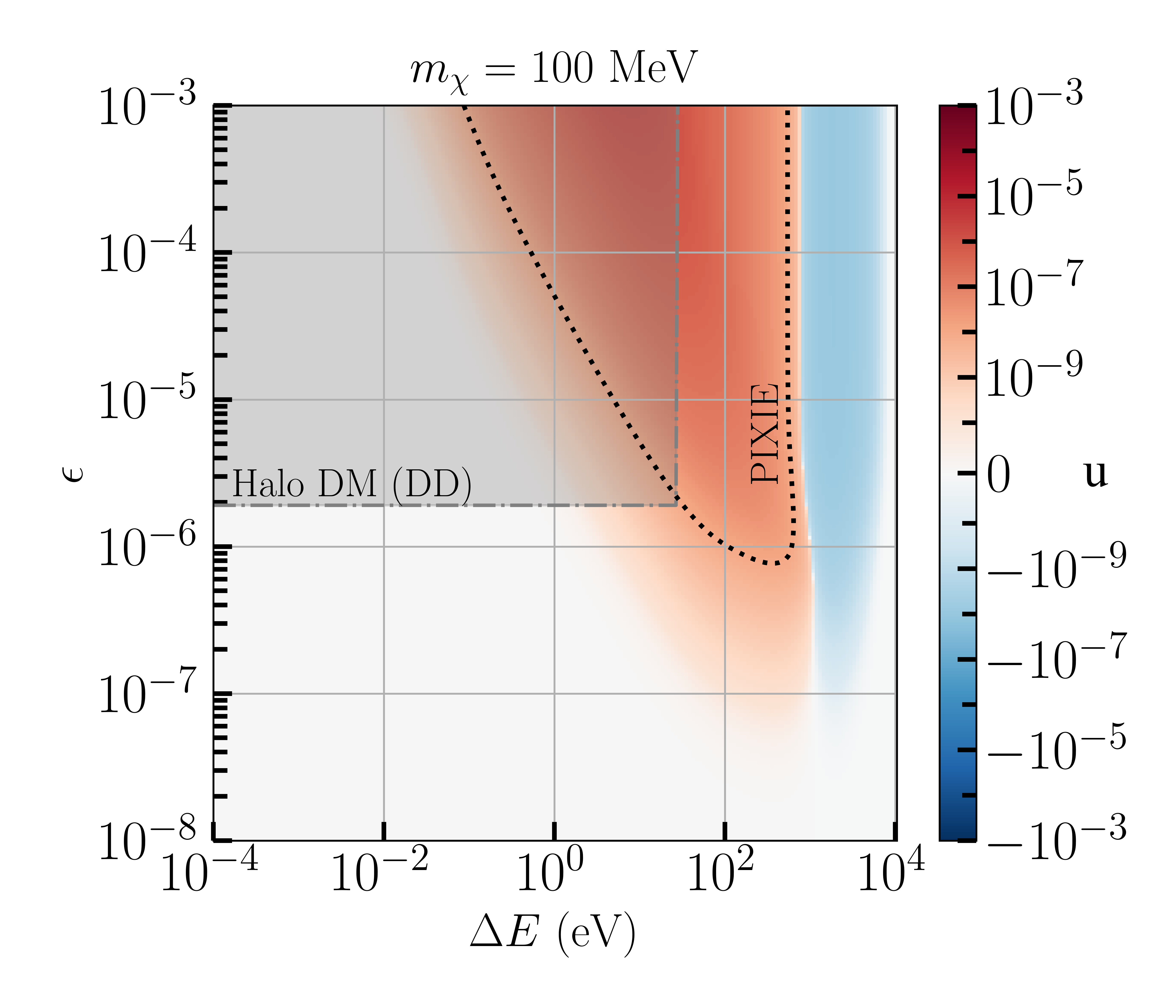}
		\caption{Constraints in $\epsilon$ versus $\Delta E$ for different choices of dark matter masses. These constraints are obtained using 2-$\sigma$ CMB spectral distortion limits from COBE (see Eq.\eqref{6.4}) and forecasted with PIXIE (see Eq.\eqref{6.5}). The grey shaded region is ruled out by the direct detection (DD) experiments from scattering due to the magnetic transition operator.}
		\label{fig:u1}
	\end{figure}
			\item $\epsilon > 10^{-2}$
                For large $\epsilon$, the physics of the CMB spectral distortions is dominated by the electromagnetic scattering of baryons with dark matter.
			At $\epsilon\gtrsim 0.01$, the electromagnetic scattering of dark matter particles with electrons and ions is efficient in keeping dark matter in thermal equilibrium with the baryons. The energy lost by baryons to dark matter is subsequently drawn from the CMB via Compton scattering. This adiabatic cooling of dark matter gives rise to negative thermal distortions as shown in the top blue region of Fig.\ref{fig:u1_phys}. Note that the energy transferred to keep the dark matter in equilibrium with the CMB only depends on the entropy density of dark matter particles which depends on its mass and is independent of $\epsilon$ and $\Delta E$. As a result, the $u$ parameter is constant in this regime denoted by `adiabatic cooling' in Fig.\ref{fig:u1_phys}. We note that for small $m_\chi$, the entropy in dark matter is higher than the entropy in baryons. Thus, the adiabatic cooling of dark matter dominates over the adiabatic cooling of baryons by many orders of magnitude.
	\end{itemize}
	\begin{figure}[t]
		\hspace{-10pt}\includegraphics[width=0.53\textwidth]{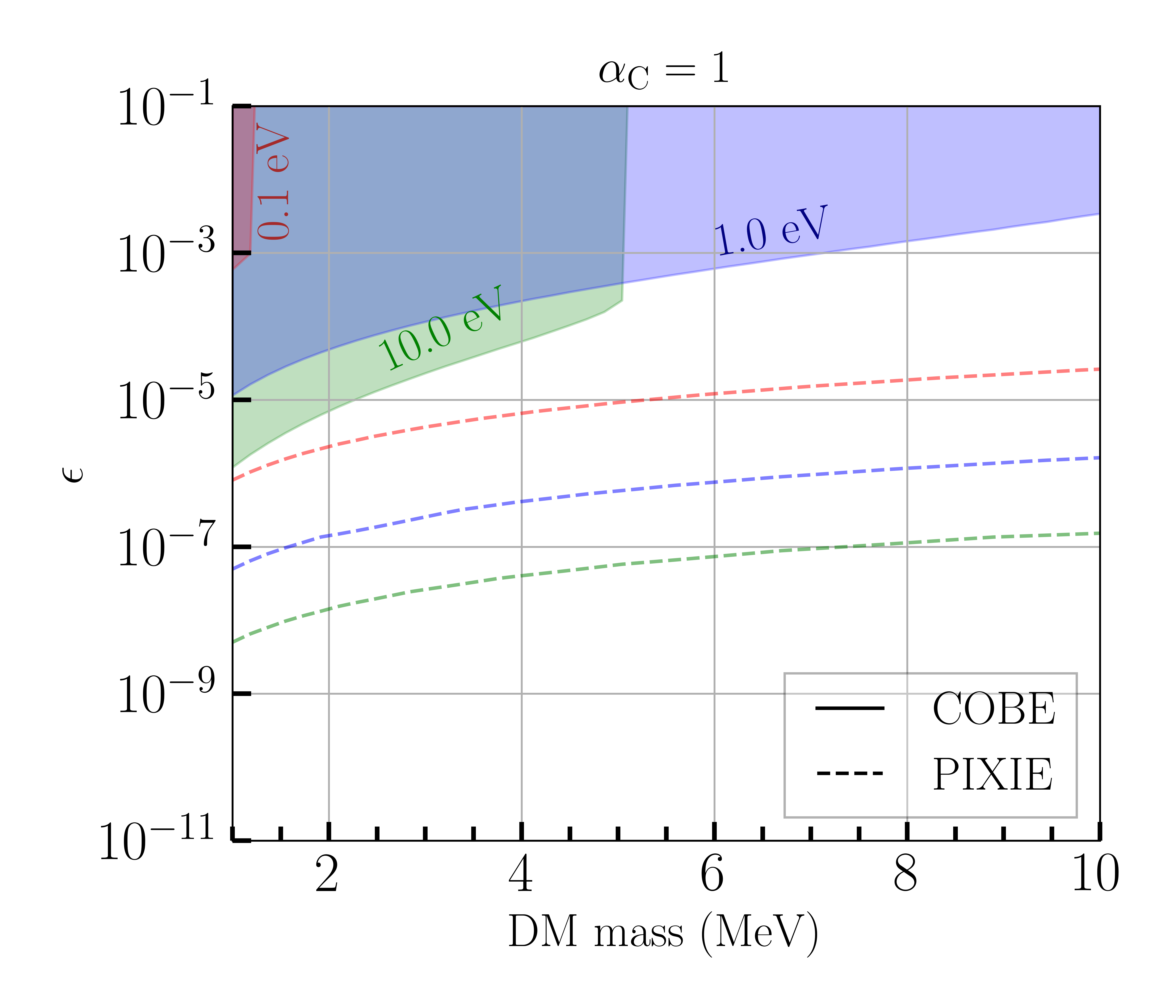}\hspace{-12pt}
		\includegraphics[width=0.53\textwidth]{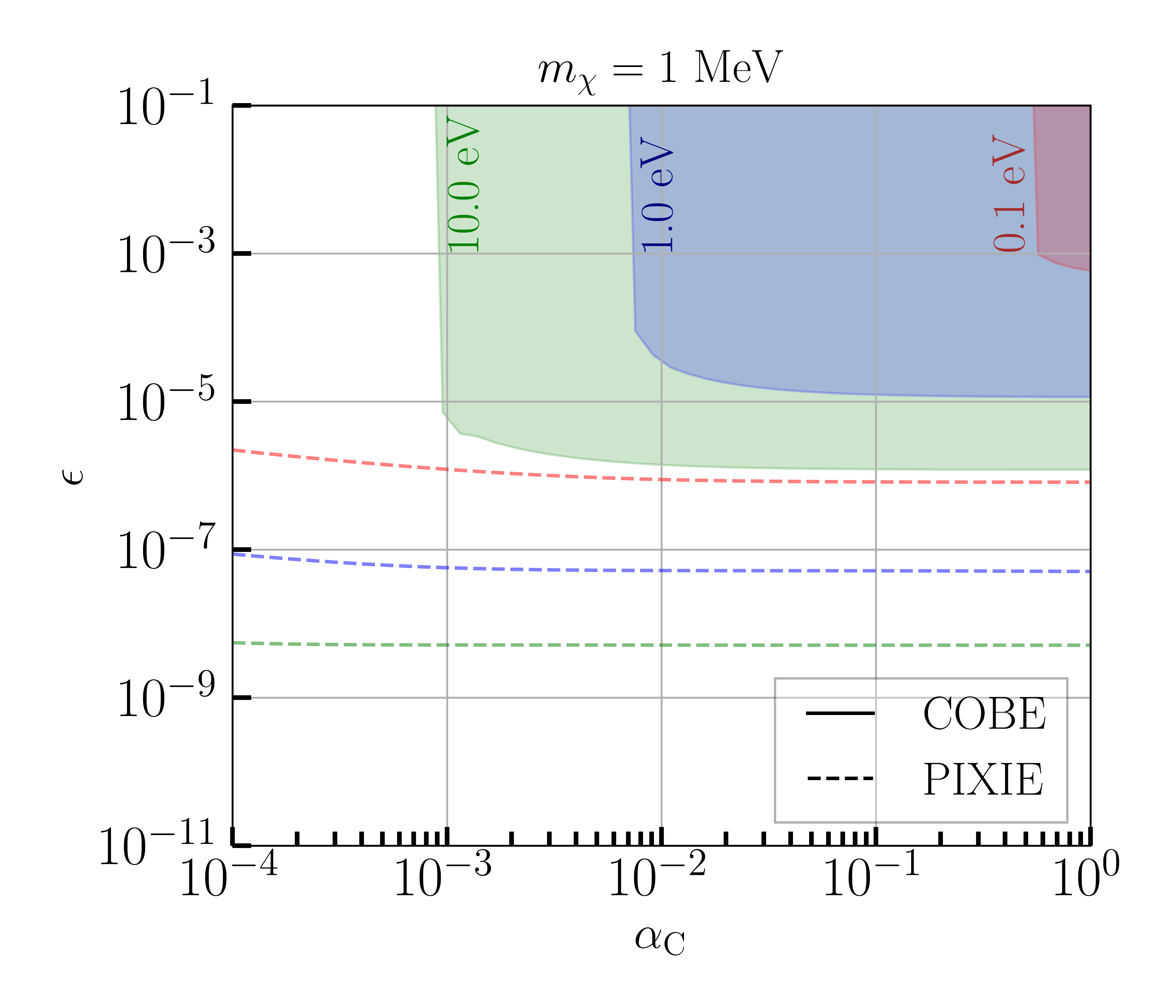}
		\caption{Spectral distortion limits from COBE-FIRAS (shaded regions) and forecasts for PIXIE (dashed lines) for $\Delta E = 0.1$ eV (red), $1.0$ eV (blue), and $10.0$ eV (green). The left (right) panel shows constraints on $\epsilon$ as a function of $m_\chi$ ($\alpha_\text{C}$) for fixed $\alpha_C$ ($m_\chi$).}
		\label{fig:mass_coll_plot}
	\end{figure}
     \textbf{Constraints from COBE-FIRAS for thermal distortions:}
	We use the $2$-$\sigma$ limits from COBE \cite{1996ApJ...473..576F} to put constraints on $\epsilon$ as a function of  $\Delta E$, keeping the collision parameter fixed $\alpha_\text{C}=1$. The constraints in the $\epsilon$ versus $\Delta E$ plane are shown for dark matter masses 1 MeV, 10 MeV, and 100 MeV in Fig.\ref{fig:u1}. As in Fig.\ref{fig:u1_phys}, the color scale varying from blue to red represents the value of the universal distortion parameter $u$. The bounded region between black solid lines is ruled out by COBE (see Eq.\eqref{6.4}). We also show the direct detection experiment constraints for inelastic scattering\footnote{The direct detection constraints in Fig.\ref{fig:u1} are for the magnetic transition operator which dominates the physics of the CMB spectral distortions. Other operators, if present, would yield additional contributions to direct detection. We refer the reader to Appendix \ref{app:es} for more details.} as derived in \cite{2024PhRvD.109f3512G} which assumes the standard halo model based on \cite{2011PhRvL.107e1301A,2016PhRvD..94i2001A, 2017PhRvD..96d3017E,2018PhRvL.121k1303A,2019PhRvL.123r1802A,2018PhRvL.121f1803C,2019PhRvL.122p1801A,2023arXiv231213342S}. Compared to the direct detection experiments, COBE puts significantly strong constraints on the dark matter parameter space for transition energy in $1-20$ eV range for MeV mass dark matter. We also forecast the constraints on $u$ from PIXIE (see Eq.\eqref{6.5}) which are shown by the regions bounded by black dashed lines.

	The effect of dark matter mass on the spectral distortion constraints comes from the dependence of $u$ on the radiative transition rate ($A_{10} \propto 1/m_\chi^2$) and the dark matter number density ($n_\chi \propto 1/m_\chi$). Since both fall with the dark matter mass, the magnitude of $u$ becomes smaller and the constraints from COBE and PIXIE become weaker at higher dark matter masses as shown in Fig.\ref{fig:u1}. We find that COBE rules out a major fraction of the parameter space for dark matter masses $\sim \mathcal{O}$(MeV). Compared to the direct-detection bounds which are at $\epsilon \sim 10^{-2}$ for $m_\chi \sim 1$ MeV, the CMB spectral distortion limits from COBE are four orders of magnitude stronger and rule out $\epsilon\gtrsim 10^{-6}$ at $\Delta E\sim 10$ eV. In the future, PIXIE with three orders of magnitude higher sensitivity than COBE will be able to probe dark matter masses up to $\sim 100$ MeV.
	
	We show the spectral distortion limits from COBE and PIXIE on $\epsilon$ as a function of dark matter mass $m_\chi$ in the first panel of Fig.\ref{fig:mass_coll_plot} for three choices of transition energies: $\Delta E = 0.1, 1.0$, and $10.0$ eV. 
	We also show the spectral distortion limits from COBE and PIXIE on $\epsilon$ as a function of the collision coefficient $\alpha_C$ in the second panel of Fig.\ref{fig:mass_coll_plot}. The constraints become weaker for smaller values of inelastic collisional coefficient. This happens because a weaker inelastic collision cross-section will decrease the redshift range for line absorption where $T_\text{CMB} > T_\text{ex}\gtrsim T_\chi$. As a result, for smaller $\alpha_C$, the line absorption of CMB photons by dark matter will happen over a smaller timescale resulting in weaker total thermal distortions integrated over the redshift. 
	\subsection{Thermal and non-thermal distortions combined}\label{subsec:combined}
	\begin{table}
		\centering
		\begin{tabular}{|l|l|} 
			\hline
			\centering
			Parameters $(\theta)$                  & Range  \\
			\hline
			$\epsilon$ &    $0-10^{-3}$    \\
			$\Delta T$    &     $-0.01-0.01$   \\
			$g_0$                    &   $-0.1-0.1$ \\
			\hline   
		\end{tabular}
		\caption{Model parameters and the corresponding range of values over which uniform prior is chosen. }
		\label{table:param}
	\end{table}
		\begin{figure}[t]
		\centering
		\hspace{-20pt}
		\includegraphics[width=0.53\textwidth]{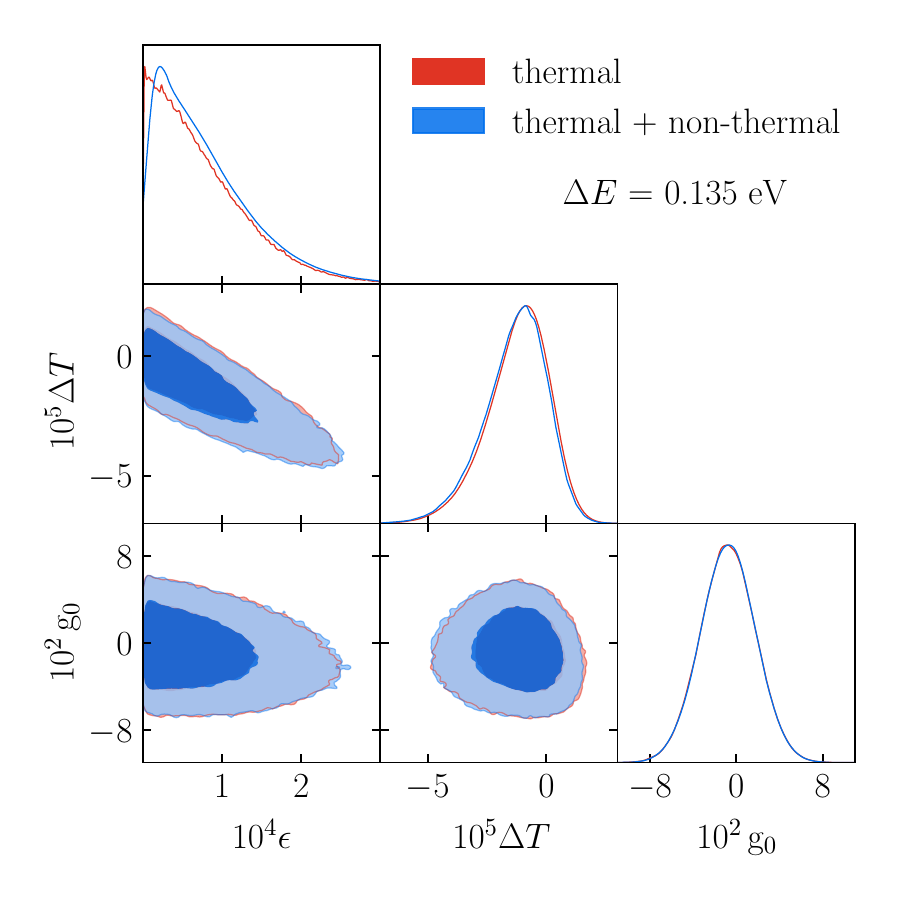}
		\hspace{-20pt}
		\includegraphics[width=0.53\textwidth]{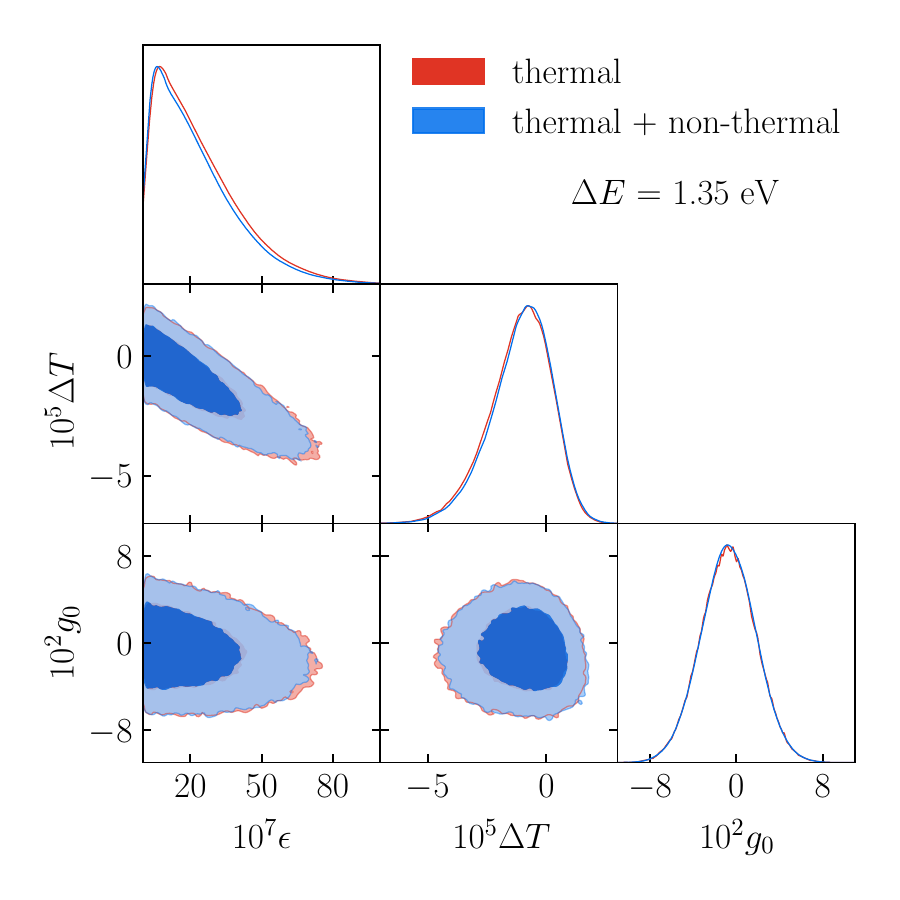}
		\includegraphics[width=0.53\textwidth]{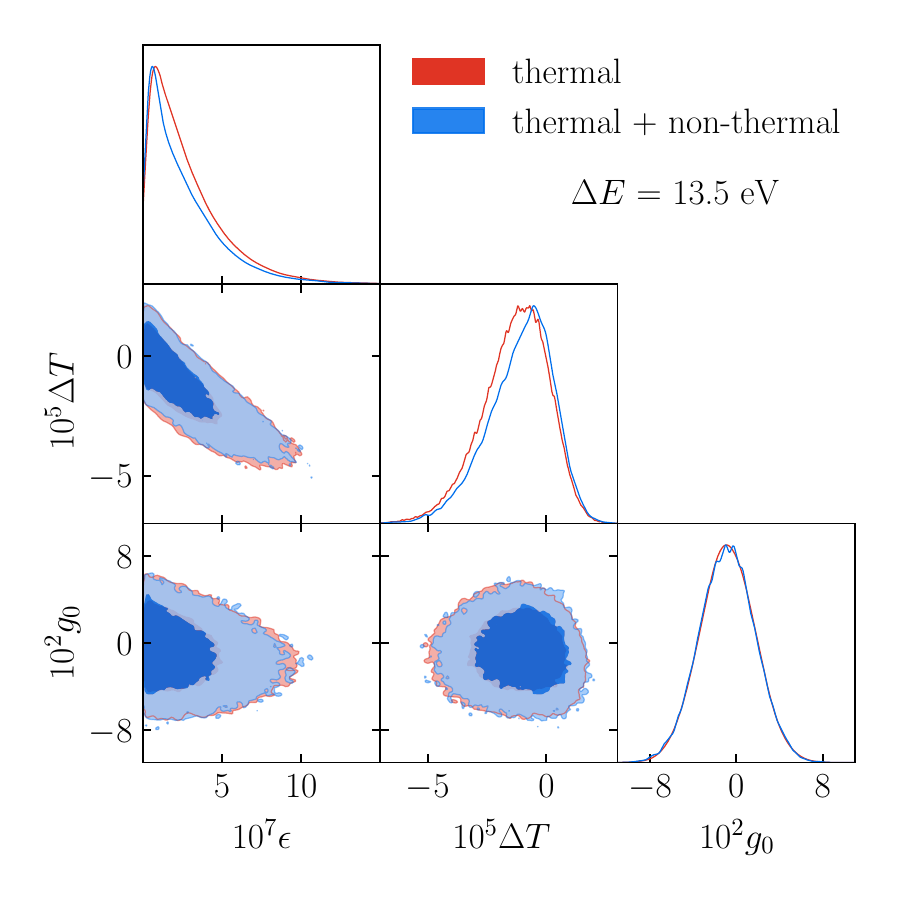}
		\caption{The 1$\sigma$ and 2$\sigma$ contours for the model parameters (strength of the magnetic transition operator $\epsilon$, temperature shift $\Delta T$, amplitude of galaxy spectrum $g_0$) from the CMB spectrum measurements of COBE-FIRAS.  }
		\label{fig:globalsignal_const}
	\end{figure}
	\begin{figure}[t]
		\centering
		\includegraphics[width=0.49\textwidth]{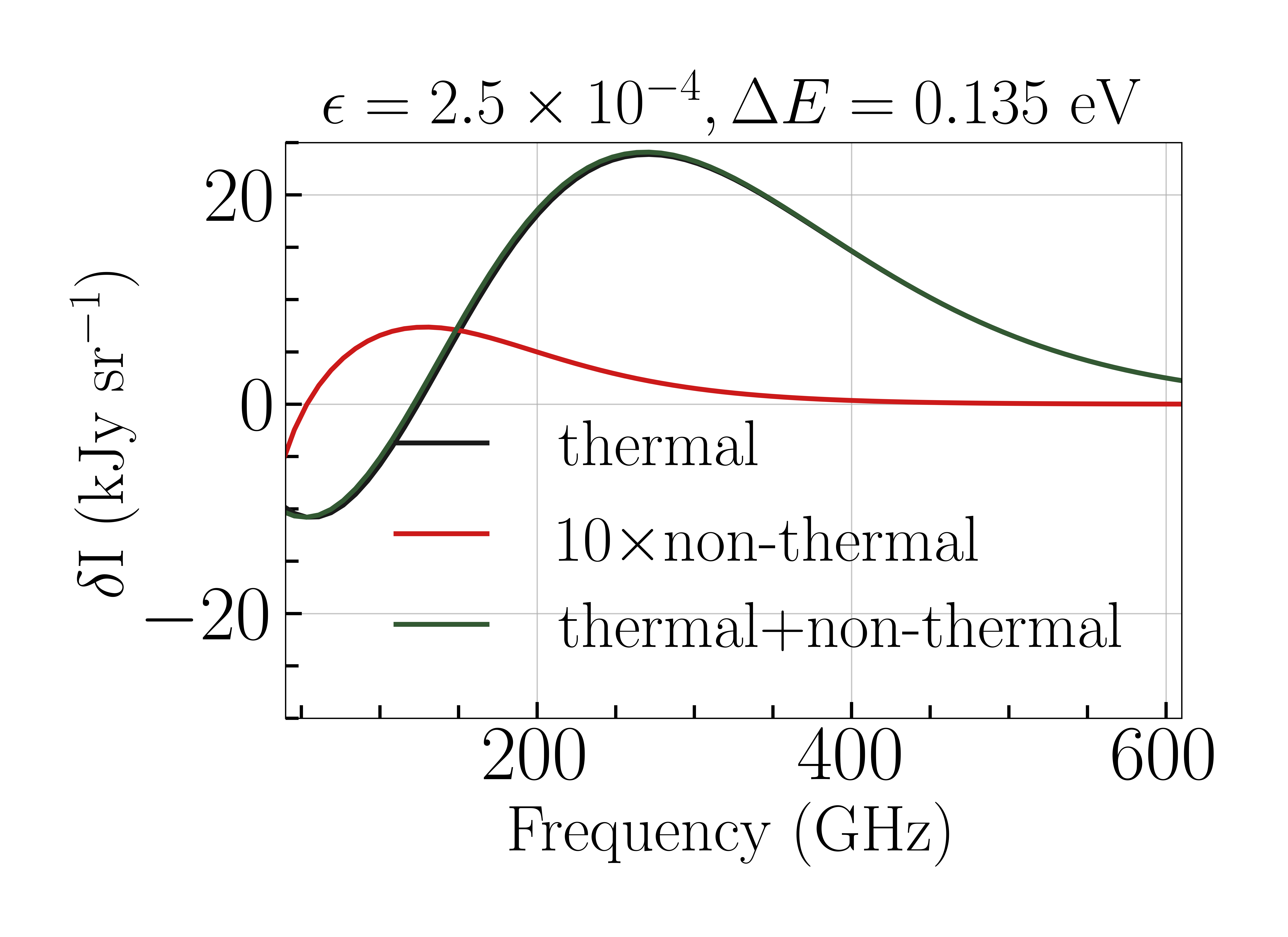}
		\includegraphics[width=0.49\textwidth]{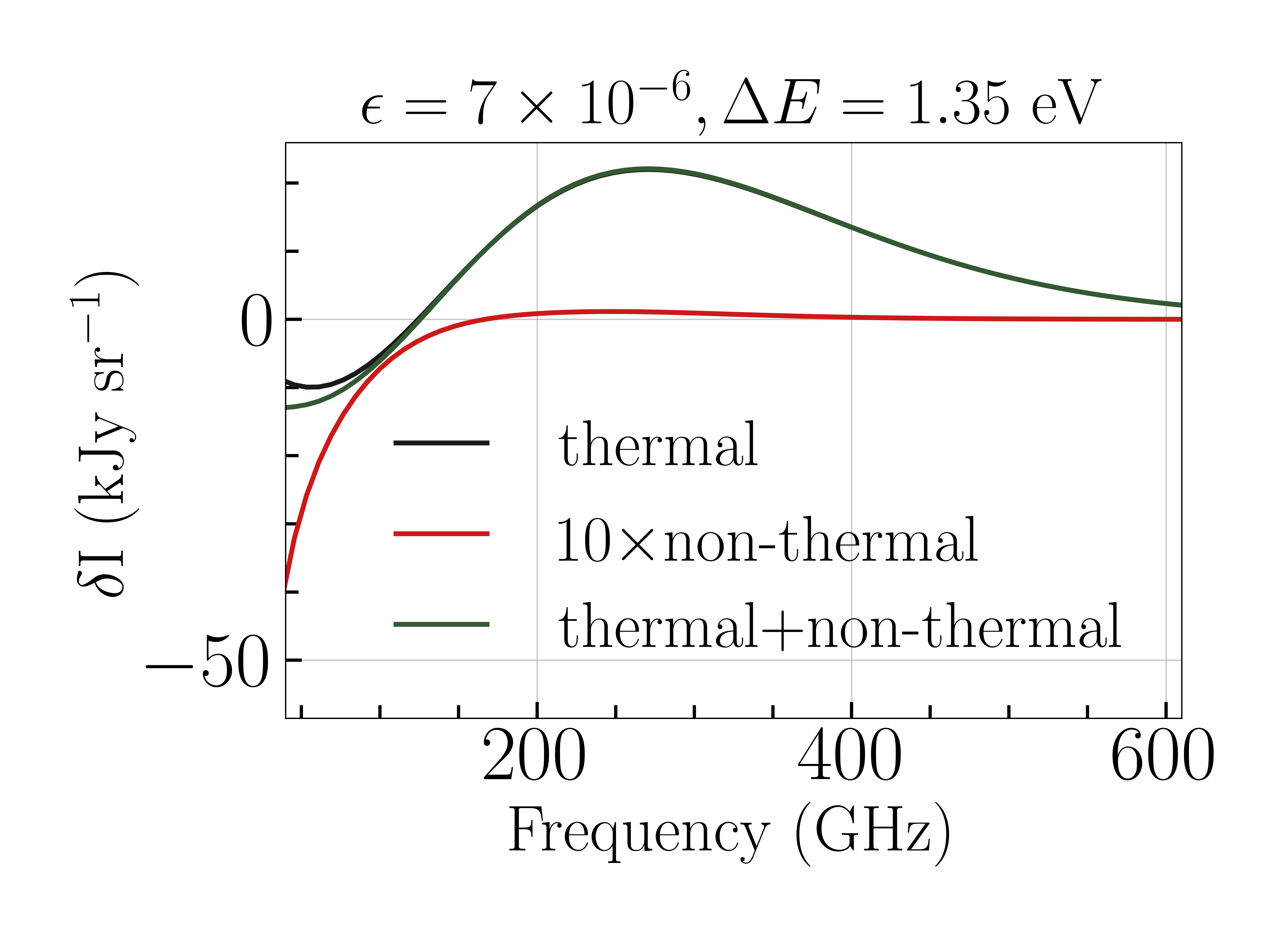}
		\includegraphics[width=0.49\textwidth]{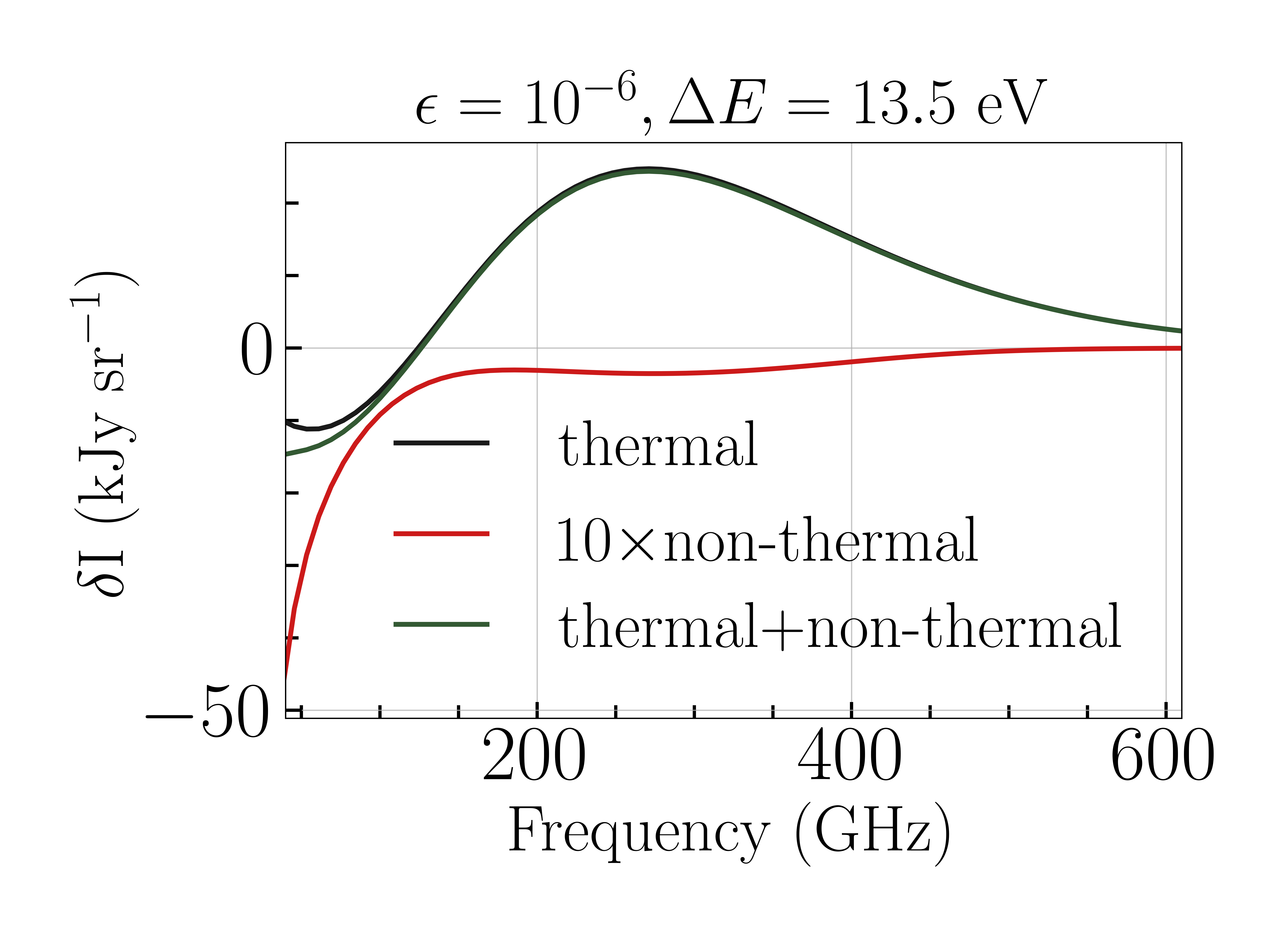}
		\caption{Thermal and non-thermal distortions in the COBE band. The choice of the model parameters corresponds to the 2-$\sigma$ limit obtained via Bayesian analysis in Fig.\ref{fig:globalsignal_const} for $m_\chi = 1$ MeV and $\alpha_\text{C}=1$. }
		\label{fig:globalsignal_plot}
	\end{figure}
	The non-thermal distortions are created roughly in the redshift range $z \sim 10^5$ to $50$, where we take $z\sim 50$ to be the lower limit because the dark matter distribution no longer stays homogeneous after $z\sim 50$ and we must take into account the clustering of dark matter into halos. Also, the first stars and galaxies would start forming further complicating the thermal history and the CMB distortions. The range of transition frequencies which would result in the non-thermal distortions in the COBE band (60-600 GHz) in this redshift range is given by,
	\begin{align}
		z_\text{min} = \dfrac{\nu^\text{min}_0}{\nu_\text{obs}^\text{max}} -1 = 50 &\implies \Delta E_\text{min} = h\nu_0^\text{min} = 0.135\, \text{eV},
		 \nonumber\\
		z_\text{max} = \dfrac{\nu^\text{max}_0}{\nu_\text{obs}^\text{min}}-1 = 10^5 &\implies  \Delta E_\text{max} = h\nu_0^\text{max} = 28.1 \,\text{eV},
	\end{align}
	where $\nu_\text{obs}^\text{min}=60$ GHz and $\nu_\text{obs}^\text{max} = 600$ GHz refer to the COBE frequency band.
	We perform a Markov-Chain Monte Carlo (MCMC) analysis using the package emcee \cite{2013PASP..125..306F} to get the posterior distribution for $\epsilon$ at three different transition energies $\Delta E = \{0.135$,  $1.35,$ and $13.5$\} eV. We fix the dark matter mass to 1 MeV and collision parameter $\alpha_\text{C}=1$. In this analysis, the data vector is the residual CMB intensity $\delta I_\nu^\text{exp}$ which is equal to the intensity of CMB measured by COBE with the best-fit blackbody intensity subtracted. The residual intensity is modelled as a sum of a shift in the temperature of the blackbody parameterized by $\Delta T$, uncertainty in the amplitude of Galaxy spectrum parameterized by $g_0$, and the sum of thermal and non-thermal distortions from dark matter denoted by $I_\nu^\text{DM}$ \cite{1996ApJ...473..576F},
	\begin{equation}
		\delta I_\nu^\text{model} \equiv I_\nu - B_\nu(T_0) =  \Delta T \frac{dB_\nu}{dT} + g_0G_\nu + I_\nu^\text{DM}(\epsilon),
	\end{equation}
	where $B_\nu$ denotes the blackbody intensity at temperature $T_0 = 2.725$ K and $G_\nu$ denotes the Galaxy spectrum.
	
	The uniform prior on the model parameters $\theta$ are given in Table \ref{table:param}. The Bayesian analysis is done using a Gaussian likelihood i.e. assuming a Gaussian noise, given by,
	\begin{equation}
		\mathcal{L}(\delta I_\nu^\text{exp}|\theta) = \prod_{i=1}^N \frac{1}{\sqrt{2\pi\sigma_i}}\exp\bigg[-\frac{(\delta I_\nu^\text{exp}-\delta I_\nu^\text{model})^2}{2\sigma_i^2}\bigg],
	\end{equation}
	where $\sigma_i$ represents the 1-$\sigma$ uncertainty in the measurement at $i^\text{th}$ frequency and $N$ denotes the total number of data points.
	We run the Markov chains with 100 walkers and $10^4$ steps for each parameter. We show the posterior distribution of $\epsilon$ in Fig.\ref{fig:globalsignal_const}. The red curve represents the posterior distribution taking into account only the thermal ($\mu$ and $y$) distortions and the blue curve takes into account both the thermal and non-thermal distortions. We find that the non-thermal distortions do not change the upper limit on $\epsilon$. This is also evident from Fig.\ref{fig:globalsignal_plot} where we show the thermal and non-thermal distortions computed at the 2-$\sigma$ bound from COBE. We find that the change in the distortions after the addition of the non-thermal signal ($\sim$ 0.1 kJy sr$^{-1}$) is negligible compared to the uncertainty in the measurements from COBE ($\sim$10 kJy sr$^{-1}$). Therefore, we need more sensitive measurements to detect the non-thermal distortions, which would be possible with future experiments like PIXIE \cite{2011JCAP...07..025K}. We note that the thermal distortions included in this analysis use the sum of actual $\mu+y$ type distortions created for each point in the parameter space. These constraints also agree with the constraints derived using the universal thermal distortion parameter $u$ in the previous subsection. This analysis thus validates the usefulness of the distortion parameter $u$ in deriving approximate constraints from COBE-FIRAS in scenarios where both $\mu$ and $y$ distortions are created. We can thus use the total energy injected to map constraints on $u$ from COBE or forecast for PIXIE to parameters of any energy injection mechanism. Using the parameter $u$, therefore, makes the separation of the total energy injected or absorbed into $\mu$, $y$, and $i$-type distortions created unnecessary as long as the distortions created are thermal. 
	
	\section{Conclusions}
  Unique signatures in the spectrum of the CMB monopole (also sometimes referred to as global signals) are a natural prediction of dark matter models with electromagnetic transitions. The resultant spectral feature is, in general, a combination of thermal i.e. $\mu$ and $y$-type distortions and non-thermal distortions. 
  	The shape of the spectral distortion encodes information about the thermal history and particle properties of dark matter. We give all our results as functions of four model parameters, namely, the dark matter mass ($m_\chi$), the transition energy ($\Delta E$), the strength of magnetic transition ($\epsilon$), and the collision parameter ($\alpha_\text{C}$). The shape of the non-thermal spectral distortions and the sign of the $\mu$-type distortions are sensitive functions of the dark matter parameters.
  The next generation experiments like PIXIE, which aim to measure spectral distortions at the level of $\sim 1$ part in $10^8$, would be sensitive to the spectral signatures of dark matter in a large volume of unexplored parameter space. Moreover, the unique shapes of non-thermal distortions enhance the prospects of distinguishing them from other standard and non-standard spectral distortions. The existing limits on spectral distortions from COBE-FIRAS rule out a large region of the parameter space of dark matter with mass $\sim $ MeV. With the next-generation experiments like PIXIE, there is a potential possibility of detecting the non-thermal distortions. The higher sensitivity of PIXIE will also extend the reach of CMB spectral distortions in probing dark matter masses upto $\sim$ 0.1 GeV. 
  
	The formalism we have developed here is general and applicable to any exotic scenario that adds or extracts photons from the CMB. We also define a universal distortion parameter $u$ which depends only on the total energy injected thermally into the CMB. The $u$ parameter can be used to derive approximate constraints from COBE-FIRAS or PIXIE on extended thermal energy injection scenarios without worrying about the separation of distortions into the $y$-type, $\mu$-type, or $i$-type distortions. This works because the constraints from COBE-FIRAS/PIXIE for $y$ and $\mu$-type distortions are comparable when expressed in terms of $u$. 
	
 Note that in case of composite dark matter, the dark spectrum naturally results in multiple states where inelastic collisions play major roles both in the direct detection experiments today and as well as in determining the thermal history of the early Universe. All these effects, however, are sensitive to all of the four dark matter parameters. Typical attempts in literature constrain composite scenarios as a function of only dark matter mass or a smaller subset of these four parameters. Such studies, therefore, only probe a small set of models in the composite dark matter paradigm. 

    Finally, we observe that if $10.2\, \text{eV} \lesssim \Delta E \lesssim 24.6\,\text{eV}$, absorption by dark matter can remove the Lyman-$\alpha$ and Lyman-continuum photons from the CMB during the recombination epoch, resulting in the speed-up of the hydrogen recombination. The net result would be a smaller sound horizon during the cosmological  hydrogen recombination which would push the Hubble constant value inferred from the CMB to higher values, potentially impacting the Hubble tension. The modification of recombination physics would also put stringent constraints on the dark matter parameters. We leave the detailed study of these interesting aspects of multi-state dark matter with electromagnetic transitions for future work.
 
	\acknowledgments
	We acknowledge the computational facilities offered by the Department of Theoretical Physics at TIFR, Mumbai. 
We also acknowledge the support from Department of Atomic Energy (DAE) Government of India under Project Identification Number RTI 4002. 

\newpage
\appendix
\section{Evolution equation for the $\mu$-parameter}
\label{app:mu}
 The photon occupation number for a Bose-Einstein spectrum at temperature $T_\text{e}$ is given by,
\begin{eqnarray}
	n_{\text{BE}}(x_\text{e}) = \frac{1}{e^{x_\text{e}+\mu}-1} \approx \frac{1}{e^{x_\text{e}}-1}-\frac{\mu e^{x_\text{e}}}{\left(e^{x_\text{e}}-1\right)^2}. \label{a1}
\end{eqnarray}
where $\mu$ denotes the dimensionless chemical potential. Since we are concerned with small distortions i.e. $\mu \ll 1$, we have approximated $n_\text{BE}$ by expanding it in a Taylor series to first order in $\mu$.

We can parameterize the two variables $\mu$ and $T_\text{e}$ in terms of the fractional change in the total number density ($N_\text{BE}$) and the total energy density ($E_{\text{BE}}$) of photons as described below.
We can use Eq.\eqref{4.7} to express the ratio of the number density of photons in the Bose-Einstein spectrum at temperature $T_\text{e}$ w.r.t. a Planck spectrum at temperature $T_\text{CMB}$ as,
\begin{align}
	\mathcal{N} \equiv \frac{N_{\text{BE}}}{N_\text{PL}}
	= \frac{1}{\mathcal{I}_2}\left(\frac{T_\text{e}}{T_\text{CMB}}\right)^3\int dx_\text{e}\,x_\text{e}^2\,n_\text{BE}
	= \left(\frac{T_\text{e}}{T_\text{CMB}}\right)^3\left(1-\mu\frac{\pi^2}{3\mathcal{I}_2}\right)\,.\label{a2}
\end{align}
The ratio for the total energy density of photons in the Bose-Einstein spectrum w.r.t. a Planck spectrum is given by,
\begin{align}
	\mathcal{E} \equiv \frac{E_{\text{BE}}}{E_\text{PL}} = \frac{1}{\mathcal{I}_3}\left(\frac{T_\text{e}}{T_\text{CMB}}\right)^4\int dx_\text{e}\,x_\text{e}^3\,n_\text{BE}=\left(\frac{T_\text{e}}{T_\text{CMB}}\right)^4\left(1-\mu\frac{6\zeta(3)}{\mathcal{I}_3}\right)\,,\label{a3}
\end{align}
where $\mathcal{I}_3 = \int dx\,x^3\,n_\text{PL}= \pi^4/15$, $\mathcal{I}_2=2\zeta(3)$, and $\zeta(3)$ denotes the Riemann zeta function. 

Taking the time derivative of Eq.\eqref{a2} and Eq.\eqref{a3} and eliminating the temperature dependence results in the following equation for the time evolution of the $\mu$-parameter,
\begin{align}
	\dfrac{\text{d}\mu}{\text{d}t} = 1.4\left(\frac{\dot{\mathcal{E}}}{\mathcal{E}} - \dfrac{4}{3}\frac{\dot{\mathcal{N}}}{\mathcal{N}}\right)\,.\label{a4}
\end{align}
For small fractional changes in CMB energy and number densities, we can approximate $E_{\text{BE}} \approx E_{\text{PL}}$ and $N \approx N_{\text{PL}}$ to the lowest order. Therefore, the Eq.\eqref{a4} becomes,
\begin{align}
	\dfrac{\text{d}\mu}{\text{d}t} \approx 1.4\left(\dot{\mathcal{E}} - \dfrac{4}{3}\dot{\mathcal{N}}\right)\,.\label{a5}
\end{align}

\section{Greens function technique for the calculation of $\mathbf{\mu}$-parameter}
\label{app:A}
We use the Green's function technique to solve the boundary value problem given in Eq.\eqref{4.22} in the frequency range $x_{\text{e}} \in [0,1]$. We first do a change of variables from $\mu_x$ to $\text{G}(x_\text{e}, x_0)$,
\begin{eqnarray}
	G(x_\text{e}, x_0) = \dfrac{K_\text{C}\mu_t}{\mathcal{I}_2}\dfrac{b_RT_\text{e}^3}{\dot{N}_{\chi\gamma}}\,\mu_x = \dfrac{\mu_x}{\alpha}\,.\label{1.9a}
\end{eqnarray}
The second order differential equation for $\text{G}(x_\text{e}, x_0)$ is given by,
\begin{eqnarray}
	\dfrac{\text{d}}{\text{d}x_\text{e}}\,x_\text{e}^2\,\dfrac{\text{dG}(x_\text{e}, x_0)}{\text{d}x_\text{e}} - \dfrac{x_\text{c}^2}{x_\text{e}^2}\,\text{G}(x_\text{e}, x_0) = \delta(x_\text{e}-x_{0}(t))\,,\label{1.9}
\end{eqnarray}
The general solution of the above equation is given by,
\begin{eqnarray}
	\text{G}(x_\text{e}, x_0) = 	\text{c}_1\left(x_0\right)e^{-x_\text{c}/x_\text{e}} + \text{c}_2\left(x_0\right)e^{x_\text{c}/x_\text{e}} \hspace{6pt} \text{for} \hspace{4 pt} x_{\text{e}}\neq x_0,
\end{eqnarray}
where c$_1$ and c$_2$ are constants.

The boundary conditions on $\mu_x$ ($\mu_x(0) = 0$ and $\mu_x(1) = 1$) imply the following boundary conditions on G: $\text{G}(0, x_0) = 0$ and $\text{G}(1, x_0) = \alpha^{-1}$. This implies,
\begin{eqnarray}
	\text{G}(x_{\text{e}}, x_0)= 
	\begin{cases}
		\text{c}_1e^{-x_\text{c}/x_{\text{e}}} & x_{\text{e}} < x_0,\\
		\text{c}_3e^{-x_\text{c}/x_{\text{e}}} + \left(\alpha^{-1} - \text{c}_{3}e^{-x_\text{c}}\right)e^{-x_\text{c}}e^{x_\text{c}/x_{\text{e}}}  & x_{\text{e}} > x_0.
	\end{cases}\label{1.12}
\end{eqnarray}
We now use different properties of G to evaluate constants c$_1$ and c$_3$. A delta function discontinuity in the second derivative of G implies that its first derivative will have a jump discontinuity while $G$ itself will be continuous at $x_{\text{e}} = x_0$. The continuity of Green's function $\text{G}(x_0^-, x_0) = \text{G}(x_0^+, x_0)$ gives,  
\begin{align}
	\text{c}_1e^{-x_\text{c}/x_0} &= \text{c}_3e^{-x_\text{c}/x_0} + \left(\alpha^{-1} - \text{c}_{3}e^{-x_\text{c}}\right)e^{-x_\text{c}}e^{x_\text{c}/x_0}, \nonumber\\
	\text{c}_1 &= c_3 + \left(\alpha^{-1} - \text{c}_{3}e^{-x_\text{c}}\right)e^{-x_\text{c}}e^{2x_\text{c}/x_0}\,.\label{1.13}
\end{align}
The jump discontinuity of the first derivative of Green's function at $x_{\text{e}} = x_0$ can be derived by integrating Eq.\eqref{1.9} from $x_0^- = x_0 - \epsilon$ to $x_0^+ = x_0 + \epsilon$ and taking the limit $\epsilon \rightarrow 0$,
\begin{align}
	\dfrac{\text{d}\text{G}(x_0^+, x_0)}{\text{d} x_{\text{e}}} - \dfrac{\text{d}\text{G}(x_0^-, x_0)}{\text{d} x_{\text{e}}} = \dfrac{1}{x_0^2}\,.\label{1.15}
\end{align}
After substituting Eq.\eqref{1.12} in \eqref{1.15} and using \eqref{1.13},
\begin{align}
	\text{c}_3e^{-x_\text{c}/x_0} - \left(\alpha^{-1} - \text{c}_{3}e^{-x_\text{c}}\right)e^{-x_\text{c}}e^{x_\text{c}/x_0} - c_3e^{-x_\text{c}/x_0} - \left(\alpha^{-1} - \text{c}_{3}e^{-x_\text{c}}\right)e^{-x_\text{c}}e^{x_\text{c}/x_0} = \dfrac{1}{x_\text{c}},
\end{align}
\begin{align}
	\text{c}_3 = \alpha^{-1}\,e^{x_\text{c}} + \dfrac{e^{2x_\text{c}}e^{-x_\text{c}/x_0}}{2x_\text{c}}. \label{1.14}
\end{align}
The solution for Green's function is given by,
\begin{align}
	\text{G}(x_{\text{e}}, x_0) = 
	\begin{cases}
		\left(\alpha^{-1}\,e^{x_\text{c}} + \dfrac{e^{2x_\text{c}}e^{-x_\text{c}/x_0}}{2 x_\text{c}} - \dfrac{e^{x_\text{c}/x_0}}{2x_\text{c}}\right)e^{-x_\text{c}/x_{\text{e}}} & x_{\text{e}} \leq x_0,\\
		\left(\alpha^{-1}\,e^{x_\text{c}} + \dfrac{e^{2x_\text{c}}e^{-x_\text{c}/x_0}}{2x_\text{c}}\right)e^{-x_\text{c}/x_{\text{e}}} - \dfrac{e^{-x_\text{c}/x_0}}{2x_\text{c}}e^{x_\text{c}/x_{\text{e}}}  & x_{\text{e}} > x_0.
	\end{cases}
\end{align}
The solution for $\mu_x$ can be found using Eq.\eqref{1.9a},
\begin{align}
	\mu(x_{\text{e}}, x_0) = 
	\begin{cases}
		\left(e^{x_\text{c}} + \alpha\dfrac{e^{2x_\text{c}}e^{-x_\text{c}/x_0}}{2 x_\text{c}} - \alpha\dfrac{e^{x_\text{c}/x_0}}{2x_\text{c}}\right)e^{-x_\text{c}/x_{\text{e}}} & x_{\text{e}} \leq x_0,\\
		\left(e^{x_\text{c}} + \alpha\dfrac{e^{2x_\text{c}}e^{-x_\text{c}/x_0}}{2x_\text{c}}\right)e^{-x_\text{c}/x_{\text{e}}} - \alpha\dfrac{e^{-x_\text{c}/x_0}}{2x_\text{c}}e^{x_\text{c}/x_{\text{e}}}  & x_{\text{e}} > x_0.
	\end{cases}\label{1.16}
\end{align}
The total number density of photons in the spectrum is given by,
\begin{equation}
	N = \dfrac{b_R T_{\text{e}}^3}{\mathcal{I}_2}\int_{0}^{\infty} dx_{\text{e}} \,x_{\text{e}}^2\, n(x_{\text{e}}, t), \label{1.18}
\end{equation}
We define $N_{\text{PL}} \equiv b_RT^3$ and take time derivative of Eq.\eqref{1.18} which gives the evolution of total photon number density,
\begin{align}
	\mathcal{I}_2\dfrac{\text{d}}{\text{d} t}\left(\dfrac{N}{N_{\text{PL}}}\right) 
	= \left(\dfrac{T_{\text{e}}}{T}\right)^3\int dx_{\text{e}} \,x_{\text{e}}^2 \,\dfrac{\partial n}{\partial t} + 3\left(\dfrac{T_{\text{e}}}{T}\right)^3\dfrac{\partial}{\partial t}\left[\ln \dfrac{T_{\text{e}}}{T}\right]\int dx_{\text{e}}\, x_{\text{e}}^2 \,n, 	
\end{align}
After some simplification and on defining $\mathcal{N} = N/N_{\text{PL}}$, we get,
\begin{align}
	\dot{\mathcal{N}} = \dfrac{1}{\mathcal{I}_2}\left(\dfrac{T_{\text{e}}}{T}\right)^3\int dx_{\text{e}}\, \left(K_{\text{br}} + 	K_{\text{dC}}\right)\dfrac{e^{-x_{\text{e}}}}{x_{\text{e}}}\left[1-n(e^{x_{\text{e}}}-1)\right] + \dfrac{\dot{N}_{\chi\gamma}}{N_{\text{PL}}}, \label{1.19}
\end{align}
Assuming $\mu\ll 1$ and $T_{\text{e}} \approx T$ and ignoring $x_{\text{e}}$ dependence of gaunt factors in $K_{\text{br}}$ and $K_{\text{dC}}$, we can substitute Eq.\eqref{4.19} into Eq.\eqref{1.19},
\begin{align}
	\dot{\mathcal{N}} &\approx \dfrac{\mu_t}{\mathcal{I}_2}\left(K_{\text{br}} + 	K_{\text{dC}}\right)\int_0^\infty dx_{\text{e}}\,\dfrac{\mu_x\,e^{-x_{\text{e}}}}{x_{\text{e}}\left(e^{x_{\text{e}}}-1\right)} + \dfrac{\dot{N}_{\chi\gamma}}{N_{\text{PL}}}\,.\label{1.20}
\end{align}
We solve the above integral by substituting the solution for $\mu_x$ in $0\leq x_{\text{e}} \leq 1$ from Eq.\eqref{1.16} and take $\mu_x = 1$ at $x_{\text{e}} \geq 1$  into Eq.\eqref{1.20},
\begin{align}
	\int_{0}^{\infty} dx_{\text{e}}\,\dfrac{\mu_x\,e^{-x_{\text{e}}}}{x_{\text{e}}\left(e^{x_{\text{e}}}-1\right)} 
	&= \int_{0}^{x_0}dx_{\text{e}}\,\dfrac{\mu_x\left(x_{\text{e}} \leq x_0\right)\,e^{-x_{\text{e}}}}{x_{\text{e}}\left(e^{x_{\text{e}}}-1\right)}  + \int_{x_0}^{1}dx_{\text{e}}\,\dfrac{\mu_x\left(x_{\text{e}} > x_0\right)\,e^{-x_{\text{e}}}}{x_{\text{e}}\left(e^{x_{\text{e}}}-1\right)} + \nonumber\\ &\;\int_{1}^{\infty}dx_{\text{e}}\, \dfrac{e^{-x_{\text{e}}}}{x_{\text{e}}\left(e^{x_{\text{e}}}-1\right)}. \label{1.21}	
\end{align}
We further approximate $\left(e^{x_{\text{e}}}-1\right)\approx x_{\text{e}}$ in the first two integrands and denote the last integral as $\delta \mathcal{N}$. This allows us to analytically solve the integrals in Eq.\eqref{1.21}.
\begin{align}
	\int_{0}^{1} dx_{\text{e}}\,\dfrac{\mu_x\,e^{-x_{\text{e}}}}{x_{\text{e}}\left(e^{x_{\text{e}}}-1\right)} 
	\approx \left(e^{x_\text{c}} + \alpha\dfrac{e^{2x_\text{c}}e^{-x_\text{c}/x_0}}{2 x_\text{c}} - \alpha\dfrac{e^{x_\text{c}/x_0}}{2x_\text{c}}\right)\dfrac{e^{-x_\text{c}/x_0}}{x_\text{c}} + \nonumber\\
	\left(e^{x_\text{c}} + \alpha\dfrac{e^{2x_\text{c}}e^{-x_\text{c}/x_0}}{2x_\text{c}}\right)\dfrac{e^{-x_\text{c}}-e^{-x_\text{c}/x_0}}{x_\text{c}} + \alpha\dfrac{e^{-x_\text{c}/x_0}}{2x_\text{c}^2}\left(e^{x_\text{c}} - e^{x_\text{c}/x_0}\right)\nonumber\\
	= \dfrac{1}{x_\text{c}}-\dfrac{\alpha}{x_\text{c}^2} + \alpha\,e^{x_\text{c}}\dfrac{e^{-x_\text{c}/x_0}}{x_\text{c}^2}.
\end{align}
After substituting the above integrals in Eq.\eqref{1.20}, we find, 
\begin{align}
	\dot{\mathcal{N}} &\approx \dfrac{\mu_t}{\mathcal{I}_2}\dfrac{\left(K_{\text{br}} + 	K_{\text{dC}}\right)}{x_\text{c}}\left(1+x_\text{c}\,\delta\mathcal{N}\right)+ \dfrac{\dot{N}_{\chi\gamma}}{N_{\text{PL}}}e^{x_\text{c}}e^{-x_\text{c}/x_0},\nonumber\\
	&\approx \dfrac{\mu_t}{\mathcal{I}_2}\dfrac{\left(K_{\text{br}} + 	K_{\text{dC}}\right)}{x_\text{c}}+ \dfrac{\dot{N}_{\chi\gamma}}{N_{\text{PL}}}e^{x_\text{c}}e^{-x_\text{c}/x_0}.
\end{align}
In the second line we ignore the small contribution from the term $x_\text{c}\,\delta \mathcal{N} \ll 1$. 
\\\\
The formula for the blackbody optical depth $\mathcal{T}$ is given by \cite{2012JCAP...06..038K},
\begin{align}
	\mathcal{T}(z) \approx \left[\left(\frac{1+z}{1+z_\text{dC}}\right)^5 + \left(\frac{1+z}{1+z_\text{br}}\right)^{5/2} \right]^{1/2} + \varepsilon \ln\left[\left(\frac{1+z}{1+z_\varepsilon}\right)^{5/4}+\sqrt{1+\left(\frac{1+z}{1+z_\varepsilon}\right)^{5/2}}\right],
	\label{bb_od}
\end{align}
where,
\begin{align}
	z_\text{dC}&=1.96 \times 10^6\nonumber\\
	z_\text{br}&=1.05 \times 10^7\nonumber\\
	z_\varepsilon&=3.67 \times 10^5\nonumber\\
	\varepsilon&=0.0151\nonumber
\end{align}
\section{Evolution equation for the $y$-parameter}
\label{app:y}
In the regime where inelastic processes are successful in negating the change in the photon number due to dark matter, the net result is a change in the energy of baryons and electrons.  This is followed by Compton scattering which transfers energy between the CMB photons and the baryons. To estimate the amplitude of $y$-type distortion, we will keep only the Compton scattering term in Eq.\eqref{4.12}. At redshifts $<z_\text{c}$, the Compton scattering between electrons and ions is not efficient ($K_\text{C}/H \ll 1$) in bringing the CMB temperature in equilibrium with the electron temperature. As a zeroth order approximation, we can assume the
resultant spectrum to be a Planck spectrum $n_{\text{PL}}$ at a temperature different from the electron temperature $T_{\text{CMB}}\neq T_\text{e}$. The solution at the next order can be obtained by plugging the zeroth order solution into the RHS of Eq.\eqref{4.12} just keeping the Compton scattering term. After doing a change of variables from $x_\text{e}$ to $x\equiv h\nu/(k_\text{B}T_\text{CMB})$ and from $dt$ to $dy_\gamma$, where $dy_\gamma  = n_\text{e}\sigma_\text{T}c k_{B}T_\text{CMB}/m_\text{e}c^2 dt$, Eq.\eqref{4.12} simplifies to \cite{1969Ap&SS...4..301Z},
\begin{align}
	\frac{\partial n}{\partial y_\gamma} \approx \frac{1}{x^2}\dfrac{\partial}{\partial x}x^4\left(n_\text{PL}+n_\text{PL}^2+\frac{T_\text{e}}{T_{\text{CMB}}}\dfrac{\partial n_\text{PL}}{\partial x}\right) = \left(\frac{T_\text{e}}{T_\text{CMB}}-1\right)n_y(x)\,,\label{b1}
\end{align}
where $n_y$ is called the $y$-type spectrum and is given by,
\begin{equation}
	n_y(x) = \frac{xe^{-x}}{(e^x-1)^2}\bigg[x\left(\frac{e^x+1}{e^x-1}\right)-4\bigg].
\end{equation}
The resultant photon spectrum can be obtained by integrating the RHS of Eq.\eqref{b1} w.r.t. $y_\gamma$, which gives,
\begin{align}
	n(x) = n_{\text{PL}}(x) + y n_y(x), \label{b3}
\end{align}
where $y$ represents the amplitude of $y$-type distortion and is defined as,
\begin{equation}
	y = \int_{0}^{y_\gamma}dy_\gamma\left(\frac{T_\text{e}}{T_\text{CMB}}-1\right)\,.\label{b4}
\end{equation}
Note that the $y$-type distortion only changes the total energy density and does not affect the total number density of CMB photons. We can express the $y$ parameter in terms of the rate of change of photon energy density on integrating Eq.\eqref{b3} over frequency after multiplying it by $d^3\nu$ giving, 
\begin{align}
	\dfrac{\text{d}y}{\text{d}t} \equiv \frac{1}{4}\dot{\mathcal{E}} \,.\label{b5}
\end{align}
\section{Thermal history for different spectral distortions}
\label{app:thermal_history}
\begin{figure}[t]
	\centering
	\includegraphics[width=0.5\textwidth]{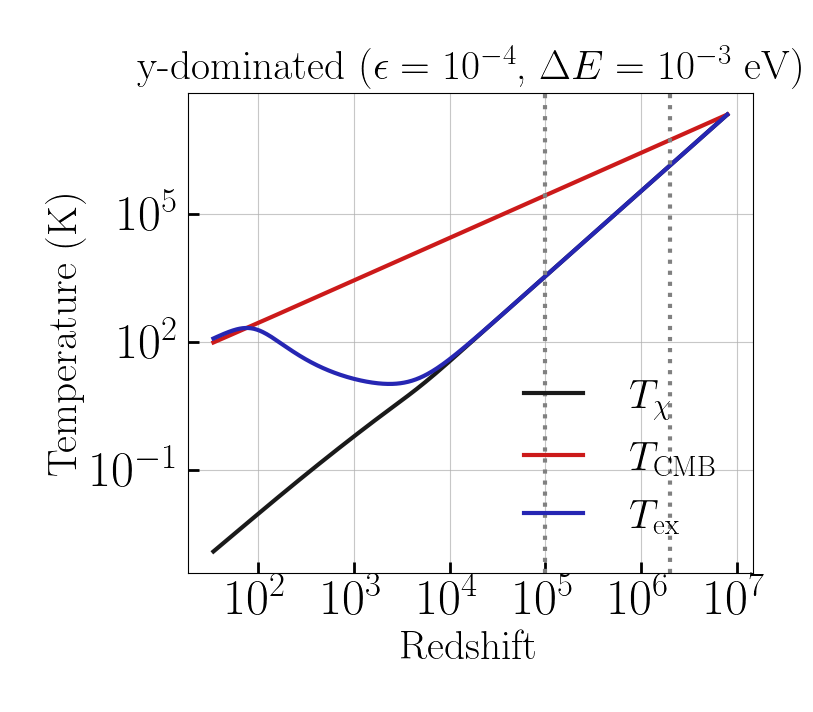}
	\caption{Thermal history of dark matter for model parameters where $y$-type distortion dominates: $m_\chi = 1$ MeV, $\epsilon = 10^{-4} $, $\Delta E = 10^{-3}$ eV, and $\alpha_\text{C}=1$. The $\mu$-era is represented by the grey dotted lines.}
	\label{fig:y_thermal_history}
\end{figure}

\begin{figure}[t]
	\hspace{-30pt}
	\includegraphics[width=1.1\textwidth]{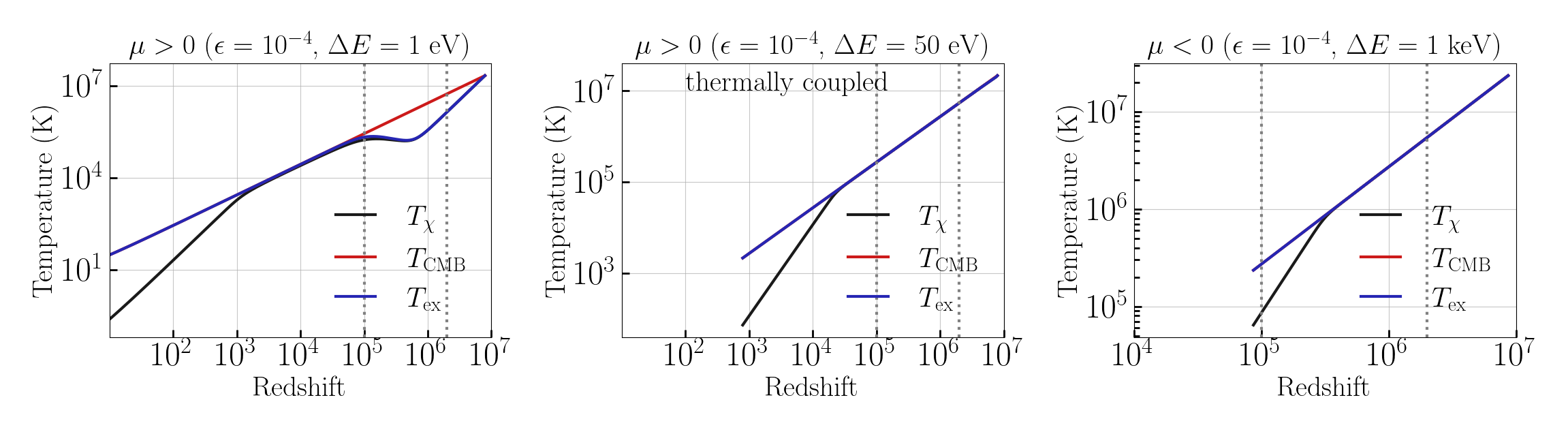}
	\caption{Thermal history of dark matter for model parameters where $\mu$-type distortion (created in the redshift range marked by the gray dotted lines) dominates for $m_\chi = 1$ MeV and $\alpha_C=1$.}
	\label{fig:mu_thermal_history}
\end{figure}

\begin{figure}[t]
	\centering
	\includegraphics[width=0.5\textwidth]{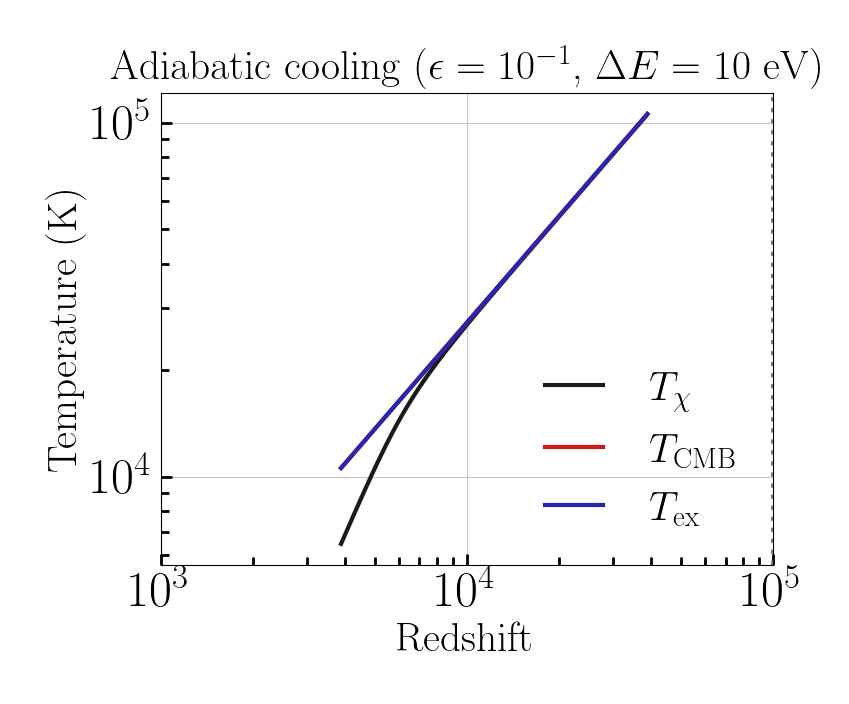}
	\caption{Thermal history of dark matter for model parameters where the electromagnetic scattering of dark matter particles with electrons/ions keeps dark matter thermally coupled to the CMB in the $\mu$-era (represented by gray dotted lines) for: $m_\chi = 1$  MeV, $\epsilon = 10^{-1} $, $\Delta E = 10$ eV, and $\alpha_C=1$.}
	\label{fig:sc_thermal_history}
\end{figure}

We show the thermal history of dark matter in the different spectral distortion regimes shown in Fig.\ref{fig:u1_phys}. To solve for the evolution of the dark matter temperature and the excitation temperature, we initialize $T_\text{ex} = T_\chi = T_\text{CMB}$ at the redshift at which the electromagnetic scattering rate between dark matter and baryons equals the Hubble expansion rate ($\Gamma_{\chi_\text{-SM}}/H\approx 1$). This implies that above the redshift of initialization at which the temperature is evolved the dark matter stays kinetically coupled to the CMB. The regime where $y$- distortions dominate the universal distortion parameter $u$ is shown in Fig.\ref{fig:y_thermal_history}. The thermal history in the case where $\mu$- distortions dominate is shown in the three panels of Fig.\ref{fig:mu_thermal_history}. The thermal evolution where the thermal distortion is created by the electromagnetic scattering of dark matter particles with baryons and electrons in shown in Fig.\ref{fig:sc_thermal_history}. 
\section{Dependence of thermal spectral distortions on the dark matter mass and the collision parameter}
	\begin{figure}[t]
	\hspace{-10pt}\includegraphics[width=0.53\textwidth]{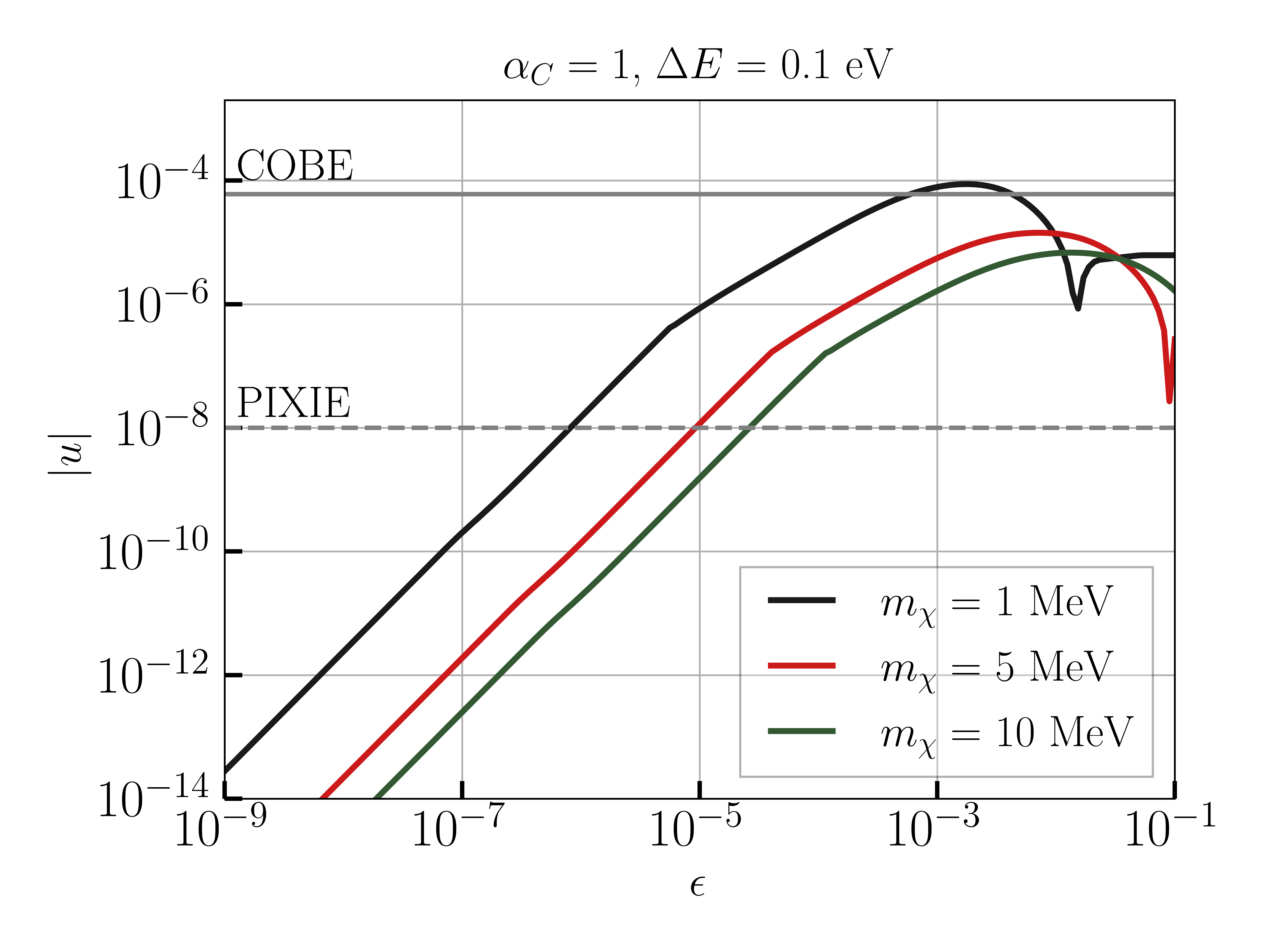}\hspace{-12pt}
	\includegraphics[width=0.53\textwidth]{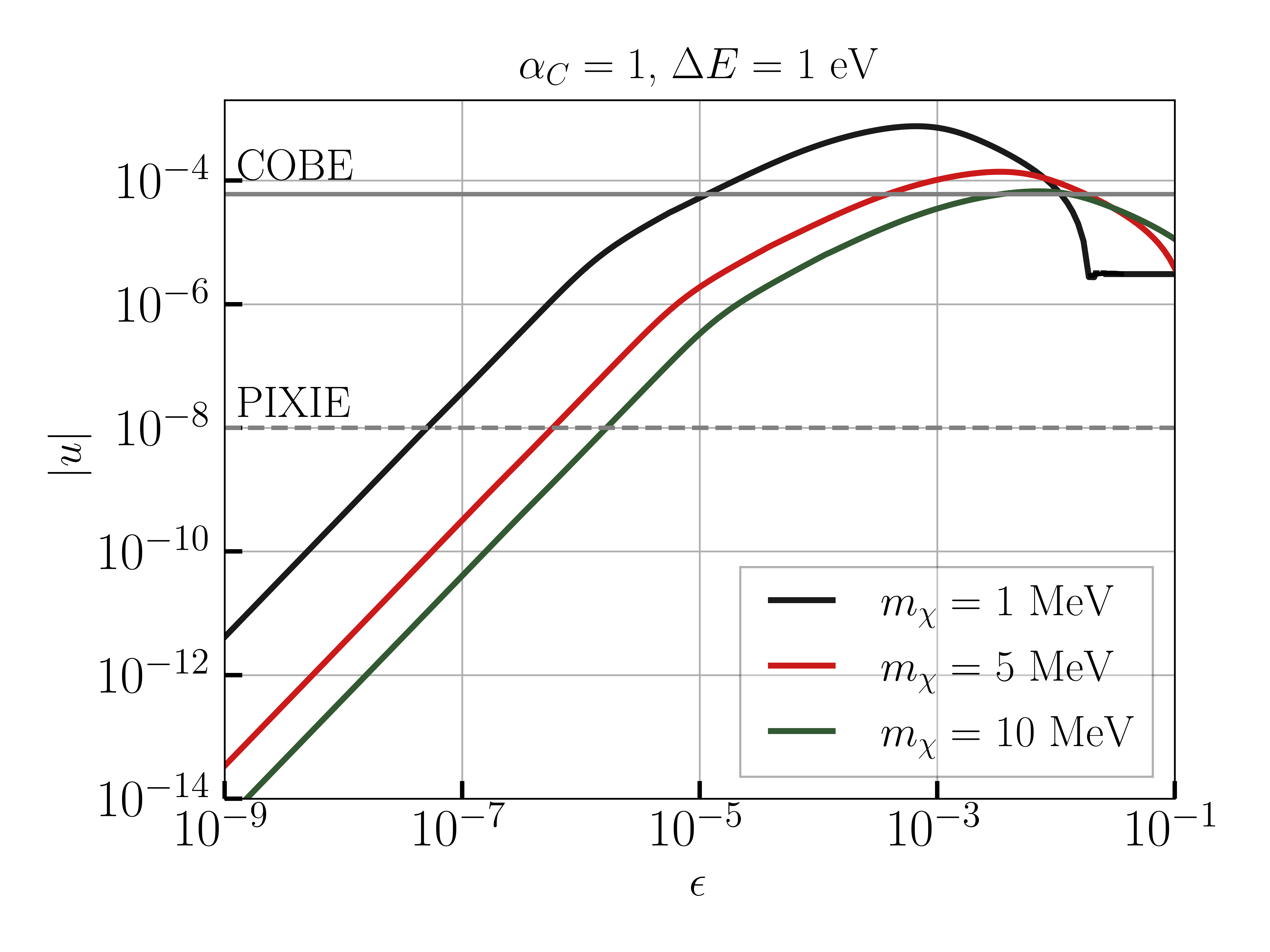}
	\centerline{\includegraphics[width=0.53\textwidth]{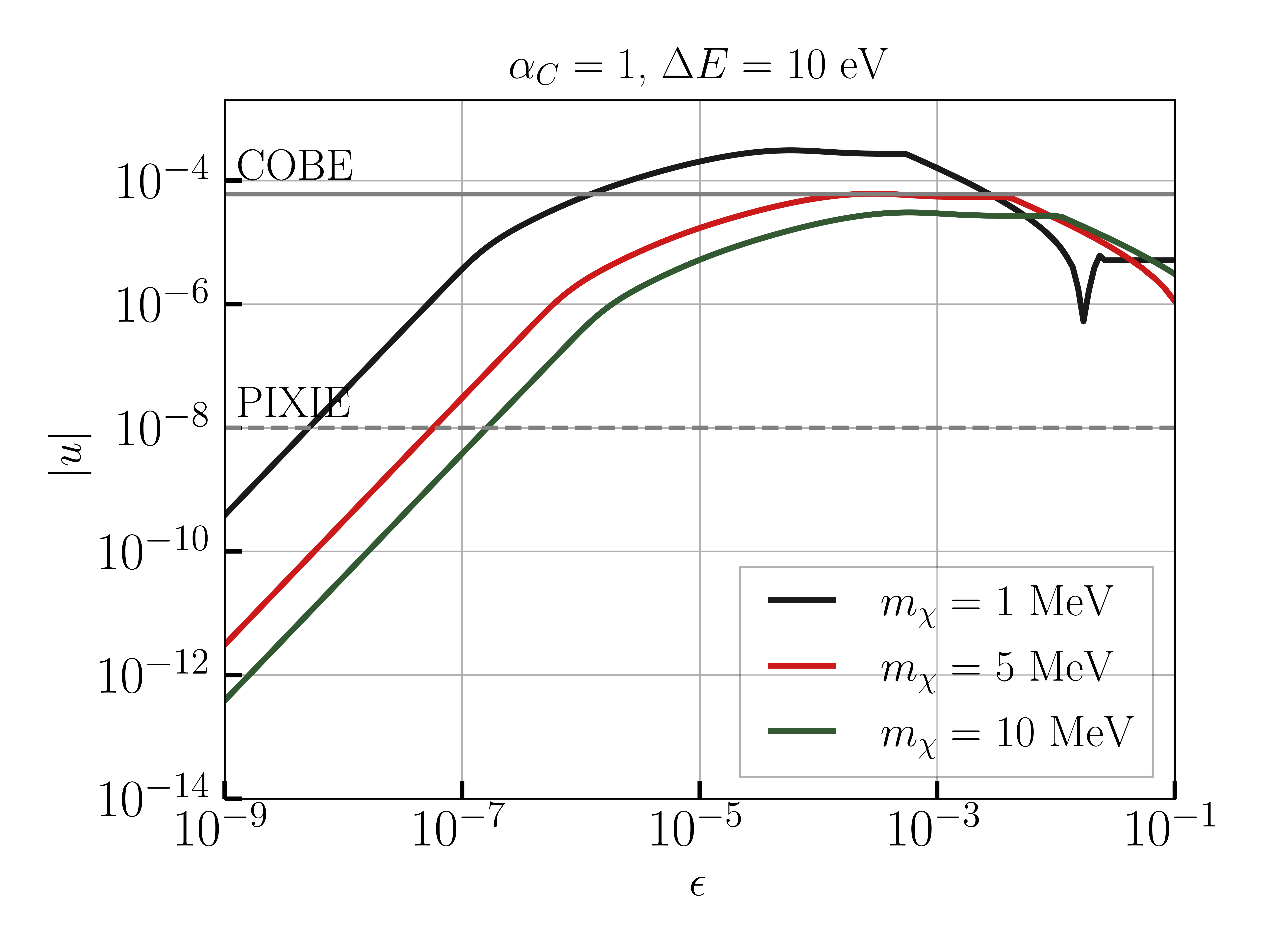}}
	\caption{Variation of the universal distortion parameter $u$ as a function of $\epsilon$ for different $m_\chi$ and $\Delta E$. The grey solid and dashed lines represent the 2-$\sigma$ limit on $u$ from COBE-FIRAS and PIXIE respectively.}
	\label{fig:mass_var_plot}
\end{figure}
We show the amplitude of the thermal distortion parameter $u$ as a function of $\epsilon$ at transition energies $\Delta E = 0.1, 1$, and $10$ eV in the three panels of Fig.\ref{fig:mass_var_plot} for $m_\chi = 1, 5$, and 10 MeV. Consider the nature of spectral distortions in the first panel of Fig.\ref{fig:mass_var_plot} for $\Delta E = 0.1$ eV. At low values of $\epsilon$, the spectral distortions are created by the line absorption of CMB photons by dark matter. The value of $|u|$ is higher for smaller dark matter masses (higher number density of dark matter particles to absorb CMB photons) as well as for a stronger radiative transition rate. The radiative transition rate being $\propto (\epsilon/m_\chi)^2$ is higher for larger $\epsilon$ and for smaller dark matter mass. This causes an initial rise in $|u|$ with $\epsilon$ for a fixed $m_\chi$. Note that $u$ is positive here as the change in the number density dominates over the change in the energy density during the $\mu$-type era. The $y$-type distortions are negligible in this case. As $\epsilon$ is increased further, the effect of energy transfer from electromagnetic scattering of baryons and dark matter starts cancelling the contribution from line absorption causing the amplitude of $u$ to decrease. For even higher $\epsilon$, the dark matter stays coupled to the CMB during $\mu$ and $y$ eras just by electromagnetic scattering which results in a constant value of $u$ independent of $\epsilon$ with the sign of $u$ becoming negative causing a kink-like feature when we plot the absolute value $|u|$. Since both dark matter number density and radiative transition rate are lower for higher dark matter masses, the resultant $u$ is smaller as $m_\chi$ is increased from 1 MeV to 10 MeV.

The second panel of Fig.\ref{fig:mass_var_plot} shows the evolution of $|u|$ w.r.t. $\epsilon$ for $\Delta E = 1$ eV. A higher radiative transition rate $\sim (\Delta E)^3$ for larger $\Delta E$ causes stronger distortions for $\Delta E = 1$ eV case compared to $\Delta E = 0.1$ eV. As $\Delta E$ is increased to 10 eV in the third panel of Fig.\ref{fig:mass_var_plot}, the spectral distortions are stronger in the rising edge compared to the top two panels. However, at intermediate values of $\epsilon$, the line absorption brings the dark matter in thermal equilibrium with the CMB and the resultant $u$ is saturated to a fixed value. It then decreases when adiabatic cooling from scattering starts contributing and changes sign.
\begin{figure}[t]
	\hspace{-10pt}\includegraphics[width=0.53\textwidth]{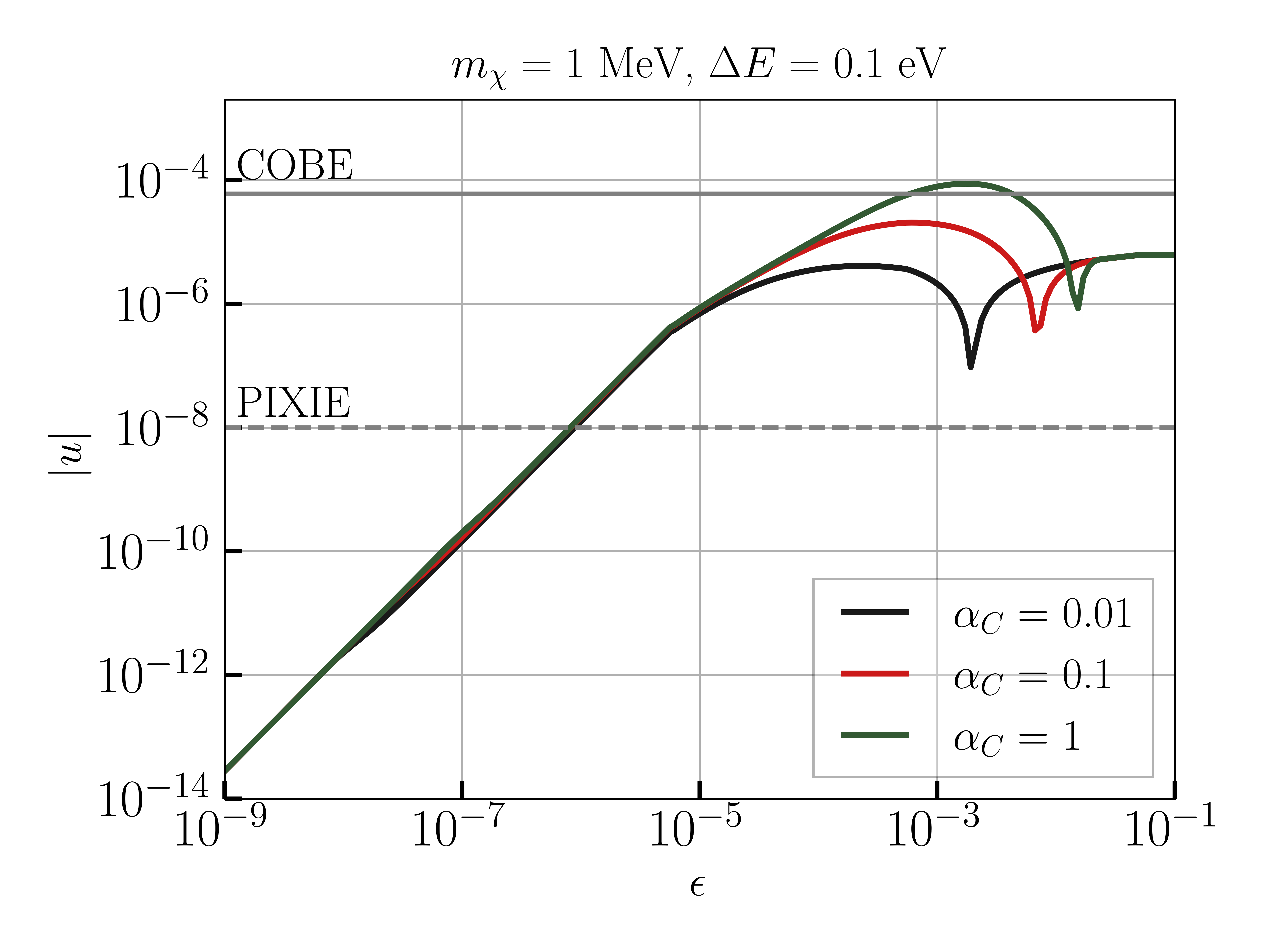}\hspace{-12pt}
	\includegraphics[width=0.53\textwidth]{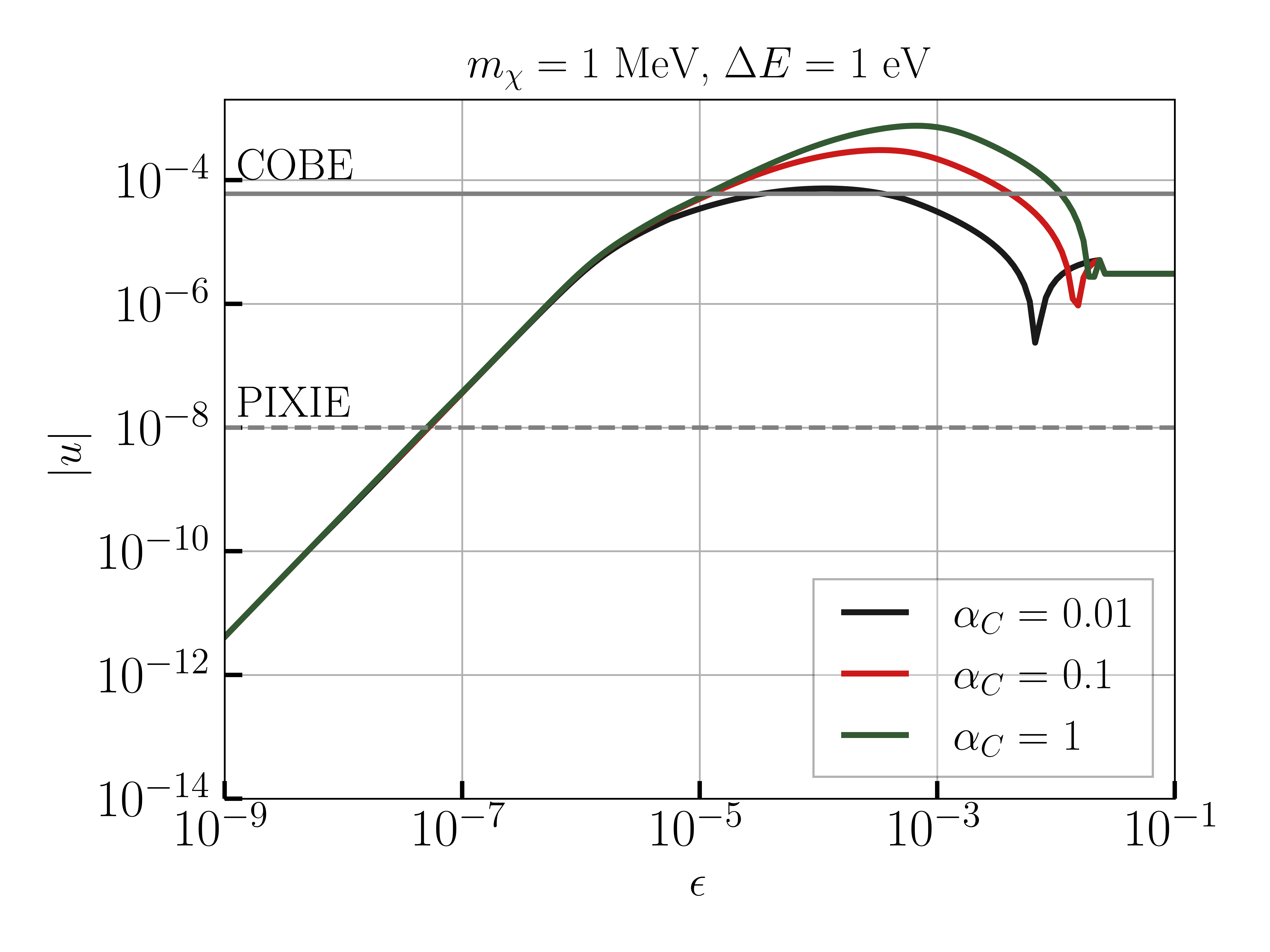}
	\centerline{\includegraphics[width=0.53\textwidth]{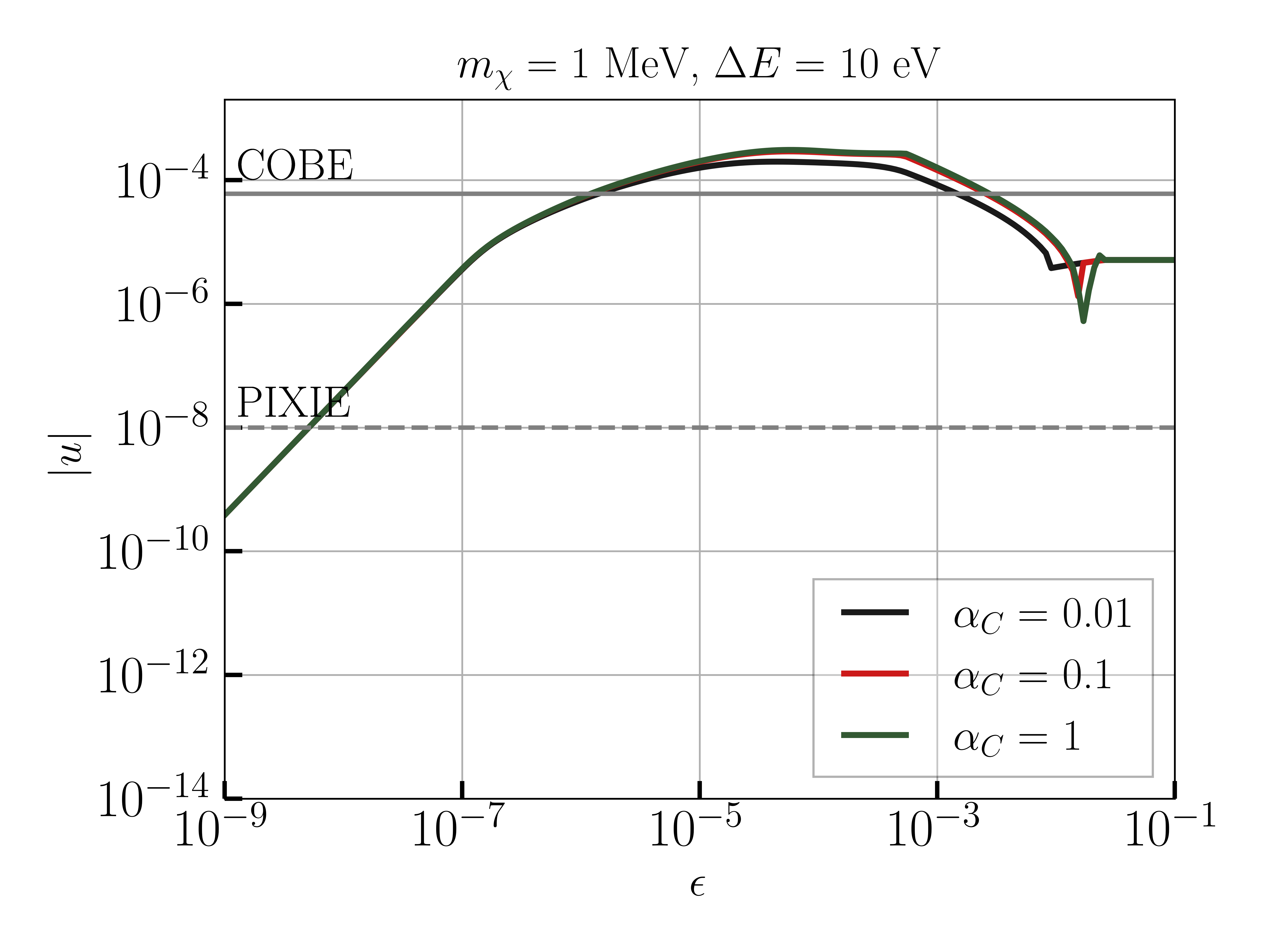}}
	\caption{Variation of the universal distortion parameter $u$ as a function of $\epsilon$ for different $\alpha_\text{C}$ and $\Delta E$. The grey solid and dashed lines represent the 2-$\sigma$ limit on $u$ from COBE-FIRAS and PIXIE respectively.}
	\label{fig:coll_var_plot}
\end{figure}

We show the amplitude of the thermal distortion parameter $u$ as a function of $\epsilon$ at transition energies $\Delta E = 0.1, 1$, and $10$ eV in the three panels of Fig.\ref{fig:coll_var_plot} for $\alpha_C = 0.01, 0.1$, and 1. At small values of $\epsilon$, the collisional transition rate dominates over the radiative transition rate keeping the excitation temperature coupled to the dark matter temperature i.e. $T_\text{ex}=T_\chi$. For small values of $\epsilon$, the change in $\alpha_C$ does not alter the thermal history and the resultant distortion $u$ therefore only depends on the radiative transition rate and increases with $\epsilon$. As $\epsilon$ increases further, the radiative transition rate starts becoming comparable to the collisional transition rate in the $\mu$-era. For smaller values of $\alpha_C$, this happens at a smaller value of $\epsilon$. In this case, the redshift range in which line absorption occurs shrinks resulting in a smaller value of $u$ with higher $\epsilon$. As $\epsilon$ is increased further, the adiabatic cooling starts cancelling the distortion due to line absorption. This makes $u$ smaller at higher $\epsilon$ values. The quantity $|u|$ becomes constant at very large $\epsilon$ when the distortion is entirely caused by adiabatic cooling of dark matter. The behavior with higher $\Delta E$ is the same as Fig.\ref{fig:mass_var_plot}.

\section{Direct detection bounds from solar reflected dark matter}
\label{app:es}
\begin{figure}[t]
	\centering
	\includegraphics[width=0.495\textwidth]{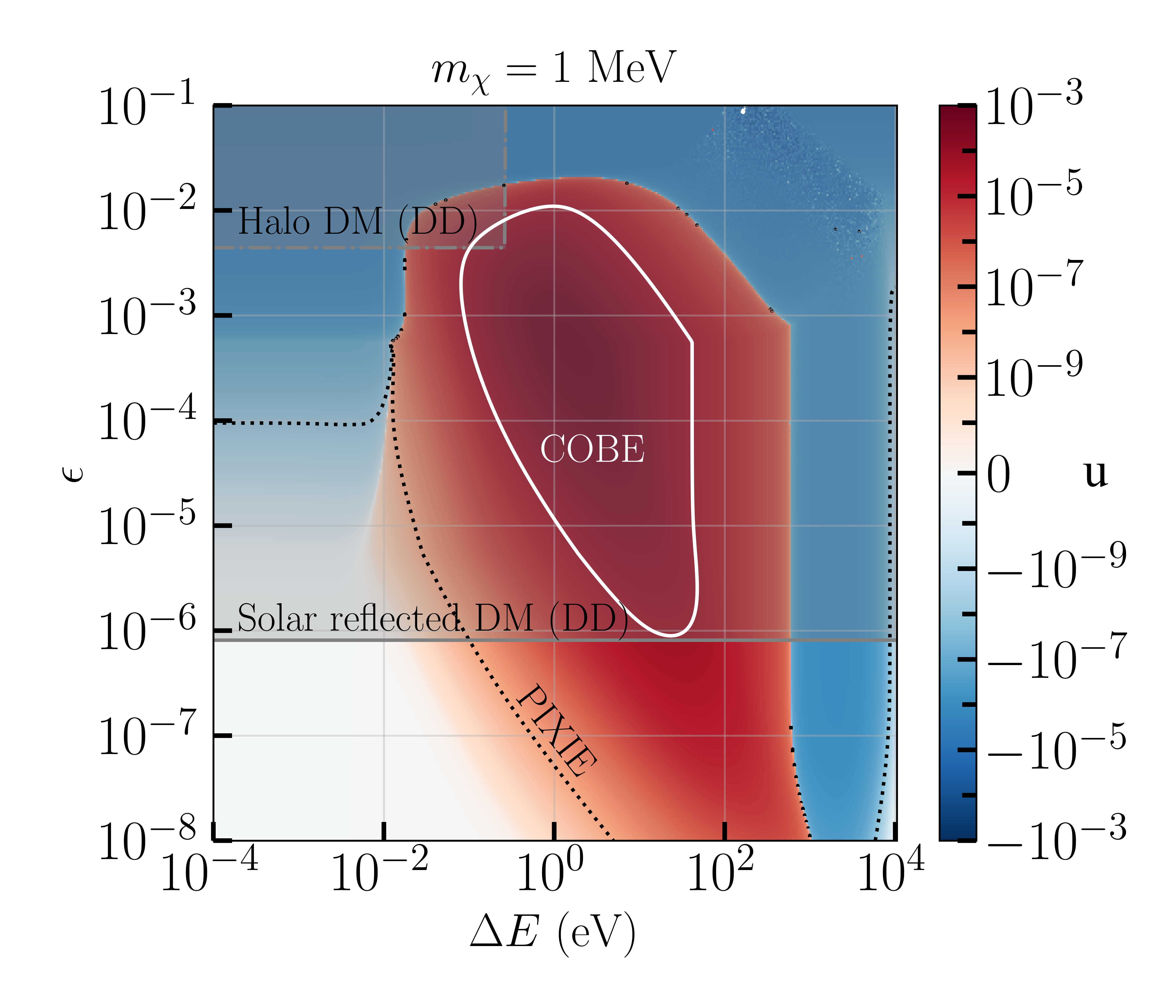}
	\includegraphics[width=0.495\textwidth]{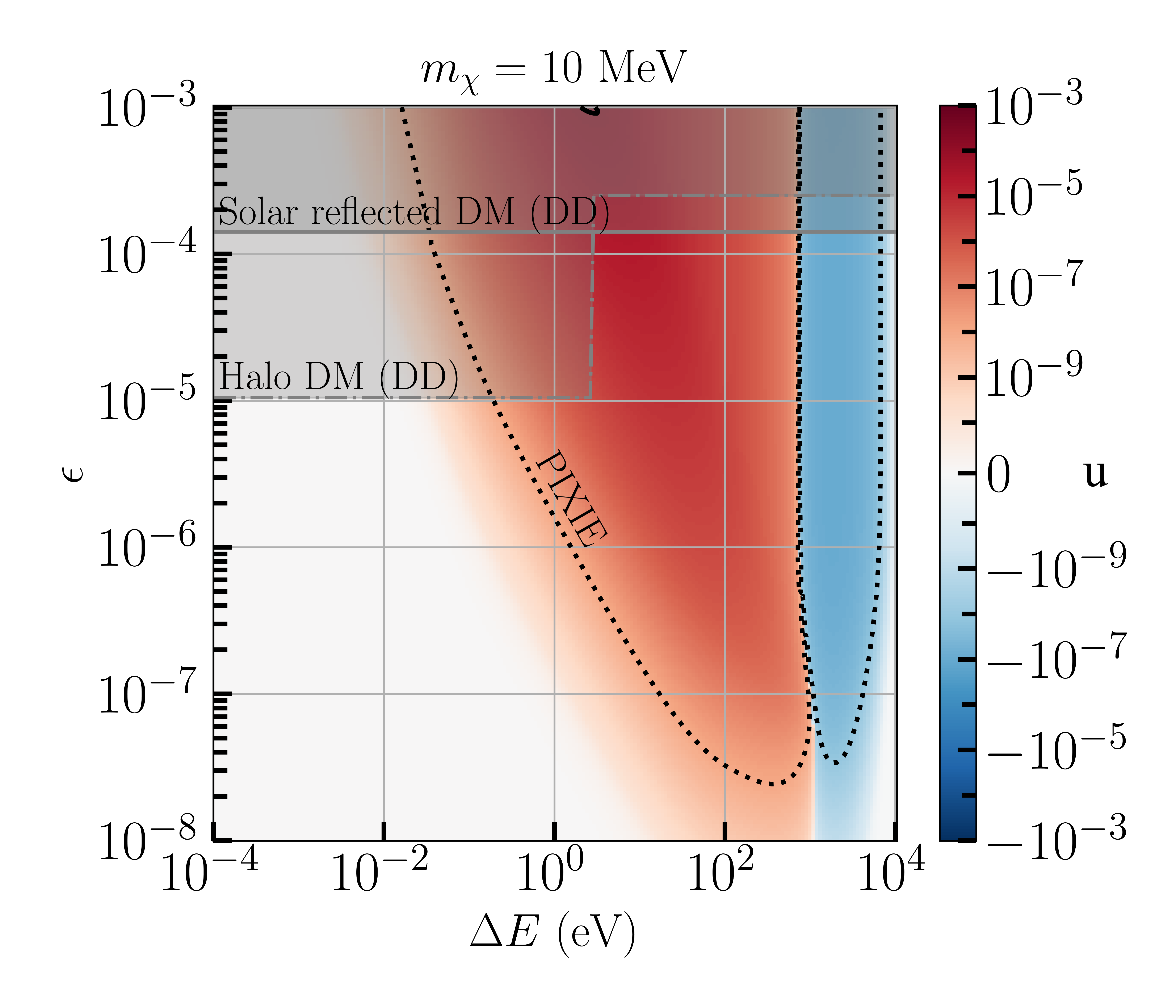}
	\includegraphics[width=0.495\textwidth]{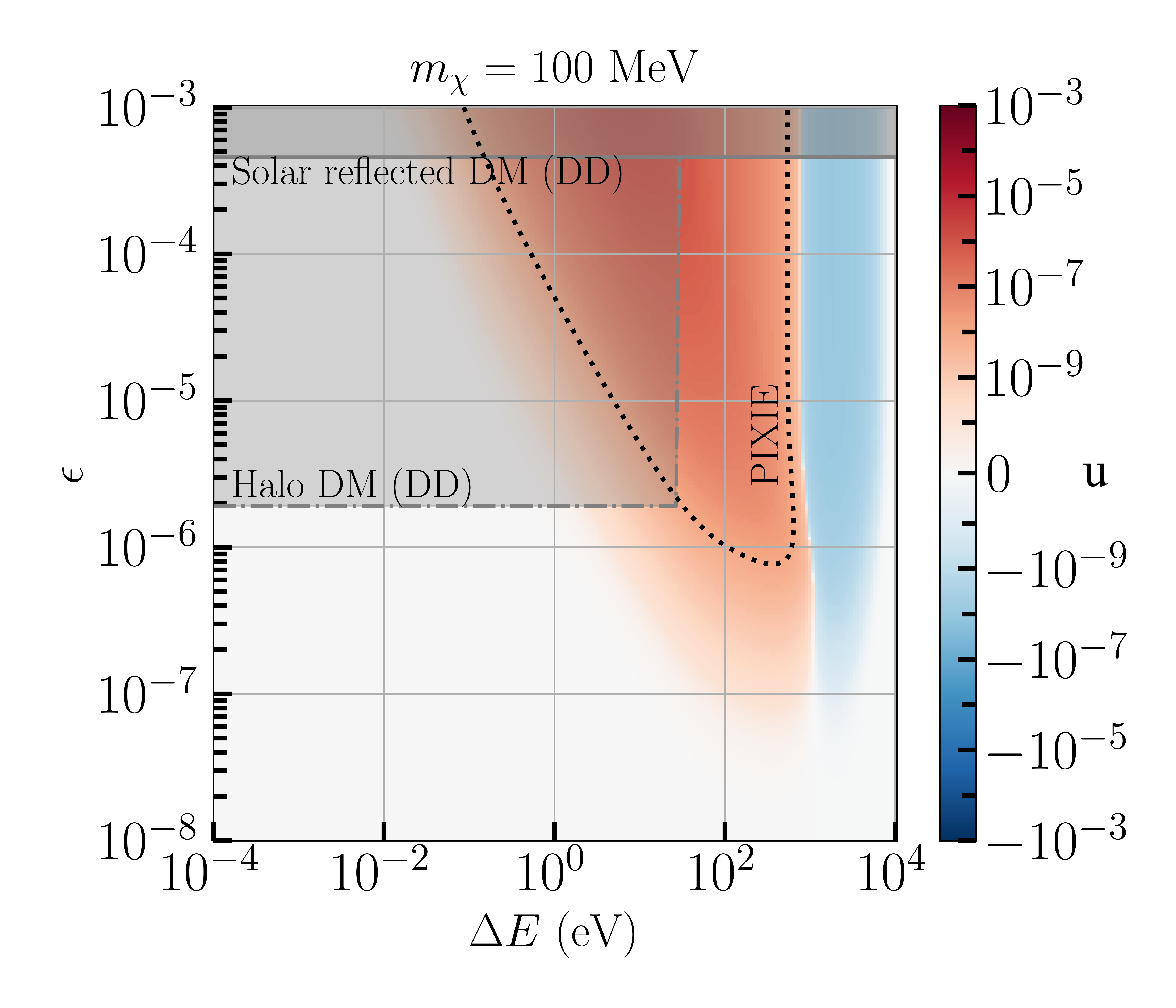}
	\caption{Same as Fig.\ref{fig:u1} with additional constraints from solar-reflected dark matter and charge radius operator.}
	\label{fig:u2}
\end{figure}

In calculating the limits from direct detection, we have considered the magnetic dipole transition operator which is responsible for both photon transition as well as inelastic scattering with electrons in direct detection experiments. The sensitivity of such experiments can reach sub-MeV masses if dark matter first scatters with highly energetic electrons/ions in the Sun and gets boosted in energy. This possibility has been considered in \cite{2017JCAP...10..037B,2018PhRvD..97f3007E,2021JHEP...04..282C,2021PhRvD.104j3026A,2022PhRvD.105f3020E} where sub-MeV dark models scatter with electrons and ions via heavy and light mediators. The boosted dark matter flux from the Sun depends on the part of the flux directed towards the Earth which in turn depends on the angular dependence of the scattering cross-section. Furthermore, in the case of inelastic scattering, some part of the dark matter energy goes into exciting the dark matter instead of boosting it. Due to these complexities, an exact calculation would require a more detailed simulation folding in the model specifications as well as the full kinematics of scattering which we leave for future work. Nonetheless, we do an approximate analysis where we consider only elastic scattering assuming the strength of the charge radius operator to be comparable to the dipole transition operator. The results are presented in Fig.\ref{fig:u2}. We do a conservative analysis and assume the normalization of both the inelastic and inelastic operators to be the same and show constraints from direct detection experiments due to solar reflection \cite{2017JCAP...10..037B,2018PhRvD..97f3007E,2021JHEP...04..282C,2021PhRvD.104j3026A,2022PhRvD.105f3020E} as well as from Milky Way halo for $m_\chi = 1, 10, 100$ MeV in the Fig.\ref{fig:u2}.
Even the much stronger solar-reflected dark matter constraints are comparable with the constraints from the COBE-FIRAS experiment in a large part of the parameter space. A future experiment like PIXIE promises to probe these constraints by many orders of magnitude.

\newpage
\bibliography{reference2.bib}
\bibliographystyle{unsrtads}
\end{document}